\documentclass[aps,pre,10pt,twocolumn,preprintnumbers,tightenlines,amsmath]{revtex4}
\usepackage{lfmath} 
\usepackage{times}
\usepackage{mathptmx}
\usepackage{amsmath}
\usepackage{amsfonts}
\usepackage{pstricks}
\usepackage{pst-text}
\psset{unit=10pt,labelsep=5pt}

\begin{document}

\title{Multi-locality and fusion rules on the generalized structure functions in two-dimensional and three-dimensional Navier-Stokes turbulence}
\author{Eleftherios Gkioulekas}
\affiliation{School of Mathematical and Statistical Sciences, University of Texas  Rio Grande Valley, Edinburg, TX, United States}
\email{drlf@hushmail.com}

\begin{abstract}
Using the fusion rules hypothesis for three-dimensional and two-dimensional Navier-Stokes turbulence, we generalize a previous non-perturbative locality proof to  multiple applications of the nonlinear interactions operator on generalized structure functions of velocity differences. We shall call this generalization of non-perturbative locality to multiple applications of the nonlinear interactions operator ``multilocality''. The resulting cross-terms pose a new challenge requiring a new argument and the introduction of a new fusion rule that takes advantage of rotational symmetry.  Our main result is that the fusion rules hypothesis implies both locality and multilocality in both the IR and UV limits for the downscale energy cascade of three-dimensional Navier-Stokes turbulence and the downscale enstrophy cascade and inverse energy cascade of two-dimensional Navier-Stokes turbulence.  We stress that these claims relate to non-perturbative locality of generalized structure functions on all orders, and not the term by term perturbative locality of diagrammatic theories or closure models that involve only two-point correlation and response functions. 
\end{abstract}
\preprint{submitted to \emph{Physical Review E}}

\maketitle

\section{Introduction}

Understanding the cascades of hydrodynamic turbulence by means of an analytical theory of the governing equations has been an ongoing effort over many decades, spearheaded by the Richardson-Kolmogorov prediction of a downscale energy cascade in three-dimensional turbulence \cite{book:Richardson:1922,article:Kolmogorov:1941,article:Kolmogorov:1941:1,article:Batchelor:1947} and Kraichnan's prediction of a downscale enstrophy cascade and an inverse energy cascade in two-dimensional  turbulence \cite{article:Kraichnan:1967:1,article:Leith:1968,article:Batchelor:1969}. Currently, the main challenge that concerns investigations of the downscale energy cascade of three-dimensional Navier-Stokes turbulence is understanding the scaling exponents of the energy spectrum and the higher-order structure functions of velocity differences and their deviation from Kolmogorov's 1941 predictions \cite{book:Frisch:1995}. These deviations are commonly known as \emph{``intermittency corrections''}, and the challenge to the theorists is two-fold: on one level to understand why they exist, and on a deeper level, to be able to predict them from first principles, using as few assumptions as possible.

 Study of two-dimensional Navier-Stokes turbulence presents us with a different set of challenges: (a) Due to the steep $k^{-3}$ slope of its energy spectrum, the downscale enstrophy cascade is only borderline local, naturally raising the question of why the dimensional analysis prediction of its energy spectrum works as well as it does; (b) the downscale enstrophy cascade does not manifest itself consistently, requiring careful tuning between forcing, small-scale dissipation, and large-scale dissipation, and as a matter of fact, it was not observed numerically until 1999 by Lindbord and Alvelius \cite{article:Alvelius:2000}, more than thirty years after Kraichnan conjectured its existence \cite{article:Kraichnan:1967:1}; (c) the inverse energy cascade appears to be initially more robust and is easier to reproduce numerically, but it tends to be disrupted by coherent structures \cite{article:Chekhlov:1999,article:Gurarie:2001,article:Gurarie:2001:1} that develop over time, raising doubts \cite{article:Danilov:2003,article:Vallgren:2011} on whether it is really a local cascade. On top of all that, there is a consensus that the cascades of two-dimensional turbulence are not subject to intermittency corrections \cite{article:Tabeling:2002}. A more detailed review of the fundamental questions facing two-dimensional turbulence research and theoretical progress on these issues was also given in a previous paper \cite{article:Tung:2006}. By contrast, the downscale energy cascade of three-dimensional turbulence is very robust and will readily manifest even in low-resolution numerical simulations, it is not subject to disruptions by coherent structures and exhibits universal scaling. Big challenge as it may be, understanding this universal scaling is the only major concern of the community. 

The first step towards a theoretical understanding of the downscale energy cascade of three-dimensional turbulence, that was pointed towards the right direction, was undertaken by Kraichnan with his formulation of the Direct Interaction Approximation (DIA) theory \cite{article:Kraichnan:1958,article:Kraichnan:1959}. Although it was not formulated as a first principles theory, and was in fact a closure modelling effort, it brought to the table the new idea of using a response function, defined as the ensemble average of the variational derivative of the velocity field with respect to the forcing field, in conjunction with the second-order velocity-velocity correlation function in formulating closure approximations. DIA did not agree with Kolmogorov's prediction of the energy cascade spectrum, and after Kolmogorov's prediction was experimentally verified for the first time in 1962 \cite{article:A.Moilliet:1962,article:Gibson:1962}, Kraichnan identified the overestimation of sweeping as the reason for the inconsistent predictions of his theory, and his next great insight was the idea that reformulating his theory in a Lagrangian representation would remove the effect of sweeping \cite{article:Kraichnan:1964}. The resulting theory is known as the Lagrangian-History Direct Interaction Approximation (LHDIA)\cite{article:Kraichnan:1965}, and it was shown to reproduce the Kolmogorov $k^{-5/3}$ energy spectrum \cite{article:Kraichnan:1966}. A detailed review of Kraichnan's work was given by Leslie \cite{book:Leslie:1972}.

Parallel to these efforts, Wyld showed that the DIA theory is a 1-loop line-renormalized diagrammatic theory, derivable from first principles directly from the Navier-Stokes equations \cite{article:Wyld:1961}. Based on this insight, it is fair to say that Kraichnan's LHDIA theory was the first successful theory of turbulence from first principles. Wyld's theory was extended to a wider range of stochastically forced dynamical systems by Martin, Siggia, and Rose \cite{article:Rose:1973} and Phythian \cite{article:Phythian:1977} used Feynman path integrals to show that the resulting Martin-Siggia-Rose formalism can be justified for all stochastically forced dynamical systems that are both local in time and first-order in time. Although the MSR theory can be used to generate higher-order versions of Kraichnan's DIA theory, it is not applicable to the improved LHDIA theory because the Navier-Stokes equations are not local in time, when written using the Lagrangian representation. This created a major obstacle towards moving forward this line of investigation, and a hiatus that lasted for a couple of decades. 

The main breakthrough that made it possible to go beyond the LHDIA theory and onto more exciting developments was the quasi-Lagrangian representation by Belinicher and L'vov  \cite{article:Lvov:1987,article:Lvov:1991} (also known as the Belinicher-L'vov representation), which makes it possible to surgically eliminate the sweeping effect directly at the level of the Navier-Stokes equations. The quasi-Lagrangian representation of the velocity field is essentially an Eulerian view using an arbitrarily chosen fluid particle as a non-inertial frame of reference. It works because the Navier-Stokes equations, written in quasi-Lagrangian form, remain local time and can be rewritten exclusively in terms of velocity differences, without it being necessary to have any velocity-velocity difference nonlinear sweeping terms. In hindsight, it is interesting that a precursor of the quasi-Lagrangian representation was used by Kolmogorov himself \cite{article:Kolmogorov:1941} in his original 1941 paper (see Ref.~\cite{article:Gkioulekas:2007} for some discussion on this point) in his definition of \emph{``homogeneous turbulence''}.

From the quasi-Lagrangian Navier-Stokes equations, L'vov and Procaccia developed a perturbative \cite{lect:Procaccia:1994,article:Procaccia:1995:1,article:Procaccia:1995:2,article:Procaccia:1996} and, in collaboration with Belinicher and Pomyalov, a non-perturbative theory \cite{article:Procaccia:1996:1,article:Procaccia:1996:2,article:Procaccia:1996:3,article:Procaccia:1997,article:Procaccia:1998,article:Procaccia:1998:1,article:Procaccia:1998:2} that culminated into a calculation of the scaling exponents of the velocity difference structure functions in the inertial range of the downscale energy cascade in three-dimensional turbulence \cite{article:Procaccia:2000}. A remarkable outcome of this calculation is that the lognormal 1962 intermittency theory by Kolmogorov \cite{article:Kolmogorov:1962,article:Oboukhov:1962} emerges as a first-order approximation, whereas the next-order approximation is sufficient to correctly estimate all scaling exponents that can be independently measured by experiment. The only shortcoming of this calculation is that it requires an independent calculation of the intermittency corrections to at least one scaling exponent in order to obtain the intermittency corrections of all pther scaling exponents. This independent calculation could be done by continuing the development of the theory of the  non-perturbative covariant closure models \cite{article:Procaccia:1998,article:Procaccia:1998:1,article:Procaccia:1998:2}, however this task remains a challenge for younger researchers. 

An alternate line of investigation, using renormalization group theory (see Ref.~\cite{article:Zhou:2010} for a recent review), combined with an operator product expansion (OPE), enabled Giles \cite{article:Giles:2001} to independently calculate the intermittency corrections to the first 10 scaling exponents, without the need to provide any experimental input. Giles used an alternate scheme that was proposed by Yakhot \cite{article:Yakhot:1981} in order to isolate the sweeping interactions from the essential part of the nonlinearity that drives the downscale energy cascade. In spite of the success, this line of investigation was subsequently abandoned. Also notable is a calculation of the Kolmogorov constant to the second-order structure function and the inertial-range skewness factor, using a two-loop approximation and the renormalization group method \cite{article:Vasilev:2002,article:Vasilev:2003}. More recently a new non-perturbative renormalization group approach has been initiated and seems to be promising \cite{article:Chate:2007,article:Delamotte:2011,article:Wschebor:2015,article:Wschebor:2016}. The combined RG with OPE approach is compelling also with respect to its wide range of applicability to shell models \cite{article:Eyink:1993:1}, the passive scalar problem \cite{article:Antonov:2006}, and many other problems (see Ref.~\cite{article:Kostenko:2015} and papers cited therein). 

Our view is that the L'vov-Procaccia theory has a lot more to offer, beyond solving the scaling exponents problem. For this reason, we have undertaken an effort to extend the theory to two-dimensional turbulence, begining with the non-perturbative portion of the theory \cite{article:Tung:2005,article:Tung:2005:1,article:Gkioulekas:2007,article:Gkioulekas:2008:1,article:Gkioulekas:p14}. This effort is still in a nascent stage, but has already provided some valuable understanding of the locality and stability of the downscale enstrophy cascade and the inverse energy cascade \cite{article:Gkioulekas:2008:1}, the transition to the dissipation range, and existence of anomalous sinks \cite{article:Gkioulekas:p14}. The scope of the present paper is to extend one aspect of the L'vov-Procaccia theory for both three-dimensional and two-dimensional turbulence. In order to explain our viewpoint and also to situate the reported results within the overall theory, we will begin with a brief overview of the logical structure of the L'vov-Procaccia theory. Readers who are already familiar and comfortable with the details and underlying logic of the L'vov-Procaccia theory can skip the next five paragraphs. 

As we already explained, the L'vov-Procaccia theory of three-dimensional Navier-Stokes turbulence is essentially two distinct but connected theories; a perturbative theory \cite{lect:Procaccia:1994,article:Procaccia:1995:1,article:Procaccia:1995:2,article:Procaccia:1996}  and a non-perturbative theory \cite{article:Procaccia:1996:1,article:Procaccia:1996:2,article:Procaccia:1996:3,article:Procaccia:1997}. The two theories make contact via the fusion rules of generalized structure functions and lead to two separate methods for calculating the structure function scaling exponents: a non-perturbative method \cite{article:Procaccia:1998,article:Procaccia:1998:1,article:Procaccia:1998:2} and a perturbative method \cite{article:Procaccia:2000}. Only the latter has been pursued to its logical conclusion. Our view is that the real underappreciated prize here is the fusion rules themselves governing the generalized structure functions. The generalized structure functions are defined as ensemble averages of velocity difference products where each velocity is evaluated at distinct points in space. They are a generalization of the standard structure functions used in the reformulation of Kolmogorov's theory by Frisch \cite{article:Frisch:1991,book:Frisch:1995}. The fusion rules govern the scaling of the generalized structure functions when some velocity difference endpoints are brought closer together, and they encapsulate, in mathematical terms, the physical understanding that turbulence cascades forget the details of random forcing within the inertial range, though they may remember the forcing length scale. 

The logical progression of the overall argument is as follows: First, it is shown that extending the LHDIA theory beyond the 1-loop approximation, into a more general finite order quasi-Lagrangian perturbative theory, represented by the Dyson-Wyld equations governing the second-order velocity difference correlation and the second-order response functions, continues to give Kolmogorov 1941 scaling \cite{article:Procaccia:1995:1}. This establishes Kolmogorov 1941 scaling as a baseline initial approximation, and indicates that intermittency corrections cannot be captured by finite-order generalizations of LHDIA. The next step of the argument is to show that without intermittency corrections, universality will be violated \cite{article:Procaccia:1995:2,article:Procaccia:1996}. Mathematically, universality means that if a perturbation is introduced to the statistics of random Gaussian forcing, the resulting perturbation to the generalized structure functions to all orders should have the same scaling and tensor structure as the overall generalized structure function. This implies that multipoint response functions, in the fusion limit, where the point-pairs associated with velocity differences come together to a smaller length scale $r$ relative to the point-pairs associated with forcing, should exhibit the same scaling exponents as the generalized structure functions. It is then shown, at the level of perturbation theory, that the diagrammatic expansion of multipoint response functions has logarithmic divergences that will add up to a power-law factor resulting in deviations from Kolmogorov scaling. Consequently, if the generalized structure functions follow Kolmogorov scaling while the corresponding response functions do not, then the generalized structure function scaling cannot be universal. An intermediate result, the so-called "rigidity" of the Feynman diagrams, is used both to derive the existence of the logarithmic divergences, and also to derive one of the fusion rules (the $p=2$ case, corresponding to the fusion of two velocity differences) from first primciples \cite{article:Procaccia:1995:2,article:Procaccia:1996}.

 The diagrammatic derivation of the fusion rules amounts to the argument that if the generalized structure functions retain universal scaling and tensor structure when subjected to forcing perturbations, then the fusion rules are satisfied. A simpler derivation was given later \cite{article:Procaccia:1996:1,article:Procaccia:1996:3} and is reviewed and extended to inverse cascades in Section II.A of the present paper, at the price of making a stronger universality hypothesis in terms of conditional ensemble averages. The conceptual similarity of the fusion rules with operator product expansions (OPE), used in renormalization group theories \cite{article:Zhou:2010}, may raise the question of whether there is an OPE reformulation of the fusion rules. L'vov and Procaccia promised to develop such a reformulation (see Reference 17 cited in Ref.~\cite{article:Procaccia:1996}) but did not publish it. Precursors of the fusion rules were derived by an additive OPE hypothesis by Eyink \cite{article:Eyink:1993,article:Eyink:1995:3}, and this approach was recently revisited by Falkovich and Zamolodchikov \cite{article:Zamolodchikov:2015}. On the other hand, such a reformulation is not needed for the further development of this theory, and the statistical argument used by L'vov and Procaccia \cite{article:Procaccia:1996:1,article:Procaccia:1996:3}, also employed  in this paper, is more transparent, in terms of the physics of universal cascades, and also with regards to the cancellation of leading-order terms in certain important cases.

Using the fusion rules, combined with the governing equations of the generalized structure functions, derived from the Navier-Stokes equations, as a point of departure, made it possible to formulate a non-perturbative theory for the downscale energy cascade \cite{article:Procaccia:1996:1,article:Procaccia:1996:2,article:Procaccia:1996:3,article:Procaccia:1997}. The main results were: establishing the non-perturbative locality and stability of the downscale energy cascade, the scaling structure of the dissipation range for higher-order generalized structure functions, bridge relations between the scaling exponents of correlations involving the energy dissipation field and the scaling exponents of generalized structure functions \cite{article:Procaccia:1996:1,article:Procaccia:1996:2,article:Procaccia:1996:3}. Another major result is the investigation of multi-time generalized structure functions, bridge relations regarding the scaling of multi-time generalized structure functions, and the key result that postulating multi-time self-similarity is a wrong assumption, as it is equivalent to axiomatically assuming the absence of intermittency corrections \cite{article:Procaccia:1997}. It is this portion of the theory that we have began extending to two-dimensional turbulence \cite{article:Tung:2005,article:Tung:2005:1,article:Gkioulekas:2007,article:Gkioulekas:2008:1,article:Gkioulekas:p14}. 

In our view, if one stands above the elaborate mathematical and technical details, we see that the main thrust of the non-perturbative theory is that it continues the argument of the perturbative theory as follows: with the requirement of universality implying the fusion rules, the next step is that the fusion rules allow us to compare the terms in the generalized balance equations (described below) and therefore determine the extent of the inertial range, as a multidimensional region. The question is thus posed and answered: if we require a cascade to have universal scaling within its inertial range, is it going to be able to have an inertial range? In three-dimensional turbulence this is not much of a problem. However, in two-dimensional turbulence, this is one of the fundamental problems that we are confronted with (see Refs. \cite{article:Gkioulekas:2008:1,article:Gkioulekas:p14} and the discussion in the conclusion of this paper). The other question that is also posed and answered mostly in the affirmative is: if we require the downscale energy cascade of three-dimensional turbulence to have universal scaling within its inertial range, does that impose enough constraints on the structure functions scaling exponents to allow us to calculate them from first principles?

In this paper we provide a generalization of one small step in the overall non-perturbative L'vov-Procaccia theory: the argument that  the fusion rules imply the non-perturbative locality of the nonlinear interactions term in the balance equations of the generalized structure functions \cite{article:Procaccia:1996:3,article:Gkioulekas:2008:1}. This generalization, which we have termed ``multi-locality'', is a major step towards broadening the range of results that can be derived by the fusion rules and also places some earlier results \cite{article:Procaccia:1996:2,article:Procaccia:1996:3,article:Procaccia:1997} on more rigorous grounds. A brief description of this notion of multilocality is given below.

We begin with the definition of the $n^{\text{th}}$-order generalized structure functions $F_n (\{\bfx,\bfxp\}_n, t)$ as
\begin{equation}
F_n^{\ga_1\ga_2\cdots\ga_n} (\{\bfx,\bfxp\}_n, t) = \avg{\prod_{k=1}^n w_{\ga_k}(\bfx_k, \bfxp_k, t)}
\end{equation}
with $w_{\ga}(\bfx, \bfxp, t)=u_{\ga}(\bfx,t)-u_{\ga}(\bfxp, t)$  and $u_{\ga}(\bfx,t)$ the Eulerian velocity field. The angle brackets represent an ensemble average over all realizations of random forcing. Differentiating with respect to time $t$ yields an equation of the form 
$\pderivin{F_n}{t}+\cO_n F_{n+1} = I_n + \ccD_n F_n + Q_n$,  with  $\cO_n F_{n+1}$ representing the nonlinear local interactions that govern the cascades, $I_n$ the sweeping interactions (see Ref.~\cite{article:Gkioulekas:2007} for details),  $\ccD_n F_n $ the dissipation terms, and $Q_n$ the forcing terms. Here $\cO_n$ and $\ccD_n$ are linear operators and a detailed account of these terms is given in Ref.~\cite{article:Gkioulekas:2008:1}. The locality of cascades is reflected mathematically in the locality of the integrals in $\cO_n F_{n+1}$, which can be deduced from the fusion rules hypothesis \cite{article:Procaccia:1996:1,article:Procaccia:1996:3,article:Gkioulekas:2008:1}. Locality implies that if $F_n$ scales as $R^{\gz_n}$ when all velocity difference endpoints are separated at length scale $R$, then the terms that comprise  $\cO_n F_{n+1}$ will scale as  $R^{\gz_{n+1}-1}$ and the extent of the inertial range can be determined by comparing them against $Q_n$, $\ccD_n F_n$ and $I_n$.

We will now consider the locality of the terms that comprise $\cO_n \cO_{n+1}\cdots\cO_{n+p-1} F_{n+p}$ . These terms arise from more general balance equations for the $p^{\text{th}}$-order time derivative $\pd^p F_n/\pd t^p$. These equations were previously used to establish bridge relations between the scaling exponents of correlations involving velocity gradients and the scaling exponents $\gz_n$ \cite{article:Procaccia:1996:3} and they are also needed to continue the previous investigation of the cross-over of generalized structure functions to the dissipation range \cite{article:Procaccia:1996:2,article:Procaccia:1996:3,article:Gkioulekas:p14}. We will show that the integrals in the general term  $\cO_n \cO_{n+1}\cdots\cO_{n+p-1} F_{n+p}$   continue to be local, implying $R^{\gz_{n+p}-p}$ scaling. We describe this generalization as \emph{multilocality}.

From the mathematical argument given below, we see that investigating  multilocality in the IR limit requires careful consideration of the fusion rule for a new geometry that was not previously needed. We call this the \emph{two-blob geometry fusion rule}, and it is discussed in detail in Section~\ref{sec:fusion-rule-two-blob-geometry}. Furthermore, we observe that, for the case of downscale cascades, the locality argument for $\cO_n F_{n+1}$ depends only on a very restricted subset of the hypothesized fusion rules which reduce to the problem of fusing two velocity differences ($p=2$) in a generalized structure function involving a large number of velocity differences. This $p=2$ case has been shown theoretically by diagrammatic techniques \cite{article:Procaccia:1995:1,article:Procaccia:1995:2,article:Procaccia:1996} for the case of three-dimensional turbulence with Kolmogorov scaling. The multilocality argument for downscale cascades, on the other hand, requires a much broader range of fusion rules, beyond what has been studied theoretically, in order to determine the relevant scaling exponents in the two-blob geometry fusion rule. 

The situation for upscale cascades, namely the inverse energy cascade of two-dimensional Navier-Stokes turbulence, is also very interesting. We find that  the fusion rules imply both locality and multilocality in the UV limit, but in the IR limit they both emerge solely as a result of  a fortunate cancellation of leading terms. The scaling of the surviving subleading terms is sufficient to ensure IR locality but whether or not it is dependent on other scaling exponents requires further investigation.  The root of the problem is traced to the predicted scaling in the two-blob geometry fusion rule for the case of upscale cascades, which is  different from the scaling claimed in my previous paper \cite{article:Gkioulekas:2008:1} in the IR locality proof for upscale cascades, which was based on an argument that was incorrect for upscale cascades.

Because the details of the overall argument are very technical, we will now provide a detailed outline of the organization of the paper. In broad strokes, we note that Section II discusses the fusion rules needed by the locality and multilocality proofs and Section III contains the main argument itself.

 More specifically, in Section II.A we review the previously reported argument \cite{article:Procaccia:1996:1,article:Procaccia:1996:3,article:Gkioulekas:2008:1} that derives most of the fusion rules, for both upscale and downscale cascades, as a consequence of the universal self-similarity hypothesis. Let $F_n^{(p)}(r,R)$ denote a generalized structure function with $p$ velocity differences reduced to length scale $r$ and the remaining $n-p$ velocity differences at length scale $R$. In the limit $r \ll R$, with both $r,R$ within the inertial range and with $2\leq p < n-1$, the main finding is that $F_n^{(p)}(r,R)$ scales as $F_n^{(p)}(r,R) \sim r^{\xi_{np}} R^{\gz_n-\xi_{np}}$ with $\xi_{np}=\gz_p$ for downscale cascades and $\xi_{np}=\gz_n-\gz_{n-p}$ for upscale cascades. 

Section II.B gives a very detailed account of the fusion rule scaling when $p=1$. First, we argue that the leading order contribution vanishes, both for upscale and downscale cascades, but for different reasons. The case $p=1$ corresponds to having $1$ velocity difference at scale $r$ while the remaining $n-1$ velocity differences remain at scale $R$, with $r \ll R$. If the minimum distance between the small velocity difference from the other velocity differences is $R_{\text{min}}$, then if $r \ll R_{\text{min}}$, we expect the scaling $F_n^{(1)} (r,R) \sim (r/R_{\text{min}}) R^{\gz_n}$, for both downscale and upscale cascades. If $r \gg R_{\text{min}}$, then the corresponding scaling is $F_n^{(1)}(r,R)\sim r^{\xi_{n,1}} R^{\gz_n-\xi_{n,1}}$, noting that $\xi_{n,1}=\gz_2$ for downscale cascades and $\xi_{n,1}=\gz_n-\gz_{n-2}$ for upscale cascades, for all $n>3$.  For $n=3$, an additional cancellation will give $\xi_{3,1}=\gz_3$ and for $n=2$ we get $\xi_{2,1}=\gz_2$. These evaluations hold  for both upscale and downscale cascades and furthermore when the small velocity difference is attached onto one of the large velocity differences, as shown in Fig.~\ref{fig:p-one-case-two-fusion-rule}, or when it is embedded in a chain of large velocity differences, as shown in Fig.~\ref{fig:p-one-double-attachment-one} or Fig.~\ref{fig:p-one-multiple-attachment}. 

Section II.C considers the $p=n-1$ fusion rule where $n-1$ velocity differences are congregated inside a small-scale blob at length scale $r$ with the remaining large velocity difference at scale $R$ and with one endpoint situated inside the small-scale blob, as shown in Fig.~\ref{fig:p-n-minus-one-geometry}. Similarly to the $p=1$ case, the leading order contribution vanishes, both for upscale and downscale cascades, resulting in $F_n^{(n-1)}(r,R)\sim r^{\gz_n} R^0$ scaling. 

Section II.D considers the new two-blob geometry, shown in Fig.~\ref{fig:the-two-blob-vel-d-geometry} where groups of velocity differences are congregated inside two small-scale blobs, separated by a large-scale distance, except for one velocity difference that straddles between the two blobs, with an endpoint inside each blob. Relevant for the locality and multilocality proofs is the fusion rule scaling exponent of the large-scale distance $R$ between the two blobs. If $\gD_{np}$ is this scaling exponent for the case where there is a total number of $n$ velocity differences, with $p$ velocity differences on one blob, $n-p-1$ velocity differences on the other blob, and one last velocity difference straddling between the two blobs, then our main result is $\gD_{np}=\gz_n-\gz_{p+1}-\gz_{n-p}<0$ for downscale cascades and $\gD_{np}=-\ga$ for upscale cascades. The scaling exponent $\ga$ is expected to satisfy $\ga>0$ and corresponds to the $(r/R)^{\ga}$ spatial decorrelation factor that results when separating a velocity difference with separation $r$ away from an $r$-scale blob of velocity differences with similar separations at a large distance $R$ (see Fig.~\ref{fig:separation-of-one-vel-d-from-one-blob}). It should be noted that all of the above scaling predictions correspond to leading-order terms, and that means that the expected scaling with respect to $R$ is established only as an upper bound. This is, of course, sufficient for the purpose of establishing locality or multilocality. 

In Section III we take on the multilocality proof. Preliminaries are given in Section III.A, where we explain the problem posed by the existence of cross-terms with regards to proving multilocality. Section III.B reviews the UV locality argument for both upscale and downscale cascades. This argument was given in previous papers \cite{article:Procaccia:1996:1,article:Procaccia:1996:3,article:Gkioulekas:2008:1}, but its particular technical details are needed in the more general multilocality argument. The multilocality proof in the UV limit is given, for the case of two operators in Section III.C and then generalized for an arbitrary number of operators on Section III.D. The main result is that multilocality holds, in the UV limit, for both upscale and downscale cascades, as long as $\xi_{n+1,1}>0$. This corresponds to the condition $\gz_2>0$ and $\gz_3>0$ for downscale cascades and the condition $\gz_n-\gz_{n-2}>0$ and $\gz_3>0$ for upscale cascades. The argument also entails a universal local homogeneity  and  isotropy assumption, in order to establish the $p=1$ fusion rule. It should be noted that these requirements are not any stronger than what is required to establish locality in the UV limit. The IR limit is discussed on Section III.E, where it becomes necessary to treat the case of upscale cascades separately from downscale cascades. For downscale cascades, our main finding is that the IR multilocality of $\cO_n\cO_{n+1}\cdots\cO_{n+p-1} F_{n+p}$ requires the condition $\gz_{n+p}-\gz_{n+p-m}-\gz_{m+1}<0$ for all $m$ with $1\leq m <p$. This condition is always satisfied, via the Holder inequalities on the generalized structure function scaling exponents. For the case of upscale cascades, however, we find that both locality and multilocality are dependent on the assumption $\ga>0$, which is both necessary and sufficient.  A detailed discussion of the underlying mathematical details is given. It should be noted that even though IR locality does not fail for upscale cascades, an equivalent difficulty can still emerge from the sweeping interactions, as was pointed out in a previous paper \cite{article:Gkioulekas:2007}. The conclusion of the paper, in Section IV, expands on this point in more detail and offers some concluding thoughts and a discussion of related theoretical and numerical work.  

\section{The fusion-rules for downscale and upscale cascades}

In this section, we review the arguments that establish the fusion rules, both for direct cascades (i.e. the energy cascade of three-dimensional Navier-Stokes turbulence and the downscale enstrophy cascade of two-dimensional Navier-Stokes turbulence) and inverse cascades (i.e. the inverse energy cascade of two-dimensional Navier-Stokes turbulence). In doing so, we provide a more careful and detailed account of the exceptional cases of the $p=1$ and $p=n-1$ fusion rules,  than was given previously \cite{article:Procaccia:1996:1,article:Procaccia:1996:3,article:Gkioulekas:2008:1}. We also introduce, both for upscale and downscale cascades, a new fusion rule for the so-called two-blob velocity difference geometry which is needed by the multilocality proof in the IR limit. 

\subsection{Fusion rules and universal symmetries}

The fusion rules are supposed to encapsulate mathematically the universality of the inertial range, that is, the notion that deep inside the inertial range and far away from the forcing range, the statistical details of random forcing are forgotten. The same dynamic plays out between the small scales $r$ and the large scales $R$, both within the inertial range, in the context of a downscale cascade; energy (or enstrophy, in two-dimensional turbulence) is passed down from length-scale $R$ to length-scale $r$, but by the time it gets to length-scale $r$, the details of the velocity field statistics at length-scale $R$ are forgotten. A similar notion applies to the inverse energy cascade of two-dimensional turbulence, where the small scales $r$ and large scales $R$ switch roles. 

To express this idea mathematically, we define the conditional generalized structure function $\Phi_{mn}$ via the following conditional ensemble average: 
\begin{multline}
\Phi_{nm}^{\ga_1\ldots\ga_n\gb_1\ldots\gb_m} (\{\bfX\}_n, \{\bfY\}_m, \{w\}_m, t) = \\ \avg{\left[\prod_{\gk=1}^n w_{\ga_k} (\bfX_k, t) \right\vert \left. w_{\gb_k} (\bfY_k, t)=w_k, \;\forall k\in\{1,\ldots, m\}  \right] 
}.
\end{multline}

Here we use lower-case vectors like $\bfx_1, \bfxp_1, \bfx_2, \bfxp_2, \ldots$ to represent the location of velocity difference endpoints, uppercase vectors like $\bfX_k = (\bfx_k, \bfxp_k)$ to represent pairs of endpoints that are used to form a velocity difference, and $\{\bfX\}_n = (\bfX_1, \bfX_2, \ldots, \bfX_n)$ to represent the geometric configuration of $n$ pairs of velocity differences. The universality hypothesis is that $\Phi_{nm}$ has the same statistical symmetries with respect to $\{\bfX\}_n$ as the generalized structure function $F_n (\{\bfX\}_n, t)$, namely: local homogeneity, local isotropy, and self-similarity \cite{article:Gkioulekas:2008:1}. This is contingent on the following assumptions: We assume that both velocity difference separations $\{\bfX\}_n$ and $\{\bfY\}_m$ are at length scales within the inertial range. For the case of a downscale cascade, we also assume that $\{\bfX\}_n$ scale as $r$ and $\{\bfY\}_m$ scale as $R$ with $r\ll R$. Likewise, for the case of an upscale cascade we assume that $\{\bfX\}_n$ scales as $R$ and $\{\bfY\}_m$ scales as $r$. In both cases, the idea is that $\{\bfY\}_m$ is closer to the forcing scale than $\{\bfX\}_n$. Under these conditions, $\Phi_{nm}$ is postulated to remain invariant upon shifting all points $\{\bfX\}_n$ by the same $\gD\bfx$ for any $\gD\bfx$ with comparable order of magnitude (universal local homogeneity) and also upon rotating all points $\{\bfX\}_n$ around their center (universal local isotropy). We also postulate universal self-similarly, that $\Phi_{nm}$ scales with respect to $\{\bfX\}_n$ with the same scaling exponent $\gz_n$ as the generalized structure functions $F_n$ according to
\begin{equation}
\Phi_{nm} (\gl\{\bfX\}_n, \{\bfY\}_m, \{w\}_m, t) = \gl^{\gz_n} \Phi_{nm} (\{\bfX\}_n, \{\bfY\}_m, \{w\}_m, t) 
\end{equation}
Together, the universality hypothesis consists of the postulates of universal local homogeneity, universal local isotropy, and universal self-similarity \cite{article:Gkioulekas:2008:1}.

The physical idea that underlies the universality hypothesis is that the conditional ensemble average, by imposing the restriction $w_{\gb_k} (\bfY_k, t)=w_k$ for all $k\in\{1,\ldots, m\}$, partitions the ensemble of all forcing histories into subensembles that are consistent with the parameters $w_k$. If the velocity difference statistics at the $\{\bfY\}_m$ scales are indeed forgotten at the $\{\bfX\}_n$ scales, then we can postulate that the statistical symmetries of the generalized structure functions are not affected by replacing the unconditional ensemble average with a restricted conditional average. 

The fusion rules hypothesis is an immediate consequence of the universality hypothesis and can be established by employing the Bayes theorem as follows: Assume that $\{\bfX\}_p \sim r$ and $\{\bfY\}_{n-p}\sim R$ with $r\ll R$, with both $r, R$ in the inertial range, and regardless of the cascade direction. Let $\cP_m^{\gb_1\ldots\gb_m}(\{\bfY\}_m, \{\bfw\}_m,t)$ be the probability of the event $w_{\gb_k} (\bfY_k, t)=w_k$ for all $k\in\{1,\ldots, m\}$, with the understanding that it is allowed to be a generalized function. For the case of a downscale cascade, $\Phi_{p,n-p} (\{\bfX\}_p, \{\bfY\}_{n-p}, \{w\}_{n-p}, t)$ is self-similar with respect to $\{\bfX\}_p$ and therefore
\begin{widetext}
\begin{align}
F_n (\gl \{\bfX\}_p, \mu \{\bfY\}_{n-p}) &= \int \left[\prod_{k=1}^{n-p} w_k\right] \cP  (\mu\{\bfY\}_{n-p}, \{\bfw_k\}_{k=1}^{n-p}) \Phi_{p,n-p} (\gl  \{\bfX\}_p, \mu \{\bfY\}_{n-p}, \{\bfw_k\}_{k=1}^{n-p}) \prod_{k=1}^{n-p} dw_k \\
&= \gl^{\gz_p}  \int \left[\prod_{k=1}^{n-p} w_k\right] \cP  (\mu\{\bfY\}_{n-p}, \{\bfw_k\}_{k=1}^{n-p}) \Phi_p (  \{\bfX\}_{p,n-p}, \mu \{\bfY\}_{n-p}, \{\bfw_k\}_{k=1}^{n-p}) \prod_{k=1}^{n-p} dw_k \\
&= \gl^{\gz_p} F_n (\{\bfX\}_p, \mu \{\bfY\}_{n-p}).
\end{align}
 For the case of an upscale cascade, we have to use a modified argument based on the self-similarly of $\Phi_{n-p,n} (\{\bfY\}_{n-p}, \{\bfX\}_q, \{w\}_p, t)$ with respect to the coordinates $\{\bfY\}_{n-p}$ which reads:
\begin{align}
F_n (\gl \{\bfX\}_p, \mu \{\bfY\}_{n-p}) &= \int \left[\prod_{k=1}^{p} w_k\right] \cP  (\gl \{\bfX\}_p, \{\bfw_k\}_{k=1}^{p}) \Phi_{n-p,n} (\mu \{\bfY\}_{n-p}, \gl  \{\bfX\}_p, \{\bfw_k\}_{k=1}^{p}) \prod_{k=1}^{p} dw_k \\
&=\mu^{\gz_{n-p}} \int \left[\prod_{k=1}^{p} w_k\right] \cP  (\gl \{\bfX\}_p, \{\bfw_k\}_{k=1}^{p}) \Phi_{n-p,n} (\{\bfY\}_{n-p}, \gl  \{\bfX\}_p, \{\bfw_k\}_{k=1}^{p}) \prod_{k=1}^{p} dw_k \\
&=\mu^{\gz_{n-p}} F_n (\gl \{\bfX\}_p,  \{\bfY\}_{n-p}).
\end{align}
\end{widetext}
The factor $F_n (\gl \{\bfX\}_p,  \{\bfY\}_{n-p})$ is now independent of $\mu$ and has to scale as $\gl^{\gz_n-\gz_{n-p}}$. In both cases, we can write the corresponding self-similar law as:
\begin{equation}
F_n (\gl \{\bfX\}_p, \mu \{\bfY\}_{n-p}) = \gl^{\xi_{np}}\mu^{\gz_n-\xi_{np}} F_n (\{\bfX\}_p, \{\bfY\}_{n-p})
\end{equation}
with $\xi_{np}=\gz_p$ for downscale cascades and $\xi_{np}=\gz_n-\gz_{n-p}$ for upscale cascades. 

A stronger version of the universality hypothesis postulates a more precise relationship between the conditional generalized structure function $\Phi_{nm}$ and the generalized structure function $F_n$ as follows: For the case of a downscale cascade, with $\{\bfX\}_n \sim r$ and $\{\bfY\}_m \sim R$ with $r \ll R$ and $r,R$ both in the inertial range, we postulate that 
\begin{equation}
\Phi_{nm} (\{\bfX\}_n, \{\bfY\}_m, \{w\}_m, t) = \tilde{F}_n (\{\bfX\}_n, t) \tilde{\Phi}_{nm} (\{\bfY\}_m,\{\bfw\}_m, t)
\end{equation}
Here, $\tilde{F}_n $ may have a different inertial range from $F_n$, dependent on the scale $R$, but is postulated to have the same tensor structure as $F_n$, as long as $\{\bfX\}_n$ is within the inertial range of $\tilde{F}_n$. For the case of the inverse energy cascade we assume that the above equation holds when $\{\bfX\}_n \sim R$ and $\{\bfY\}_m \sim r$, with $r \ll R$ and $r, R$ both in the inertial range. 

Using the same argument, via the Bayes theorem, we can show that for $\{\bfX\}_{k=1}^p \sim r$ and $\{\bfX\}_{k=p+1}^n \sim R$ with $r \ll R$ and both $r, R$ in the inertial range, the generalized structure function $F_n (\{\bfX\}_n, t)$ will give
\begin{equation}
F_n (\{\bfX\}_n, t) = \tilde{F}_p (\{\bfX\}_{k=1}^p, t) \Psi_{n,p} (\{\bfX\}_{k=p+1}^n, t)
\end{equation}
 for a downscale cascade and
\begin{equation}
F_n (\{\bfX\}_n, t) = \tilde{F}_{n-p} (\{\bfX\}_{k=p+1}^n, t) \Psi_{n,n-p} (\{\bfX\}_{k=1}^p, t)
\end{equation}
for an upscale cascade, which leads to the evaluation of the fusion rule scaling exponents $\xi_{np}$ given above. 

\subsection{Fusion rules for $p=1$}

The case $p=1$ where the velocity difference $(\bfx_1, \bfxp_1)$ fuses to the small scale $r$ while the other $n-1$ velocity differences remain at scale $R$ with $r\ll R$ requires special consideration, because the leading order term vanishes. The reason for that depends on whether the cascade direction  is upscale or downscale. For a downscale cascade, the corresponding fusion rule would be $F_n (\{\bfX\}_n,t)=\tilde{F}_1 (\bfx_1,\bfxp_1,t) \Psi_{n,1}(\{\bfX_k\}_{k=2}^n,t)$ and from spatial homogeneity, the $F_1$ factor vanishes.

 For an upscale cascade, the corresponding fusion rule reads $F_n (\{\bfX\}_n,t)=\tilde{F}_{n-1}(\{\bfX_k\}_{k=2}^n,t) \Psi_{n,n-1}(\bfX_1,t)$ and it should result in the scaling $F_n \sim R^{\gz_{n-1}} r^{\gz_n-\gz_{n-1}}$. Nevertheless, this leading order contribution also vanishes, but for a different reason. From the universality postulate, we expect that $F_n$ is an isotropic tensor with respect to rotating the velocity differences at the endpoints $\{\bfX_k\}_{k=2}^n\sim R$ since at scales $R$, the dynamics of the upscale cascade process forgets what is happening at scale $r$. Equivalently, $\Psi_{n,n-1}$ should also be an isotropic tensor with respect to $(\bfx_1,\bfxp_1)$, allowing us to rewrite it in the form
\begin{align}
F_n^{\ga_1\ldots\ga_n} (\{\bfX\}_n,t) &= C F_{n-1}^{\ga_2\ldots\ga_n} (\{\bfX_k\}_{k=2}^n,t) \Psi_{n,n-1}^{\ga_1} (\bfX_1,t) \\
&= C F_{n-1}^{\ga_2\ldots\ga_n} (\{\bfX_k\}_{k=2}^n,t) \frac{r_{\ga_1}}{\nrm{\bfr}} \gy (\nrm{\bfr},t)
\end{align}
with $\bfr=\bfx_1-\bfxp_1$ and $C$ a numerical coefficient. The incompressibility condition implies that $\pd_{\ga_1, \bfx_1}F_n^{\ga_1\ldots\ga_n}(\{\bfX\}_n,t)=0$ and we will show that, in turn, it implies that $C=0$, thereby annihilating the leading order contribution to the fusion rule. 

Noting in general that $\pd_{\ga}\nrm{\bfx}=x_{\ga}/\nrm{\bfx}$ (with $\pd_{\ga}$ a spatial partial derivative with respect to the $x_{\ga}$ component of $\bfx$) and $\pd_\ga (x_\ga/\nrm{\bfx})=(d-1)/\nrm{\bfx}$, with $d$ the dimension of $\bfx$, we derive the identity
\begin{equation}
\pd_\ga\left(\frac{x_\ga}{\nrm{\bfx}}f(\nrm{\bfx})\right) =\frac{d-1}{\nrm{\bfx}}f(\nrm{\bfx})+f'(\nrm{\bfx})
\end{equation}
for any differentiable scalar function $f$. Using this identity, the divergence of the leading order fusion rule contribution to the generalized structure function $F_n$ reads 
\begin{align}
\pd_{\ga_1, \bfx_1} &F_n^{\ga_1\ldots\ga_n}(\{\bfX\}_n,t) = \\ &= C F_{n-1}^{\ga_2\ldots\ga_n} (\{\bfX_k\}_{k=2}^n,t) \pd_{\ga_1, \bfx_1} \left[ \frac{r_{\ga_1}}{\nrm{\bfr}} \gy (\nrm{\bfr},t) \right] \\
&= C F_{n-1}^{\ga_2\ldots\ga_n} (\{\bfX_k\}_{k=2}^n,t) \left[\frac{d-1}{r}+\pderiv{}{r}\right]\gy (r,t)
\end{align}
 with $r=\nrm{\bfr}$. The divergence condition gives a differential equation with respect to $r$ that can be rewritten in equidimensional form as $[d-1+r\pderivin{}{r}]\gy (r,t)=0$. This equation has only one independent solution $\gy (r,t)=Cr^{1-d}$, that is inconsistent with the $r^{\gz_n-\gz_{n-1}}$ scaling predicted by the universality postulate, therefore $C=0$ and the leading order term vanishes. It is worth noting that this is really the same argument that is used to eliminate the leading contribution in the $p=n-1$ fusion rule for downscale cascades (see below and also Section II.C.3 of Ref.~\cite{article:Procaccia:1996:3}). We deviate from the previous version of the argument \cite{article:Procaccia:1996:3} in that we apply it only to the leading order fusion rule contribution to $F_n$, and not to $F_n$ as a whole. 

In both cases of an upscale or downscale cascade, we need to find the next-order term. We distinguish between the following two cases:

\textbf{Case 1:} Let us assume that the fused velocity difference $w_{\ga_1}(\bfx_1,\bfxp_1,t)$ has endpoints that are far away from all endpoints of all other velocity differences. Since $\bfx_1$ and $\bfxp_1$ are close together, we use a Taylor expansion to write
\begin{align}
w_{\ga_1}(\bfx_1,\bfxp_1,t)&=u_{\ga_1}(\bfx_1,t)-u_{\ga_1}(\bfxp_1,t)\\
&=(\bfx_1-\bfxp_1)_\gb \pd_{\gb,\bfx_1}u_{\ga_1}(\bfxp_1,t)+\cdots
\end{align}
and the generalized structure function $F_n$, to leading order, reads:
\begin{multline}
F_n(\{\bfX\}_n,t)= \\ (\bfx_1-\bfxp_1)_\gb \pd_{\gb,\bfx_1}\avg{u_{\ga_1}(\bfxp_1,t)\left[\prod_{k=2}^{n} w_{\ga_k}(\bfx_k,\bfxp_k,t)\right]}+\cdots
\end{multline}
The ensemble average of the derivative of the velocity/velocity difference product in the above equation depends on all endpoint separations but retains symmetry with respect to shifting all endpoints equally, since the original generalized structure function $F_n$ satisfies local homogeneity. The derivative $\pd_{\gb,\bfx_1}$, in shaking the point $\bfx_1$, is also shaking all separations between $\bfx_1$ and $\bfxp_1,\bfx_2,\bfxp_2,\cdots,\bfx_n,\bfxp_n$. Consequently, the derivative $\pd_{\gb,\bfx_1}$ will result in multiple contributions, with the dominant contribution scaling as 
\begin{equation}
F_n^{(1)}(r,R)\sim (r/R_{\mathrm{min}}) \psi_n (R) 
\label{eq:p-one-case-one-fusion-rule}
\end{equation}
where $R_{\mathrm{min}}$ is the minimum distance between $\bfx_1$ and $\bfxp_1,\bfx_2,\bfxp_2,\cdots,\bfx_n,\bfxp_n$. All other contributions will also give the same scaling exponent with respect to $r$. This argument was given previously for downscale cascades \cite{article:Procaccia:1996:3}, and it also applies without modifications to upscale cascades, except that a new argument, given above, is needed to justify eliminating the leading fusion rule contribution.

\begin{figure}[tb]\begin{center}
\begin{align}
&
\begin{pspicture}[shift=*](-1,-1)(12,5)
\psline{*-*}(1,1)(1,3)
\psline{*-*}(1,3)(10,3)
\uput[d](1,1){$\bfxp_1$}
\uput[u](1,3){$\bfx_1=\bfx_2$}
\uput[u](10,3){$\bfxp_2$}
\end{pspicture} = \\
&\quad
\begin{pspicture}[shift=*](-1,-1)(5,5)
\psline{*-*}(1,1)(1,3)
\psline{*-*}(1.4,1)(1.4,3)
\uput[d](1,1){$\bfxp_1$}
\uput[u](1,3){$\bfx_1=\bfx_2$}
\end{pspicture}
+
\begin{pspicture}[shift=*](-1,-1)(12,5)
\psline{*-*}(1,1)(1,3)
\psline{*-*}(1,1)(10,3)
\uput[d](1,1){$\bfxp_1$}
\uput[u](1,3){$\bfx_1=\bfx_2$}
\uput[u](10,3){$\bfxp_2$}
\end{pspicture}
\end{align}
\caption{\label{fig:p-one-case-two-fusion-rule}\small Graphical representation of Eq.~\eqref{eq:p-one-case-two-fusion-rule}. Velocity differences not associated with the decomposition $w_{\ga_2}(\bfx_2,\bfxp_2)=w_{\ga_2}(\bfx_1,\bfxp_1)+w_{\ga_2}(\bfxp_1,\bfxp_2)$ are omitted}
\end{center}\end{figure}

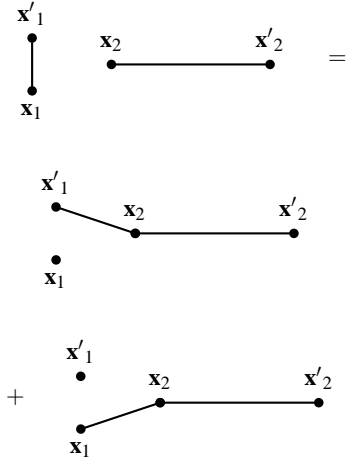
\begin{figure}[tb]\begin{center}
\begin{align*}
&
\begin{pspicture}[shift=*](-1,-1)(12,5)
\psline{*-*}(1,1)(1,3)
\psline{*-*}(4,2)(10,2)
\uput[d](1,1){$\bfx_1$}
\uput[u](1,3){$\bfxp_1$}
\uput[u](4,2){$\bfx_2$}
\uput[u](10,2){$\bfxp_2$}
\end{pspicture}
= \\
&\quad
\begin{pspicture}[shift=*](-1,-1)(12,5)
\psdots[dotstyle=*](1,1)
\psline{*-*}(4,2)(1,3)
\psline{*-*}(4,2)(10,2)
\uput[d](1,1){$\bfx_1$}
\uput[u](1,3){$\bfxp_1$}
\uput[u](4,2){$\bfx_2$}
\uput[u](10,2){$\bfxp_2$}
\end{pspicture}
\\
&\quad +
\begin{pspicture}[shift=*](-1,-1)(12,5)
\psdots[dotstyle=*](1,3)
\psline{*-*}(4,2)(1,1)
\psline{*-*}(4,2)(10,2)
\uput[d](1,1){$\bfx_1$}
\uput[u](1,3){$\bfxp_1$}
\uput[u](4,2){$\bfx_2$}
\uput[u](10,2){$\bfxp_2$}
\end{pspicture}
\end{align*}
\caption{\label{fig:p-one-case-two-detached-fusion-rule}\small Graphical representation of Eq.~\eqref{eq:p-one-case-two-detached-fusion-rule}. Velocity differences not associated with the decomposition $w_{\ga_1}(\bfx_1,\bfxp_1,t)=w_{\ga_1}(\bfx_1,\bfx_2,t)+w_{\ga_1}(\bfx_2,\bfxp_1)$ are omitted} 
\end{center}\end{figure}

\textbf{Case 2:} As was previously noted \cite{article:Procaccia:1996:3}, the scaling claimed by Eq.~\eqref{eq:p-one-case-one-fusion-rule} should break down in the limit $R_{\mathrm{min}}\to 0^{+}$, as it does not make sense for $R_{\mathrm{min}}$ to maintain a negative scaling exponent all the way to $0^{+}$. To find the correct scaling, we begin with the case $R_{\mathrm{min}}=0$. Assume, with no loss of generality, that $\bfx_1=\bfx_2$ and write $w_{\ga_2}(\bfx_2,\bfxp_2)=w_{\ga_2}(\bfx_1,\bfxp_1)+w_{\ga_2}(\bfxp_1,\bfxp_2)$. With $\bfY$ defined as $\bfY=(\bfxp_1,\bfxp_2)$, substituting this equation to the generalized structure function $F_n$ results in the following balance between the three velocity difference geometries, shown in Fig.~\ref{fig:p-one-case-two-fusion-rule} and given by
\begin{multline}
F_n (\bfX_1,\bfX_2,\{\bfX_k\}_{k=3}^n,t)=\\ F_n (\bfX_1,\bfX_1,\{\bfX_k\}_{k=3}^n,t)+F_n (\bfX_1,\bfY,\{\bfX_k\}_{k=3}^n,t)
\label{eq:p-one-case-two-fusion-rule}
\end{multline}
 If the leading fusion rule contribution had not been eliminated, then as we fuse $(\bfx_1,\bfxp_1)$ the dominant balance would have been between the two $p=1$ geometry terms. However, the same universal local isotropy postulate responsible for eliminating the leading contribution, also allows us to argue that the two $p=1$ terms differ only by a proportionality constant. Furthermore, eliminating the leading-order contributions shifts the dominant balance so that the $p=1$ terms and the $p=2$ terms have the same scaling. This argument applies both to downscale and upscale cascades, so in both cases, the $p=1$ fusion rule becomes $F_n^{(1)}(r,R)\sim r^{\xi_{n,1}}R^{\gz_n-\xi_{n,1}}$ with $\xi_{n,1}=\gz_2$ for a downscale cascade and $\xi_{n,1}=\gz_n-\gz_{n-2}$ for an upscale cascade, for all $n>3$. For the special case $n=3$, after the removal of the leading order contribution, we have a velocity difference geometry with two small velocity differences at scale $r$ and one large velocity difference at scale $R$. According to the argument of the next section, this particular velocity difference geometry results in additional cancellation, with the next-order contribution consisting of a velocity difference geometry with three small velocity differences at scale $r$. The corresponding fusion scaling exponents are $\xi_{3,1}=\gz_3$ for both downscale and upscale cascades. Another special case is $n=2$, where there are simply no additional velocity differences, so the resulting fusion scaling exponent is $\xi_{2,1}=\gz_2$, both for downscale and upscale cascades.  It is also worth noting that decomposing $w_{\ga_1}$ instead of $w_{\ga_2}$ in the above argument results in 3 distinct velocity difference geometries, leading to a dead end. 

Now, let us consider the more general case where $R_{\mathrm{min}}\sim r$. This case can be reduced to the $R_{\mathrm{min}}=0$ case by reattaching the fused velocity difference $w_{\ga_1}$ back onto the nearest neighboring velocity difference endpoint. This is done by the decomposition $w_{\ga_1}(\bfx_1,\bfxp_1,t)=w_{\ga_1}(\bfx_1,\bfx_2,t)+w_{\ga_1}(\bfx_2,\bfxp_1)$, as shown on Fig.~\ref{fig:p-one-case-two-detached-fusion-rule}, where the other unfused velocity differences are omitted. This results in breaking the generalized structure function $F_n$ into two contributions given by
\begin{multline}
F_n (\{\bfX\}_n, t)=F_n (\bfx_1,\bfx_2, \bfx_2,\bfxp_2,\{\bfX_k\}_{k=3}^n,t)\\+F_n (\bfx_2,\bfxp_1,\bfx_2,\bfxp_2,\{\bfX_k\}_{k=3}^n,t)
\label{eq:p-one-case-two-detached-fusion-rule}
\end{multline}
Both of these contributions correspond to the case $R_{\mathrm{min}}=0$, discussed previously, and therefore both will scale as described above. Consequently, the overall generalized structure function $F_n$ maintains the same scaling. 

\begin{figure}[t]\begin{center}
\begin{pspicture}[shift=*](-1,-1)(12,5)
\psline{*-*}(1,1)(10,1)
\psline{*-*}(1,1)(1,3)
\psline{*-*}(1,3)(10,3)
\uput[d](5,1){$R$}
\uput[u](5,3){$R$}
\uput[l](1,2){$r$}
\end{pspicture}
\caption{\label{fig:p-one-double-attachment-one}\small A $p=1$ velocity difference geometry where the fused velocity difference is attached to unfused vrlocity differences from both sides. }
\end{center}\end{figure}
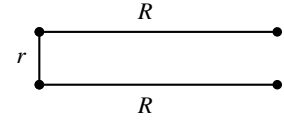

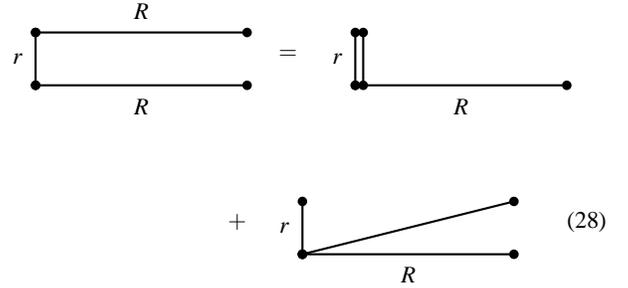
\begin{figure}[t]\begin{center}
\begin{multline}
\begin{pspicture}[shift=*](-1,-1)(10,5)
\psline{*-*}(1,1)(9,1)
\psline{*-*}(1,1)(1,3)
\psline{*-*}(1,3)(9,3)
\uput[d](5,1){$R$}
\uput[u](5,3){$R$}
\uput[l](1,2){$r$}
\end{pspicture}
=
\begin{pspicture}[shift=*](-1,-1)(10,5)
\psline{*-*}(1,1)(9,1)
\psline{*-*}(1,1)(1,3)
\psline{*-*}(1.3,1)(1.3,3)
\uput[d](5,1){$R$}
\uput[l](1,2){$r$}
\end{pspicture}\\
+
\begin{pspicture}[shift=*](-1,-1)(10,5)
\psline{*-*}(1,1)(9,1)
\psline{*-*}(1,1)(1,3)
\psline{*-*}(1,1)(9,3)
\uput[d](5,1){$R$}
\uput[l](1,2){$r$}
\end{pspicture} 
\end{multline}
\caption{\label{fig:p-one-double-attachment-two}\small Decomposing the velocity difference geometry of Fig~\ref{fig:p-one-double-attachment-one} into two contributions.}
\end{center}\end{figure}

\begin{figure}[tb]\begin{center}
\begin{align*}
\begin{pspicture}[shift=*](-1,-1)(5,10)
\psline{*-*}(1,1)(1,8)
\psline{*-*}(1,1)(3,1)
\psline{*-*}(1,1)(3,8)
\uput[l](1,5){$R$}
\uput[d](2,1){$r$}
\end{pspicture}
&=
\begin{pspicture}[shift=*](-1,-1)(5,10)
\psline{*-*}(1,1)(1,8)
\psline{*-*}(1,1)(3,1)
\psline{*-*}(1,1.3)(3,1.3)
\uput[l](1,5){$R$}
\uput[d](2,1){$r$}
\end{pspicture}
+
\begin{pspicture}[shift=*](-1,-1)(5,10)
\psline{*-*}(1,1)(1,8)
\psline{*-*}(1,1)(3,1)
\psline{*-*}(3,1)(3,8)
\uput[l](1,5){$R$}
\uput[r](3,5){$R$}
\uput[d](2,1){$r$}
\end{pspicture}
\\
&=
\begin{pspicture}[shift=*](-1,-1)(5,10)
\psline{*-*}(1,1)(1,8)
\psline{*-*}(1,1)(3,1)
\psline{*-*}(1,1.3)(3,1.3)
\uput[l](1,5){$R$}
\uput[d](2,1){$r$}
\end{pspicture}
+
\begin{pspicture}[shift=*](-1,-1)(5,10)
\psline{*-*}(3,1)(3,8)
\psline{*-*}(1,1)(3,1)
\psline{*-*}(1,1.3)(3,1.3)
\uput[l](1,5){$R$}
\uput[d](2,1){$r$}
\end{pspicture} \\
&\qquad +
\begin{pspicture}[shift=*](-1,-1)(5,10)
\psline{*-*}(3,1)(3,8)
\psline{*-*}(1,1)(3,1)
\psline{*-*}(3,1)(1,8)
\uput[l](1,5){$R$}
\uput[d](2,1){$r$}
\end{pspicture}
\end{align*}
\caption{\label{fig:p-one-double-attachment-three}\small Decomposing the two unfused velocity differences through the fused velocity difference, effectively moving them to the other endpoint of the fused velocity difference}
\end{center}\end{figure}
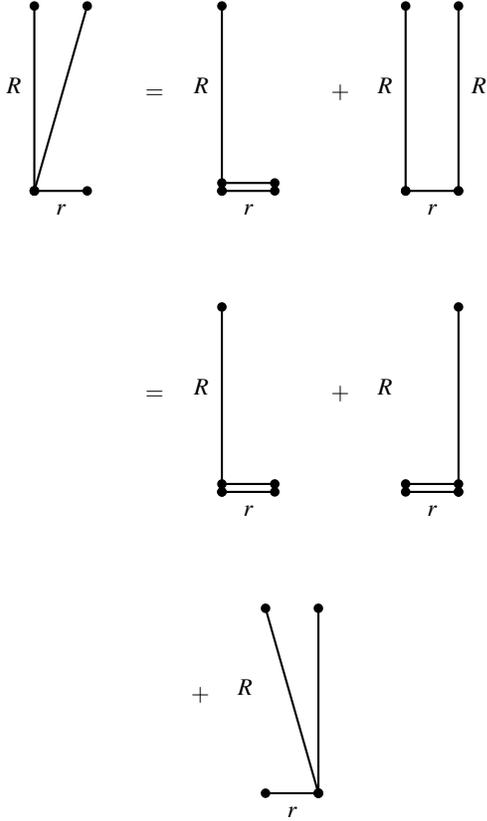

The multilocality proof requires us to consider an additional geometry in which the fused velocity difference is attached to two unfused velocity differences as shown in Fig.~\ref{fig:p-one-double-attachment-one}. Decomposing one of the fused velocity differences into a sum of two unfused velocity differences similarly to the argument that we gave previously for the $R_{\mathrm{min}}=0$ case gives two contributions that are shown in Fig.~\ref{fig:p-one-double-attachment-two}. The first contribution follows the scaling of the $p=2$ fusion rule. The second contribution is a variation of the $R_{\mathrm{min}}=0$ case of the $p=1$ fusion rule where two unfused velocity differences are attached to the same endpoint of the fused velocity difference. We can now argue that attaching the additional unfused velocity difference has no effect on the overall scaling which will still follow the $p=2$ fusion rule. This can be done by decomposing both unfused velocity differences and running them through the fused velocity difference, as shown in Fig.~\ref{fig:p-one-double-attachment-three}. This has the effect of reattaching the two unfused velocity differences to the other endpoint of the fused velocity difference, and results in three terms. The first two terms have two fused velocity differences and scale according to the $p=2$ fusion rule. For the third term, we note that it can be obtained by rotating the fused velocity difference relative to the two unfused velocity differences. There is the problem that the angle between the two unfused velocity differences also changes by reattaching them to the other endpoint of the fused velocity difference. On the other hand, from the universal local isotropy assumption, the left hand side term in Fig.~\ref{fig:p-one-double-attachment-three} and the third right hand side term have the same tensor structure with respect to all velocity difference separations, except for a numerical constant dependent on all angles. It follows that the second contribution in Fig.~\ref{fig:p-one-double-attachment-two} also scales according to the $p=2$ fusion rule, by repeating the previous dominant balance argument.

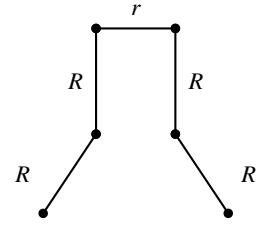
\begin{figure}[tb]\begin{center}
\begin{pspicture}[shift=*](-1,-1)(8,9)
\psline{*-*}(3,8)(6,8)
\uput[u](4.5,8){$r$}
\psline{*-*}(3,8)(3,4)
\psline{*-*}(6,8)(6,4)
\uput[l](3,6){$R$}
\uput[r](6,6){$R$}
\psline{*-*}(3,4)(1,1)
\psline{*-*}(6,4)(8,1)
\uput[l](1,2.5){$R$}
\uput[r](8,2.5){$R$}
\end{pspicture}
\caption{\label{fig:p-one-multiple-attachment}\small The fused velocity difference embedded within a chain of unfused velocity differences}
\end{center}\end{figure}

It is easy to see that in Fig.~\ref{fig:p-one-double-attachment-three}, attaching additional unfused velocity differences on the same endpoint as the fused velocity difference, does not change the scaling with respect to $r$ and $R$. We simply use the same procedure to reattach all of the unfused velocity differences onto the other side of the fused velocity difference one by one, and obtain a series of terms with two fused velocity differences and an additional term with the same tensor structure as the left hand side. Likewise, in the argument of Fig.~\ref{fig:p-one-double-attachment-two}, it makes no difference if the fused velocity difference is embedded within a chain of unfused velocity differences as shown in Fig.~\ref{fig:p-one-multiple-attachment}. The argument of Fig.~\ref{fig:p-one-double-attachment-two} still carries through with no need for any additional considerations. 

\subsection{Fusion rule for $p=n-1$}

\begin{figure}[tb]\begin{center}
\begin{pspicture}[shift=*](-3,-6)(5,3)
\pscircle(0,0){2}
\psline{*-*}(0,1)(1,-1)
\psline{*-*}(-1,0)(1,1)
\psline{*-*}(0,-1)(0,-6)
\uput[l](0,-4){$R$}
\psline{|<->|}(3,2)(3,-2)
\uput[l](3,0){$r$}
\end{pspicture}
\caption{\label{fig:p-n-minus-one-geometry}\small The $p=n-1$ fusion rule velocity difference geometry where one endpoint of the unfused velocity difference is within the $r$-blob where the other $n-1$ velocity differences are gathered}
\end{center}\end{figure}
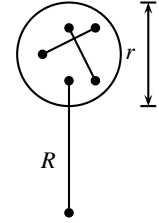

The fusion rule for the case $p=n-1$ also has the same predicament as the previously considered case $p=1$ in that the leading order contribution vanishes and we need to determine the next subleading contribution. The velocity difference geometry under consideration is shown on Fig.~\ref{fig:p-n-minus-one-geometry}  and consists of $n-1$ fused velocity differences at scale $r$ congregated together inside a blob with size $r$ and one unfused velocity difference at scale $R$ with one endpoint inside the $r$-blob and the other endpoint outside the $r$-blob. Although we do not encounter this particular velocity difference geometry in either the locality or multilocality proofs, it is a necessary stepping stone for analysing the two-blob fusion rule described in the next subsection. It is also relevant with regards to the additional cancellation that underlies the claim $\xi_{3,1}=\gz_{3}$.

We begin with noting that the leading contributions predicted by the $p=n-1$ fusion rule vanish for the same reasons as in the case of the $p=1$ fusion rule. For an upscale cascade, the $p=n-1$ fusion rule predicts that the generalized structure function $F_n$ is given by
\begin{equation}
F_n (\{\bfX\}_n,t)=F_1 (\bfx_1,\bfxp_1,t)\Psi_{n,n-1}(\{\bfX\}_{k=2}^n,t)
\end{equation}
and the $F_1$ factor vanishes immediately by homogeneity. For a downscale cascade, the universal local isotropy hypothesis suggests that $F_n$ should remain invariant upon rotating the fused velocity differences around their geometric center. Equivalently, we expect invariance upon fixing the fused velocity differences and rotating the unfused velocity difference instead. It follows that the tensor structure of the leading fusion rule contribution should take the form
\begin{align}
F_n (\{\bfX\}_n,t) &= \tilde{F}_{n-1}(\{\bfX\}_{k=2}^n,t)\Psi_{n,n-1}(\bfX_1,t) \\
&= \tilde{F}_{n-1}(\{\bfX\}_{k=2}^n,t)C\frac{R_{\ga_1}}{R}\gy (R,t)
\end{align}
with $\bfR=\bfx_1-\bfxp_1$, and $R=\nrm{R}$ the norm of $\bfR$. From the incompressibility condition we have $\pd_{\bfx_1,\ga_1} F_n^{\ga_1\ldots\ga_n}(\{\bfX\}_n,t)=0$, and using the same mathematical argument as in the case $p=1$ for an upscale cascade, we find that $\gy (R,t)=CR^{1-d}$, with $d$ the dimension of space. Since $R$  was supposed to scale as $R^{\gz_n-\gz_{n-1}}$, it follows that $C=0$, and the leading term, once again, vanishes.

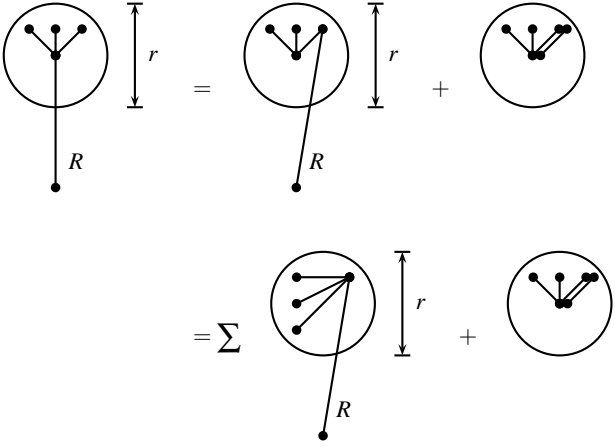
\begin{figure}[tb]\begin{center}
\begin{align*}
\begin{pspicture}[shift=*](-3,-6)(5,3)
\pscircle(0,0){2}
\psline{*-*}(0,0)(0,1)
\psline{*-*}(0,0)(-1,1)
\psline{*-*}(0,0)(1,1)
\psline{*-*}(0,0)(0,-5)
\uput[r](0,-4){$R$}
\psline{|<->|}(3,2)(3,-2)
\uput[r](3,0){$r$}
\end{pspicture}
&=
\begin{pspicture}[shift=*](-3,-6)(5,3)
\pscircle(0,0){2}
\psline{*-*}(0,0)(0,1)
\psline{*-*}(0,0)(-1,1)
\psline{*-*}(0,0)(1,1)
\psline{*-*}(1,1)(0,-5)
\uput[r](0,-4){$R$}
\psline{|<->|}(3,2)(3,-2)
\uput[r](3,0){$r$}
\end{pspicture}
+
\begin{pspicture}[shift=*](-3,-6)(5,3)
\pscircle(0,0){2}
\psline{*-*}(0,0)(0,1)
\psline{*-*}(0,0)(-1,1)
\psline{*-*}(0,0)(1,1)
\psline{*-*}(0.3,0)(1.3,1)
\end{pspicture} \\
&= \sum
\begin{pspicture}[shift=*](-3,-6)(5,3)
\pscircle(0,0){2}
\psline{*-*}(1,1)(-1,1)
\psline{*-*}(1,1)(-1,0)
\psline{*-*}(1,1)(-1,-1)
\psline{*-*}(1,1)(0,-5)
\uput[r](0,-4){$R$}
\psline{|<->|}(3,2)(3,-2)
\uput[r](3,0){$r$}
\end{pspicture}
+
\begin{pspicture}[shift=*](-3,-6)(5,3)
\pscircle(0,0){2}
\psline{*-*}(0,0)(0,1)
\psline{*-*}(0,0)(-1,1)
\psline{*-*}(0,0)(1,1)
\psline{*-*}(0.3,0)(1.3,1)
\end{pspicture}
\end{align*}
\caption{\label{fig:p-n-minus-one-next-order-contribution}\small Dominant balance argument for determining the next-order contribution for the $p=n-1$ fusion rule.}
\end{center}\end{figure}

To determine the scaling of the subleading contribution to $F_n$, we assume with no loss of generality that the unfused velocity difference $w_{\ga_1}$ is attached to all of the fused velocity differences inside the $r$-blob. We decompose $w_{\ga_1}$ into a sum of one fused velocity difference and an unfused velocity difference, as shown in Fig.~\ref{fig:p-n-minus-one-next-order-contribution}. Then we reattach the other fused velocity differences onto the same endpoint as the unfused velocity difference $w_{\ga_1}$. This results in a series of terms that have the same configuration of velocity differences as the left hand side, albeit with different angles between the velocity differences, and one additional term where all velocity differences lie within the $r$-blob. Given the elimination of the leading order contribution, we predict via a dominant balance argument that the subleading contribution will scale according to $r^{\gz_n}$. This results in the evaluation $\xi_{n,n-1}=\gz_n$, which holds both for downscale and upscale cascades. For more general velocity difference geometries within the $r$-blob, it is straightforward to reattach all velocity differences onto the same endpoint, resulting in a sum of terms whose geometry is similar to the left hand side of Fig.~\ref{fig:p-n-minus-one-next-order-contribution}.

\subsection{Fusion rule for the two-blob geometry}
\label{sec:fusion-rule-two-blob-geometry}

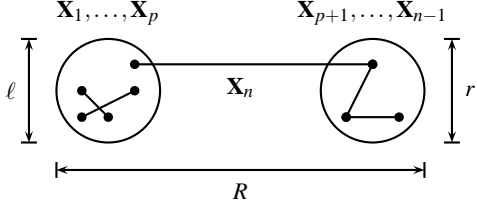
\begin{figure}[tb]\begin{center}
\begin{pspicture}[shift=*](-3,-4.5)(13,4)
\pscircle(0,0){2}
\psline{*-*}(-1,0)(0,-1)
\psline{*-*}(-1,-1)(1,0)
\psline{|<->|}(-3,-2)(-3,2)
\uput[l](-3,0){$\ell$}
\uput[u](0,2){$\bfX_1,\ldots,\bfX_p$}
\pscircle(10,0){2}
\psline{*-*}(10,1)(9,-1)
\psline{*-*}(9,-1)(11,-1)
\psline{|<->|}(13,-2)(13,2)
\uput[r](13,0){$r$}
\uput[u](10,2){$\bfX_{p+1},\ldots,\bfX_{n-1}$}
\psline{*-*}(1,1)(10,1)
\uput[d](5,1){$\bfX_n$}
\psline{|<->|}(-2,-3)(12,-3)
\uput[d](5,-3){$R$}
\end{pspicture}
\caption{\label{fig:the-two-blob-vel-d-geometry}\small The two-blob velocity difference geometry with $\ell\ll R$ and $r\ll R$ and $\ell, r, R$ all within the inertial range. }
\end{center}\end{figure}

To formulate a locality proof for the general terms of $\cO_n \cO_{n+1}\cdots\cO_{n+p-1} F_{n+p}$ it  becomes now necessary to give special consideration to a new velocity difference geometry shown in Fig.~\ref{fig:the-two-blob-vel-d-geometry}: there are two blobs with length scales $r$ and $l$ separated by a distance scale $R$. The $r$-blob holds a congregation of $n-p-1$ velocity differences with scale $r$ point separations, the $l$-blob holds $p$ velocity differences with scale $l$ point separations, and the remaining velocity difference has length scale $R$ with one end inside the $r$-blob and the other end inside the $l$-blob. We take the intermediate asymptotic limits $r \ll R$ and $l\ll R$ with $r$, $l$, and $R$ all within the inertial range. Note that the velocity differences in the $l$-blob will be attached to each other in the geometries that arise from the locality integrals, however they can be detached with no consequence to the fusion rule scaling exponents, as long as all endpoints remain separated at length scale $l$. 

The case of upscale cascades presents us with a technical difficulty that will be discussed below. For the case of downscale cascades, the  rotational invariance argument that was used in the previous subsection  can be now repeated from the viewpoint of either blob.  For the special case $p=1$, this reduces to the fusion geometry needed to establish the IR locality of $\cO_n F_{n+1}$. However the resulting scaling with respect to $R$ is stronger, in the sense that it helps establish a faster vanishing of the integrals in the IR limit, than the fusion scaling that was used previously to establish the IR locality of $\cO_n F_{n+1}$ \cite{article:Procaccia:1996:1,article:Procaccia:1996:3,article:Gkioulekas:2008:1}. The previous argument was to begin with all velocity differences at scale $R$, reduce one velocity difference to scale $\ell$ and another group of $n-2$ velocity differences to scale $r$, with one velocity difference remaining at scale $R$ \cite{article:Gkioulekas:2008:1}. The problem is that this two-step process does not require congregating the velocity differences into two small-scale blobs, separated from each other at scale $R$, as the endpoint separation of the velocity differences is reduced. The resulting scaling with respect to $R$ gives us IR locality only marginally, requiring an additional workaround to obtain better scaling.  Taking advantage of the assumption that, aside from the $R$ velocity difference, the remaining velocity differences are confined within their respective blobs allows us to directly obtain stronger scaling with respect to $R$. This is a valid assumption for the velocity difference geometries that arise in the local interaction integrals, but none of the previous arguments \cite{article:Procaccia:1996:1,article:Procaccia:1996:3,article:Gkioulekas:2008:1} took full advantage of it. 

First we will explain why the leading fusion rule contribution vanishes. Then we obtain the scaling of the subleading contribution. For an upscale cascade, the same argument that was used in the previous subsection for the $p=n-1$ fusion rule to show that the leading fusion rule contribution vanishes, carries over without any need for modifications. For a downscale cascade, a more careful argument is needed. The universal local isotropy hypothesis implies that the generalized structure function $F_n$ should remain invariant upon rotating the velocity differences within either the $r$-blob or the $\ell$-blob, or both around the corresponding geometric centers (see Fig.\ref{fig:the-two-blob-vel-d-geometry}). We note that rotating the velocity differences within both blobs  with the same angle and direction is equivalent to rotating the $R$-scale velocity difference around either one blob, while carrying along the opposite blob. If the vector $\bfR$ represents the endpoint separation in the $R$-scale velocity difference $w_{\ga_n}(\bfX_n, t)$, then, from the point of view  of the $\ell$-blob, we can assume that the velocity differences of the $r$-blob have their endpoints shifted by $\bfR$. It follows that the leading fusion rule contribution from the point of view of the $\ell$-blob is given by:
\begin{multline}
F_n^{\ldots\ga_n}(\{\bfX\}_n, t) = C F_p (\{\bfX_k\}_{k=1}^p, t) \Psi_{n,p}^{\ga_n}(\bfR, \{\bfX\}_{k=p+1}^{n-1}+\bfR, t) \\
=  C F_p (\{\bfX_k\}_{k=1}^p, t) (R_{\ga_n}/R)\tilde{\Psi}_{n,p}(R, \{\bfX\}_{k=p+1}^{n-1}+\bfR, t)
\end{multline}
Here, the notation $\{\bfX\}_{k=p+1}^{n-1}+\bfR$ represents shifting all endpoints of velocity differences in $\{\bfX\}_{k=p+1}^{n-1}$ equally by the vector $\bfR$. The universal local homogeneity hypothesis \cite{article:Gkioulekas:2008:1}  implies that if we shift the velocity differences of one blob by a small distance, in some direction, relative to the velocity differences of the other blob and the $w_{\ga_n}$ velocity difference, then the overall generalized structure function should remain invariant. This, in turn, implies that 
\begin{widetext}
\begin{equation}
\sum_{m=p+1}^{n-1}(\pd_{\ga_n,\bfx_m}+\pd_{\ga_n,\bfxp_m}) \Psi_{n,p}^{\ga_n}(\bfR, \{\bfX\}_{k=p+1}^{n-1}+\bfR, t) =0
\end{equation}
which reduces to
\begin{equation}
\frac{R_{\ga_n}}{R}\sum_{m=p+1}^{n-1}(\pd_{\ga_n,\bfx_m}+\pd_{\ga_n,\bfxp_m}) \tilde{\Psi}_{n,p}(R, \{\bfX\}_{k=p+1}^{n-1}+\bfR, t) =0 \label{eq:incompressibility-trick}
\end{equation}
since both $R_{\ga_n}$ and $R$ are constant with respect to the derivatives $\pd_{\ga_n,\bfx_m}$ and $\pd_{\ga_n,\bfxp_m}$ for all $m\in \{p+1, \ldots, n-1\}$.  We can use this equation to write the divergence of the generalized structure function with respect to $(\ga_n,\bfx_n)$ as: 
\begin{align}
\pd_{\ga_n,\bfx_n} F_n^{\ldots\ga_n}(\{\bfX\}_n, t) 
&= C F_p (\{\bfX_k\}_{k=1}^p, t) \pd_{\ga_n,\bfx_n} \left[\frac{R_{\ga_n}}{R} \tilde{\Psi}_{n,p}(R, \{\bfX\}_{k=p+1}^{n-1}+\bfR, t) \right] \\
&= C F_p (\{\bfX_k\}_{k=1}^p, t) \left[  \tilde{\Psi}_{n,p}(R, \{\bfX\}_{k=p+1}^{n-1}+\bfR, t) \pd_{\ga_n,\bfx_n} \fracp{R_{\ga_n}}{R} + \frac{R_{\ga_n}}{R} \pd_{\ga_n,\bfx_n} \tilde{\Psi}_{n,p}(R, \{\bfX\}_{k=p+1}^{n-1}+\bfR, t) \right] \\
&= C F_p (\{\bfX_k\}_{k=1}^p, t) \left[ \frac{d-1}{R}\tilde{\Psi}_{n,p}(R, \{\bfX\}_{k=p+1}^{n-1}+\bfR, t) + \frac{R_{\ga_n}}{R} \left( \pderiv{}{R} \tilde{\Psi}_{n,p}(R, \{\bfX\}_{k=p+1}^{n-1}+\bfR, t) \right) (\pd_{\ga_n,\bfx_n} R) \right. \\
&\qquad\left. +\frac{R_{\ga_n}}{R}\sum_{m=p+1}^{n-1}(\pd_{\ga_n,\bfx_m}+\pd_{\ga_n,\bfxp_m}) \tilde{\Psi}_{n,p}(R, \{\bfX\}_{k=p+1}^{n-1}+\bfR, t) \right] \label{eq:incompressibility-trick-used-here} \\
&= C F_p (\{\bfX_k\}_{k=1}^p, t) \left[ \frac{d-1}{R}+\frac{R_{\ga_n}R_{\ga_n}}{R^2}\pderiv{}{R}\right] \tilde{\Psi}_{n,p}(R, \{\bfX\}_{k=p+1}^{n-1}+\bfR, t) \\
&= C F_p (\{\bfX_k\}_{k=1}^p, t) \left[ \frac{d-1}{R}+\pderiv{}{R}\right] \tilde{\Psi}_{n,p}(R, \{\bfX\}_{k=p+1}^{n-1}+\bfR, t)
\end{align}
\end{widetext}
Here, $\pderivin{}{R}$ represents a scalar partial derivative with respect to the scalar argument of $\tilde{\Psi}_{n,p}$, notwithstanding the dependence of other arguments of $\tilde{\Psi}_{n,p}$ on the vector $\bfR$. We also use $\pd_{\ga_n,\bfx_n} R = R_{\ga_n}/R$ and $R_{\ga_n}R_{\ga_n}=R^2$and Eq.~\eqref{eq:incompressibility-trick} to eliminate the third term of Eq.~\eqref{eq:incompressibility-trick-used-here}. From the incompressibility condition $\pd_{\ga_n,\bfx_n} F_n^{\ldots\ga_n}(\{\bfX\}_n, t) =0$, we obtain once again the equation
\begin{equation}
C\left[ \frac{d-1}{R} + \pderiv{}{R}\right] \tilde{\Psi}_{n,p}(R, \{\bfX\}_{k=p+1}^{n-1}+\bfR, t) =0
\end{equation}
and using the same argument as in the preceding fusion rules with $p=1$ and $p=n-1$, we conclude that $C=0$ and that the leading contribution to the two-blob geometry fusion rule vanishes, so we must consider the subleading terms. 

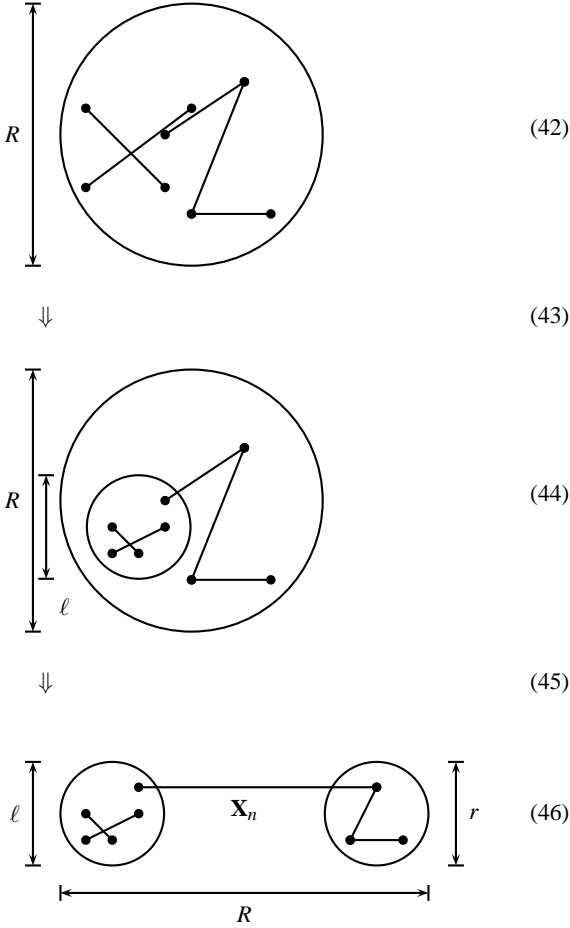
\begin{figure}[tb]\begin{center}
\begin{align}
& \begin{pspicture}[shift=*](-6,-6)(6,6)
\pscircle(0,0){5}
\psline{|<->|}(-6,-5)(-6,5)
\uput[l](-6,0){$R$}
\psline{*-*}(-1,0)(2,2)
\psline{*-*}(2,2)(0,-3)
\psline{*-*}(0,-3)(3,-3)
\psline{*-*}(-4,1)(-1,-2)
\psline{*-*}(-4,-2)(0,1)
\end{pspicture} \\
&\Downarrow \\
&\begin{pspicture}[shift=*](-6,-6)(6,6)
\pscircle(0,0){5}
\psline{|<->|}(-6,-5)(-6,5)
\uput[l](-6,0){$R$}
\psline{*-*}(-1,0)(2,2)
\psline{*-*}(2,2)(0,-3)
\psline{*-*}(0,-3)(3,-3)
\pscircle(-2,-1){2}
\psline{*-*}(-3,-1)(-2,-2)
\psline{*-*}(-3,-2)(-1,-1)
\psline{|<->|}(-5.5,-3)(-5.5,1)
\uput[r](-5.5,-4){$\ell$}
\end{pspicture} \\
&\Downarrow \\
&\begin{pspicture}[shift=*](-3,-4.5)(13,4)
\pscircle(0,0){2}
\psline{*-*}(-1,0)(0,-1)
\psline{*-*}(-1,-1)(1,0)
\psline{|<->|}(-3,-2)(-3,2)
\uput[l](-3,0){$\ell$}
\pscircle(10,0){2}
\psline{*-*}(10,1)(9,-1)
\psline{*-*}(9,-1)(11,-1)
\psline{|<->|}(13,-2)(13,2)
\uput[r](13,0){$r$}
\psline{*-*}(1,1)(10,1)
\uput[d](5,1){$\bfX_n$}
\psline{|<->|}(-2,-3)(12,-3)
\uput[d](5,-3){$R$}
\end{pspicture}
\end{align}
\caption{\label{fig:construction-two-blob-geometry-downscale}\small Construction of the two-blob geometry for an downscale cascade in two steps}
\end{center}\end{figure}

For the next step of determining the subleading contribution to the fusion rule, we distinguish and treat separately the cases of a downscale vs an upscale cascade. For downscale cascades, the two-blob geometry can be constructed in two steps, as shown in Fig.~\ref{fig:construction-two-blob-geometry-downscale}. We begin with $n$ velocity differences at scale $R$. We reduce $p<n$ velocity differences to scale $\ell\ll R$, thereby creating a smaller $\ell$-blob inside the larger $R$ blob. We then reduce another $n-p-1$ velocity differences to scale $r\ll R$, also concentrating them within a separate small $r$-blob. One velocity difference remains at scale $R$ straddling between the two small blobs. The leading contribution to the corresponding fusion rule is $\ell^{\gz_p} r^{\gz_{n-p-1}}R^{\gz_n-\gz_{n-p-1}-\gz_p}$. However, in the second step we can invoke our previous argument and claim that this leading contribution with respect to $r$ vanishes. To pick up the next-order contribution, we rearrange the velocity differences inside the $r$-blob so that one of them shares an endpoint with the $w_{\ga_n}$ velocity difference straddling between the two blobs. Using an argument similar to the one used for the $p=1$ fusion rule (see Fig.~\ref{fig:p-one-case-two-fusion-rule}), we find that the next order contribution scales as $r^{\gz_{n-p}}$. Repeating the argument by reversing the sequence by which the two small blobs are created gives the next-order scaling with respect to $\ell$ as $\ell^{\gz_{p+1}}$. The overall scaling for the two-blob geometry is thus $\ell^{\gz_{p+1}} r^{\gz_{n-p}} R^{\gz_n-\gz_{p+1}-\gz_{n-p}}$. Note that the next-order contributions correspond to geometries where the velocity difference between the two blobs is no longer present, so the previous incompressibility argument cannot be repeated. In terms of the generalized fusion-rule scaling exponents, the resulting scaling law is $F_n \sim \ell^{\xi_{n,p+1}} r^{\xi_{n,n-p}} R^{\gz_n-\xi_{n,p+1}-\xi_{n,n-p}}$.

\begin{figure}[tb]\begin{center}
\begin{align*}
&\begin{pspicture}[shift=*](-6,-5)(5,5)
\pscircle(0,0){5}
\psline{*-*}(-3,1)(-1,-1)
\psline{*-*}(-3,-1)(1,1)
\psline{*-*}(1,2)(3,2)
\psline{*-*}(4,2)(2,-1)
\psline{*-*}(2,-1)(4,-1)
\psline{|<->|}(-5.5,-5)(-5.5,5)
\uput[l](-5.5,0){$r$}
\end{pspicture} \\
&\Downarrow \\
&\begin{pspicture}[shift=*](-3.5,-3)(13,3.5)
\pscircle(0,0){3}
\psline{*-*}(-1,0)(0,-1)
\psline{*-*}(-1,-1)(1,0)
\psline{*-*}(1,1)(2,1)
\psline{|<->|}(-3.5,-3)(-3.5,3)
\uput[l](-3.5,0){$r$}
\pscircle(10,0){2}
\psline{*-*}(10,1)(9,-1)
\psline{*-*}(9,-1)(11,-1)
\psline{|<->|}(12.5,-2)(12.5,2)
\uput[r](12.5,0){$r$}
\psline{|<->|}(-3,3.5)(13,3.5)
\uput[u](5,3.5){$R$}
\end{pspicture} \\
&\Downarrow \\
&\begin{pspicture}[shift=*](-3,-4.5)(13,4)
\pscircle(0,0){2}
\psline{*-*}(-1,0)(0,-1)
\psline{*-*}(-1,-1)(1,0)
\psline{|<->|}(-3,-2)(-3,2)
\uput[l](-3,0){$r$}
\pscircle(10,0){2}
\psline{*-*}(10,1)(9,-1)
\psline{*-*}(9,-1)(11,-1)
\psline{|<->|}(13,-2)(13,2)
\uput[r](13,0){$r$}
\psline{*-*}(1,1)(10,1)
\psline{|<->|}(-2,-3)(12,-3)
\uput[d](5,-3){$R$}
\end{pspicture}
\end{align*}
\caption{\label{fig:construction-two-blob-geometry-upscale}\small Construction of the two-blob geometry for an upscale cascade requires a different sequence of steps than it does for a downscale cascade}
\end{center}\end{figure}
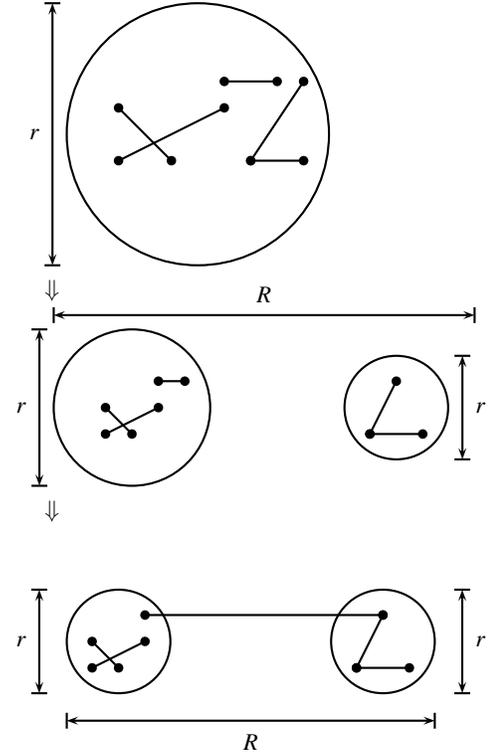

The case of an upscale cascade is very subtle because the same two-blob geometry is constructed by a different sequence of fusion events: we begin with all velocity differences inside a blob at scale $r$. We separate $p$ velocity differences by shifting them across a large distance $R$ with $r \ll R$, without resizing them, resulting in two separated blobs, each with size $r$. Then we pick one velocity difference from either blob and expand it to size $R$ so that it straddles between the two blobs. This process is shown in Fig.~\ref{fig:construction-two-blob-geometry-upscale}. Recall that in upscale cascades ``fusing'' corresponds to expanding a velocity difference from small scales to large scales, that being the direction away from the forcing range. So, the overall process of constructing the two-blob geometry involves only one fusion event, preceded by the blob separation. We will show that the resulting scaling is $r^{\gz_n+\ga} R^{-\ga}$, with $\ga$ a scaling exponent that is expected to satisfy $\ga>0$, noting that the scaling exponent of $R$ is the one relevant to the multilocality proof.  

\begin{figure}[tb]\begin{center}
\begin{equation*}
\begin{pspicture}[shift=*](-2,-7)(2.5,2)
\pscircle(0,0){2}
\psline{*-*}(-1,0)(0,-1)
\psline{*-*}(-1,-1)(1,0)
\psline{*-*}(-1,-6)(1,-6)
\end{pspicture}
=
\begin{pspicture}[shift=*](-2,-7)(2.5,2)
\pscircle(0,0){2}
\psline{*-*}(-1,0)(0,-1)
\psline{*-*}(-1,-1)(1,0)
\psline[arrowsize=2pt 5,arrowlength=2]{*->}(-1,-6)(0,-1)
\psdots[dotstyle=*](1,-6)
\end{pspicture}
+
\begin{pspicture}[shift=*](-2,-7)(2.5,2)
\pscircle(0,0){2}
\psline{*-*}(-1,0)(0,-1)
\psline{*-*}(-1,-1)(1,0)
\psline[arrowsize=2pt 5,arrowlength=2]{<-*}(1,-6)(0,-1)
\psdots[dotstyle=*](1,-6)
\psdots[dotstyle=*](-1,-6)
\end{pspicture}
\end{equation*}
\caption{\label{fig:separation-of-one-vel-d-from-one-blob}\small Separating one velocity difference away from a group of velocity differences congregated inside a small-scale blob. The arrows indicate the direction of the velocity differences involved in the major cancellation of the leading $r^{\gz_n}R^0$ contribution. }
\end{center}\end{figure}
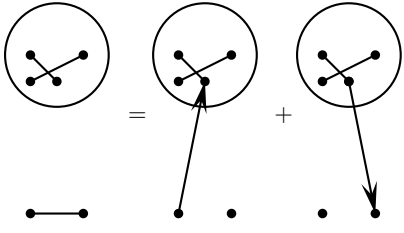

We begin with considering the special case in which only one velocity difference is pulled away at distance $R$ from the rest of the group, as shown in Fig.~\ref{fig:separation-of-one-vel-d-from-one-blob}. We can decompose the remote velocity difference by bouncing it off the endpoint of one of the other velocity differences inside the blob and obtain the sum of two generalized structure functions that correpond to the $p=n-1$ fusion rule discussed previously. Since they scale as $r^{\gz_n} R^0$, the separated geometry shown in Fig.~\ref{fig:separation-of-one-vel-d-from-one-blob}  can be expected to also scale as $r^{\gz_n} R^0$. However, the situation is in fact much better, because in the limit $R\to +\infty$, the two terms in the right-hand-side of Fig.~\ref{fig:separation-of-one-vel-d-from-one-blob}  become nearly identical and are being in fact subtracted from each other, as the large velocity difference has opposite direction between the two terms. This should result in a leading-order term cancellation, with the next order term scaling as $r^{\gz_n} R^0 (r/R)^{\ga} \sim r^{\gz_n+\ga} R^{-\ga}$. Since there are no fusion events in perturbing the first term onto the second term of the right-hand-side of the equation in Fig.~\ref{fig:separation-of-one-vel-d-from-one-blob}, we should expect a basic Taylor expansion with at least $\ga=1$; however all we need for the locality and multilocality proof is the assumption $\ga > 0$.

\begin{figure}[tb]\begin{center}
\begin{equation*}
\begin{pspicture}[shift=*](-2.5,-9)(2.5,2)
\pscircle(0,0){2}
\pscircle(0,-7){2}
\uput[l](0,-3.5){$F_n$}
\psline{*-*}(-1,0)(1,-1)
\psline{*-*}(0,1)(-1,-1)
\psline{*-*}(-1,-1)(1,-7)
\psline{*-*}(-1,-7)(1,-7)
\end{pspicture}
=
\begin{pspicture}[shift=*](-2.5,-9)(2.5,2)
\pscircle(0,0){2}
\pscircle(0,-7){2}
\uput[l](0,-3.5){$F_n^{(1)}$}
\psline{*-*}(-1,0)(1,-1)
\psline{*-*}(0,1)(-1,-1)
\psline{*-*}(0.2,1)(-0.8,-1)
\psline{*-*}(-1,-7)(1,-7)
\end{pspicture}
+
\begin{pspicture}[shift=*](-2.5,-9)(2.5,2)
\pscircle(0,0){2}
\pscircle(0,-7){2}
\uput[l](0,-3.5){$F_n^{(2)}$}
\psline{*-*}(-1,0)(1,-1)
\psline{*-*}(0,1)(-1,-1)
\psline{*-*}(0,1)(1,-7)
\psline{*-*}(-1,-7)(1,-7)
\end{pspicture}
\end{equation*}
\caption{\label{fig:two-blob-upscale-induction-first-step}\small Generalized structure function decomposition used to initialize the first step of the inductive argument for the two-blob velocity difference geometry fusion rule in upscale cascades}
\end{center}\end{figure}
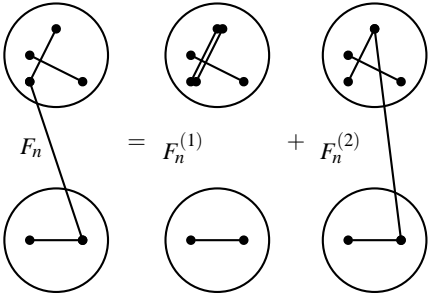

Now let us expand one velocity difference inside the blob so that one of its endpoints comes in contact with an endpoint of the remote velocity difference. In order for that to occur, it is assumed that previously the remote velocity difference was shifted in the correct direction. The leading term in the corresponding fusion rule should scale as $R^{\gz_1} r^{\gz_n-\gz_1}$, but the $R^{\gz_1}$ factor immediately vanishes, as it involves the ensemble average of one velocity difference, by homogeneity. Since in upscale cascades the relevant limit is $R\to +\infty$, the next order contribution needs to have a smaller scaling exponent with respect to $R$. To show this, we employ a dominant balance  argument similar to the one that we used previously in the context of the $p=1$ fusion rule. As shown in Fig.~\ref{fig:two-blob-upscale-induction-first-step}, we decompose the velocity difference that straddles between blobs across the triangle that it forms with one of the velocity differences on the original $r$-blob. As a result, the generalized structure function $F_n$ corresponding to the two blob geometry breaks into two contributions $F_n = F_n^{(1)}+F_n^{(2)}$, where $F_n$, $ F_n^{(2)}$ have the the same scaling and where $F_n^{(1)}$ scales as $r^{\gz_n+\ga}R^{-\ga}$, according to the previous argument. It is worth noting that if the leading-order contribution to the fusion rule of $F_n$ did not vanish, we would have a dominant balance between $F_n$ and $F_n^{(2)}$, with $F_n^{(1)}$ negligible in the limit $R\to +\infty$. However, given that the leading contribution to $F_n$ vanishes, in the subleading contribution all three terms have the same scaling, and it follows that $F_n$ scales as $r^{\gz_n+\ga}R^{-\ga}$.

\begin{figure}[t]\begin{center}
\begin{align*}
&\begin{pspicture}[shift=*](-3,-11)(7,2)
\pscircle(0,0){2}
\pscircle(0,-7){2}
\psline{*-*}(-1,0)(1,-1)
\psline{*-*}(-1,-1)(0,1)
\psline{*-*}(-1,-1)(1,-7)
\psline{*-*}(1,-7)(-1,-6)
\psline{*-*}(-1,-6)(-1,-8)
\uput[l](0,-4){$F_n$}
\uput[r](1,-3){$n-p-2$}
\uput[r](1,-10){$p+1$}
\end{pspicture}
\implies
\begin{pspicture}[shift=*](-3,-11)(7,2)
\pscircle(0,0){2}
\pscircle(0,-7){2}
\psline{*-*}(-1,0)(1,-1)
\psline{*-*}(-1,-1)(0,1)
\psline{*-*}(-0.8,-1)(0.2,1)
\psline{*-*}(1,-7)(-1,-6)
\psline{*-*}(-1,-6)(-1,-8)
\uput[l](0,-4){$F_n^{(1)}$}
\uput[r](1,-3){$n-p-1$}
\uput[r](1,-10){$p+1$}
\end{pspicture} \\
&=
\begin{pspicture}[shift=*](-3,-11)(7,2)
\pscircle(0,0){2}
\pscircle(0,-7){2}
\psline{*-*}(-1,0)(1,-1)
\psline{*-*}(-1,-1)(0,1)
\psline{*-*}(-0.8,-1)(0.2,1)
\psline{*-*}(-1,-6)(1,-1)
\psdots[dotstyle=*](1,-7)
\psline{*-*}(-1,-6)(-1,-8)
\uput[l](0,-4){$F_n^{(2)}$}
\uput[r](1,-3){$n-p-1$}
\uput[r](1,-10){$p$}
\end{pspicture}
+
\begin{pspicture}[shift=*](-3,-11)(7,2)
\pscircle(0,0){2}
\pscircle(0,-7){2}
\psline{*-*}(-1,0)(1,-1)
\psline{*-*}(-1,-1)(0,1)
\psline{*-*}(-0.8,-1)(0.2,1)
\psline{*-*}(1,-7)(1,-1)
\psline{*-*}(-1,-6)(-1,-8)
\uput[l](0,-4){$F_n^{(3)}$}
\uput[r](1,-3){$n-p-1$}
\uput[r](1,-10){$p$}
\end{pspicture}
\end{align*}
\caption{\label{fig:two-blob-upscale-induction-general-step}\small  Generalized structure function decomposition used for the general step of the inductive argument for the two-blob velocity difference geometry fusion rule in upscale cascades}
\end{center}\end{figure}
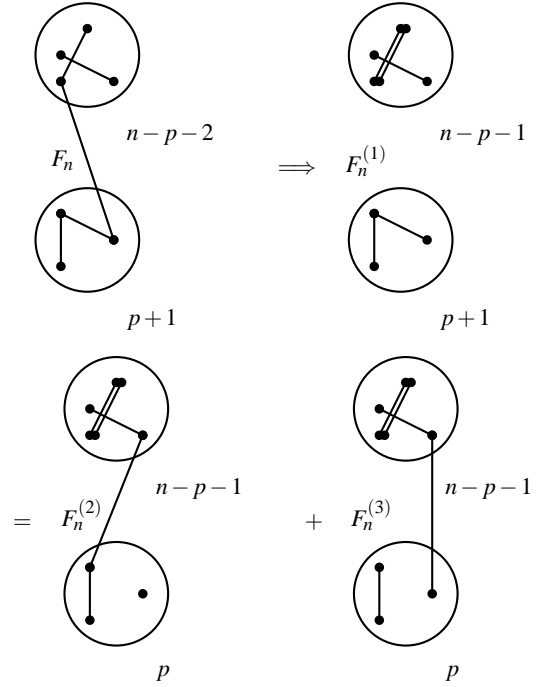

The above argument can be now generalized inductively for the more general case where more than one velocity difference is separated away from the original $r$-blob during the formation of the two-blob geometry. Let us assume that  we have already shown that a two-blob geometry with $p$ velocity differences in one blob, $n-p-1$ velocity differences in the other blob, and one velocity difference straddling in between also scales as $r^{\gz_n+\ga}R^{-\ga}$. Replacing $p$ with $p+1$, let us now consider a two-blob geometry consisting of one blob with $p+1$ velocity differences, a separate blob with $n-p-2$ velocity differences and an additional velocity difference straddling between the two blobs. Similarly to the $p=1$ case, the leading order contribution to the fusion rule for $F_n$  is $R^{\gz_1} r^{\gz_n-\gz_1}$ and vanishes for the same reasons. The next-order contribution is the generalized structure function $F_n^{(1)}$, shown in Fig.~\ref{fig:two-blob-upscale-induction-general-step}, and to obtain its scaling, we break it down as $F_n^{(1)} = F_n^{(2)}+F_n^{(3)}$ by decomposing one of the velocity differences on the $p+1$ blob across some triangle with an endpoint of a velocity difference from the opposite blob. The two contributions $F_n^{(2)}$, $F_n^{(3)}$ that result from this decomposition correspond to the preceding two-blob geometry that was assumed to scale as $r^{\gz_n+\ga}R^{-\ga}$ via the induction hypothesis. This concludes the inductive argument, and we find thus $r^{\gz_n+\ga}R^{-\ga}$ scaling for all two-blob velocity difference geometries regardless of how the velocity differences are distributed between blobs.

\section{Multilocality proof}

We now turn to the main problem of establishing multilocality in the IR and UV limit, and establishing corresponding locality conditions in terms of the generalized structure function scaling exponents $\gz_n$ and the fusion scaling exponents $\xi_{np}$. The arguments are based on the fusion rules that were discussed in the preceding section, and both upscale and downscale cascades are investigated, making the analysis relevant to the cascades of both three-dimensional and two-dimensional Navier-Stokes turbulence. A condensed summary of the main findings has already been given in the paper's introduction. 

\subsection{Preliminaries}

The main challenge with extending the locality proof to the terms of $\cO_n \cO_{n+1}\cdots\cO_{n+p-1} F_{n+p}$, is the existence of cross-terms. The mathematical structure of $\cO_n F_{n+1}$ takes the form
\begin{widetext}
\begin{equation}
\cO_n F_{n+1}(\{\bfX\}_n, t) = \sum_{k=1}^n \iint\df{\bfY_1}\df{\bfY_2} \cO (\bfX_k, \bfY_1, \bfY_2) F_{n+1} (\{\bfX\}_n^k,\bfY_1, \bfY_2).
\end{equation}
 Here, $\{\bfX\}_n^k=(\bfX_1, \ldots, \bfX_{k-1}, \bfX_{k+1},\ldots, \bfX_n)$ and $\cO_n (\bfX_k, \bfY_1, \bfY_2)$ is a generalized function representing the appropriate integrodifferential operator, encapsulating the nonlinear interactions that drive the turbulence cascades, sans the sweeping interactions. Note that in $F_{n+1}$, $\bfX_k$ is replaced with $\bfY_1$ and $\bfY_2$ is added thereafter. A detailed account of the terms of the balance equations was given in my previous paper \cite{article:Gkioulekas:2008:1}. It is easy to show that  $\cO_n\cO_{n+1}F_{n+2}$ takes the form 
\begin{align}
\cO_n \cO_{n+1}& F_{n+2} (\{\bfX\}_n,t) 
= 
\sum_{l=1}^n \iint\df{\bfZ_1}\df{\bfZ_2}\;\cO (\bfX_l, \bfZ_1, \bfZ_2) \cO_{n+1}F_{n+2}(\{\bfX\}_n^l, \bfZ_1, \bfZ_2) \\
&=
\sum_{l=1}^n \iint\df{\bfZ_1}\df{\bfZ_2}\;\cO (\bfX_l, \bfZ_1, \bfZ_2) \biggl[ \sum_{\substack{k=1\\ k\neq l}}^n \iint\df{\bfY_1}\df{\bfY_2} \cO (\bfX_k, \bfY_1, \bfY_2) F_{n+2}(\{\bfX\}_n^{kl}, \bfY_1, \bfY_2, \bfZ_1, \bfZ_2)  \biggr. \\
&\qquad +  \iint\df{\bfY_1}\df{\bfY_2} \cO (\bfZ_1, \bfY_1, \bfY_2) F_{n+2}(\{\bfX\}_n^l, \bfY_1,\bfY_2,\bfZ_2)  + \iint\df{\bfY_1}\df{\bfY_2} \cO(\bfZ_2, \bfY_1, \bfY_2) F_{n+2}(\{\bfX\}_n^l, \bfY_1,\bfY_2,\bfZ_1) \biggr] \\
&=
\sum_{l=1}^n \sum_{\substack{k=1\\ k\neq l}}^n \iint\df{\bfZ_1}\df{\bfZ_2}\iint\df{\bfY_1}\df{\bfY_2} \cO (\bfX_l, \bfZ_1, \bfZ_2) \cO (\bfX_k, \bfY_1, \bfY_2) F_{n+2}(\{\bfX\}_n^{kl}, \bfY_1, \bfY_2, \bfZ_1, \bfZ_2) \\
&\qquad + \sum_{l=1}^n \iint\df{\bfZ_1}\df{\bfZ_2}\iint\df{\bfY_1}\df{\bfY_2} \cO (\bfX_l, \bfZ_1, \bfZ_2) \cO (\bfZ_1, \bfY_1, \bfY_2) F_{n+2}(\{\bfX\}_n^l, \bfY_1,\bfY_2,\bfZ_2) \\
&\qquad + \sum_{l=1}^n \iint\df{\bfZ_1}\df{\bfZ_2}\iint\df{\bfY_1}\df{\bfY_2} \cO (\bfX_l, \bfZ_1, \bfZ_2) \cO(\bfZ_2, \bfY_1, \bfY_2) F_{n+2}(\{\bfX\}_n^l, \bfY_1,\bfY_2,\bfZ_1)
\end{align}
Here  $\{\bfX\}_n^{kl}$ consists of $n-2$ velocity difference endpoint pairs, with $\bfX_k$ and $\bfX_l$ deleted from the original velocity difference geometry $\{\bfX\}_n$. Obviously, the locality of the first term in the above expression follows from the same argument that establishes locality in $\cO_n F_{n+1}$. The problem is that a separate argument is needed for the other two terms, corresponding to the case $k = l$ in the double summation above. This argument was not previously given \cite{article:Procaccia:1996:2,article:Procaccia:1996:3} in the derivation of the bridge relations, with no explanation as to how one dispatches the cross-terms. We will now show that it is indeed possible to extend the locality proof to these cross-terms.

We begin with the observation that a typical contribution to  $\cO_n F_{n+1}$ takes the forms
\begin{equation}
\int\df{\bfy}\; P_{\ga_k \gb}(\bfy) \pd_{\gc, \bfx_k} \avg{\biggl[ \prod_{\substack{l=1 \\ l\neq k}}^n w_{\ga_l} (\bfX_l) \biggr]w_\gb (\bfx_k-\bfy, \bfxp_k-\bfy) w_\gc (\bfx_k-\bfy, \bfs)}
\label{eq:simple-interaction-integral-one}
\end{equation}
or
\begin{equation}
\int\df{\bfy}\; P_{\ga_k \gb}(\bfy) \pd_{\gc, \bfxp_k} \avg{\biggl[ \prod_{\substack{l=1 \\ l\neq k}}^n w_{\ga_l} (\bfX_l) \biggr]w_\gb (\bfx_k-\bfy, \bfxp_k-\bfy) w_\gc (\bfxp_k-\bfy, \bfs)}
\label{eq:simple-interaction-integral-two}
\end{equation}
\end{widetext}
with possible values of $\bfs$ being $\bfs\in\{\bfx_1, \ldots,\bfx_n, \bfxp_1, \ldots, \bfxp_n\}$. Repeated tensorial indices in product forms (e.g. the index $\gc$) imply a summation over all components. Here $\pd_{\gc,\bfx_k}$ is the spatial partial derivative with respect to the $\gc$ component of $\bfx_k$ and $P_{\ga\gb}(\bfx)$  is the kernel of the operator $\ccP_{\ga\gb}=\gd_{\ga\gb}-\pd_\ga \pd_\gb \del^{-2}$  with $\gd_{\ga\gb}$ the Kronecker delta. It is given by 
\begin{align}
P_{\ga\gb}(\bfx) &= \gd_{\ga\gb}\gd (\bfx) - g'' (\nrm{\bfx})\frac{x_\ga x_\gb}{\nrm{\bfx}} \\ & \qquad - g' (\nrm{\bfx})\left[ \frac{\gd_{\ga\gb}}{\nrm{\bfx}} -  \frac{x_\ga x_\gb}{\nrm{\bfx}^3} \right] \\
&\equiv \gd_{\ga\gb}\gd (\bfx)-P_{\ab}^{\parallel}(\bfx)
\end{align}
 with $g(r)$ the Green's function for the inverse Laplacian $\del^{-2}$ , which is $g(r)=-1/(4\pi r)$  in three dimensions and $g(r) = \ln r/(2\pi)$  in two dimensions, and it scales as $P_{\ab}(\bfx)\sim \nrm{\bfx}^{-d}$ with $d$ the number of dimensions. An immediate consequence of the incompressibility of the velocity field is that $\cP_{\ab} u_\gb = u_\ga$ or equivalently
\begin{equation}
\int_{\bbR^d} \df{\bfy} P_{\ab}^{\parallel}(\bfy) u_\gb (\bfy) = 0
\end{equation}
We also note that both $P_{\ab}(\bfy) $ and $P_{\ab}^{\parallel}(\bfy) $ are even, in the sense that $P_{\ab}(-\bfy)=P_{\ab}(\bfy)$ and $P_{\ab}^{\parallel}(-\bfy)=P_{\ab}^{\parallel}(\bfy)$ for all $\bfy\in\bbR^d$.

It is worth noting how $F_n (\{\bfX\}_n, t)$ is transformed into the above contributions to $\cO_n F_{n+1}(\{\bfX\}_n, t)$, in order to understand the more general case: (a) The index $\ga_k$ is moved to  $P_{\ga_k \gb}$ and in the velocity product we replace $w_{\ga_k}(\bfx_k,\bfxp_k)$ with $w_\gb(\bfx_k,\bfxp_k)$; (b) both arguments of  $w_\gb(\bfx_k,\bfxp_k)$ are shifted by $\bfy$, giving $w_\gb (\bfx_k-\bfy, \bfxp_k-\bfy)$; (c) A new velocity factor $w_{\gc}(\bfx_k-\bfy, \bfs)$  is introduced in the velocity product for the $\pd_{\gc, \bfx_k}$  terms and correspondingly we introduce $w_{\gc}(\bfxp_k-\bfy, \bfs)$ for  the $\pd_{\gc, \bfxp_k}$ terms.

\subsection{Review of UV locality proof}
\label{sec:uv-locality}

UV locality corresponds to vanishing integral contributions in the limits $\bfy\to\bfzero$, $\bfx_k-\bfy\to\bfx_l$, $\bfx_k-\bfy\to\bfxp_l$, $\bfxp_k-\bfy\to\bfx_l$, $\bfxp_k-\bfy\to\bfxp_l$. It is sufficient to consider only the limit $\bfy\to\bfzero$ where $P_{\ab}(\bfy)$ becomes singular. The other limits present with similar situations, but the absence of the singularity in the projection function makes convergence even easier. Furthermore, it is sufficient to focus only on the integral contribution that corresponds to $P_{\ab}^{\parallel}(\bfy)$. The integral over the delta function contribution to $P_{\ab}(\bfy)$ is formal, and can be eliminated; the result involves a correlation with no fusions and involves no non-local interactions that may require concern in either the UV or IR limits. The multilocality proof in the UV limit builds upon the previous UV locality proof for $\cO_n F_{n+1}$, so we begin with a review of that argument. Then we extend it, first to $\cO_n\cO_{n+1} F_{n+2}$ , then to the more general case of $\cO_n\cO_{n+1}\cdots\cO_{n+p-1} F_{n+p}$. Consider, with no loss of generality, the integral given by Eq.~\eqref{eq:simple-interaction-integral-one} and distinguish between the following cases:

\textbf{Case 1:} Assume that $\bfx_k\neq\bfs$. Then, there are no fusions of velocity differences as $\bfy\to\bfzero$, so the derivative of the ensemble average in Eq.~\eqref{eq:simple-interaction-integral-one} is analytic with respect to $\bfy$ and can be Taylor expanded as: 
\begin{align}
I &= \int_{0^+}\df{\bfy}\; P_{\ga_k\gb}(\bfy)[\cA_\gb + \cB_{\bc}y_\gc + \cC_{\bcd} y_\gc y_\gd+\cdots ] \\
&= \cA_\gb \int_{0^+}\df{\bfy}\; P_{\ga_k\gb}(\bfy) + \cB_{\bc} \int_{0^+}\df{\bfy}\; P_{\ga_k\gb}(\bfy) y_\gc \\
\qquad &+ \cC_{\bcd} \int_{0^+}\df{\bfy}\; P_{\ga_k\gb}(\bfy) y_\gc y_\gd + \cdots
\end{align}
The first integral consists of the delta function contribution to $P_{\ga_k\gb}(\bfy)$, which is local and does not contribute anything for $\bfy\neq\bfzero$, and a $P_{\ga_k\gb}^{\parallel}(\bfy)$ contribution that vanishes, as the $\cP_{\ga_k\gb}^{\parallel}$ operator is applied on a constant field that is trivially incompressible. The second integral also vanishes because the integrand is odd with respect to replacing $\bfy$ with $-\bfy$. In the third integral, using $\rho=\nrm{\bfy}$ the differential contributes $\rho^{d-1}\df{\rho}$, the projection operator $P_{\ga_k\gb}^{\parallel}(\bfy) $ contributes $\rho^{-d}$ and $y_\gc y_\gd$ contributes $\rho^2$. Overall, the integral scales as $I\sim \rho^2$, which is independent of the dimension $d$, and is clearly local.

\begin{figure}[tb]\begin{center}
\begin{pspicture}(16,8)
\psset{unit=10pt}
\psline{*-*}(1,6)(3,3)
\psline{*-*}(3,3)(13,5)
\uput[d](3,3){$\bfx_k-\bfy$}
\uput[r](1,6){$\bfx_k$}
\uput[r](13,5){$\bfxp_k-\bfy$}
\uput[l](1,4){$\gc$}
\uput[l](9,3){$\gb$}
\end{pspicture}
\caption{\label{fig:locality-uv}\small UV limit for the case $\bfx_k = \bfs$. We employ the fusion rule shown in Fig.~\ref{fig:p-one-case-two-fusion-rule}}
\end{center}\end{figure}
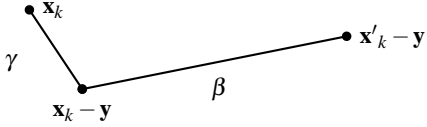

\textbf{Case 2:} Assume that $\bfx_k=\bfs$. Then there is a fusion in the product $w_\gb (\bfx_k-\bfy, \bfxp_k-\bfy) w_\gc (\bfx_k-\bfy, \bfx_k)$ as $\bfy\to\bfzero$ that breaks the regularity preventing the Taylor expansion that we did in the previous case. The corresponding velocity difference configuration is shown in Fig.~\ref{fig:locality-uv}, and if we let $F_{n+1}$ denote the velocity difference product ensemble average, then the leading order scaling in the inertial range as $\bfy\to\bfzero$ is given by the $p=1$ fusion rule and reads 
\begin{equation}
F_{n+1} \sim \Phi_2 (\bfx_k-\bfy, \bfx_k, \bfx_k-\bfy, \bfx_k) \Phi_{n-1}(\{\bfX\}_n^k)
\end{equation}
and the integral $I$ can be rewritten as: 
\begin{equation}
I \sim \Phi_{n-1}(\{\bfX\}_n^k) \int_{\bbR^d}\df{\bfy} P_{\ga_k\gb}(\bfy) \pd_{\gc, \bfx_k} \Phi_2 (\bfx_k-\bfy, \bfx_k, \bfx_k-\bfy, \bfx_k) 
\end{equation}
In terms of $\rho$, $\df{\bfy} $ still contributes $\rho^{d-1}\df{\rho}$, the projection function $P_{\ga_k\gb}(\bfy)$ contributes $\rho^{-d}$, and  $\Phi_2$ contributes $\rho^{\xi_{n+1,1}}$. The derivative $\pd_{\gc, \bfx_k}$ does not contribute to the $\rho$ scaling exponent, due to the geometric configuration of $\Phi_2(\bfx_k-\bfy, \bfx_k, \bfx_k-\bfy, \bfx_k)$ and its local homogeneity. In particular, when  we ``shake'' $\bfx_k$, it shakes both $\bfx_k$ and $\bfx_k-\bfy$ equally, but the distance $\rho$ between these two points remains constant, so the derivative has no effect on $\rho$. The shaking itself has no effect on $\Phi_2 (\bfx_k-\bfy, \bfx_k, \bfx_k-\bfy, \bfx_k)$ due to universal local homogeneity. Altogether, the integral scales as $I\sim \rho^{\xi_{n+1,1}}$, and the corresponding locality condition is $\xi_{n+1,1}>0$.

From the above argument we see that in Case 1 the integral has UV locality unconditionally, whereas in Case 2 we have a UV locality condition $\xi_{n+1,1}>0$. Assuming $n>2$, for the three-dimensional downscale energy cascade we expect $\xi_{n+1,1}=\gz_2>2/3>0$, for the two-dimensional downscale enstrophy cascade we expect $\xi_{n+1,1}=\gz_2=2>0$, and for the two-dimensional inverse energy cascade we expect $\xi_{n+1,1}=\gz_{n+1}-\gz_{n-1}>0$. For $n=2$, we have the evaluation  $\xi_{n+1,1}=\xi_{3,1}=\gz_3$, both for upscale and downscale cascades, and it also satisfies the condition $\xi_{n+1,1}>0$ needed for UV locality. Thus, under the fusion rules hypothesis, all integrals in $\cO_n F_{n+1}$ are UV local.

\subsection{UV multilocality proof for two operators}
\label{sec:UV-multilocality-proof-two-operators}

Now, let us consider the  locality of the cross-terms in $\cO_n \cO_{n+1} F_{n+2}$. First, we note that any terms involving the derivatives $\pd_{\gc_1,\bfx_k}\pd_{\gc_2,\bfx_l}$, $\pd_{\gc_1,\bfx_k}\pd_{\gc_2,\bfxp_l}$, $\pd_{\gc_1,\bfxp_k}\pd_{\gc_2,\bfxp_l}$ with $k\neq l$ are not cross-terms, and their locality is an immediate consequence of the previously shown locality for $\cO_n F_{n+1}$. For $k=l$, typical $\pd_{\gc_1, \bfx_k}\pd_{\gc_2, \bfx_k}$ cross-terms take the form
\begin{multline}
I_1 = \int_{\bbR^d}\df{\bfy_2}\int_{\bbR^d}\df{\bfy_1}\; P_{\ga_k\gb_1}(\bfy_2) P_{\gb_1\gb_2}(\bfy_1) \pd_{\gc_2,\bfx_k}\pd_{\gc_1,\bfx_k}\\
\times\oavg\biggl[ \prod_{\substack{l=1 \\ l\neq k}}^n w_{\ga_l} (\bfX_l) \biggr] w_{\gb_2}(\bfx_k-\bfy_1-\bfy_2, \bfxp_k-\bfy_1-\bfy_2)\cpha \\ \opha \times w_{\gc_1}(\bfx_k-\bfy_1-\bfy_2, \bfs_1) w_{\gc_2}(\bfx_k-\bfy_2, \bfs_2)  \cavg
\label{eq:interaction-double-integral-one}
\end{multline}
 with $\bfs_1, \bfs_2\in\{\bfx_1,\ldots,\bfx_n,\bfxp_1,\ldots,\bfxp_n\}$ and  $\bfs_1 \neq \bfx_k$. Note that the $\bfy_1$ integral comes from $\cO_{n+1}$ and the $\bfy_2$ integral comes from $\cO_n$. Starting from $F_n (\{\bfX\}_n, t)$, in  the cross-terms of $\cO_n F_{n+1}$, in the velocity product, $w_{\ga_k}$ is replaced with $w_{\gb_1}(\bfx_k-\bfy_1, \bfxp_k-\bfy_1)$ . We also append the factor $w_{\gc_1} (\bfx_k-\bfy_1, \bfs_1)$ to the velocity product. Moving on to $\cO_n \cO_{n+1} F_{n+2}$, we introduce the  $\pd_{\gc_2, \bfx_k}$ derivative and the $\bfy_2$  integral, at which point all previous occurrences of $\bfx_k$  are shifted by $\bfy_2$. As a result  $w_{\gc_1}$ becomes $w_{\gc_1}(\bfx_k-\bfy_1-\bfy_2, \bfs_1)$,  $w_{\gb_1}$ becomes $w_{\gb_2} (\bfx_k-\bfy_1-\bfy_2, \bfxp_k-\bfy_1-\bfy_2)$ , the $\gb_1$  index goes to the projection function of $\bfy_1$ and the $\ga_k$  index is pushed onto the projection function of $\bfy_2$. A new factor $w_{\gc_2} (\bfx_k-\bfy_2, \bfs_2)$ is also introduced. Note that if $\bfs_1=\bfx_k$, then $\bfs_1$ is also shifted by $\bfy_2$, yielding a cross-term of the form
\begin{multline}
I_2 = \int_{\bbR^d}\df{\bfy_2}\int_{\bbR^d}\df{\bfy_1}\; P_{\ga_k\gb_1}(\bfy_2) P_{\gb_1\gb_2}(\bfy_1) \pd_{\gc_2,\bfx_k}\pd_{\gc_1,\bfx_k}\\
\times\oavg\biggl[ \prod_{\substack{l=1\\ l\neq k}}^n w_{\ga_l} (\bfX_l) \biggr] w_{\gb_2}(\bfx_k-\bfy_1-\bfy_2, \bfxp_k-\bfy_1-\bfy_2)\cpha \\ \opha \times w_{\gc_1}(\bfx_k-\bfy_1-\bfy_2, \bfx_k-\bfy_2) w_{\gc_2}(\bfx_k-\bfy_2, \bfs_2)  \cavg
\label{eq:interaction-double-integral-two}
\end{multline}
Similar integrals arise from cross-terms that involve an $\pd_{\gc_2,\bfxp_k} \pd_{\gc_1, \bfx_k} $ combination of derivatives that read 
\begin{multline}
I_3 = \int_{\bbR^d}\df{\bfy_2}\int_{\bbR^d}\df{\bfy_1}\; P_{\ga_k\gb_1}(\bfy_2) P_{\gb_1\gb_2}(\bfy_1) \pd_{\gc_2,\bfxp_k}\pd_{\gc_1,\bfx_k}\\
\times\oavg\biggl[ \prod_{\substack{l=1 \\ l\neq k}}^n w_{\ga_l} (\bfX_l) \biggr] w_{\gb_2}(\bfx_k-\bfy_1-\bfy_2, \bfxp_k-\bfy_1-\bfy_2)\cpha \\ \opha \times w_{\gc_1}(\bfx_k-\bfy_1-\bfy_2, \bfs_1) w_{\gc_2}(\bfxp_k-\bfy_2, \bfs_2)  \cavg
\label{eq:interaction-double-integral-three}
\end{multline}
 for $\bfs_1, \bfs_2 \in\{\bfx_1,\ldots,\bfx_n,\bfxp_1,\ldots,\bfxp_n\}$ with $\bfs_1\neq\bfx_k$. For the special case $\bfs_1=\bfx_k$, upon introducing the $\bfy_2$ integral, $\bfs_1=\bfx_k$ is also shifted by $\bfy_2$ resulting in an integral that reads 
\begin{multline}
I_4=\int_{\bbR^d}\df{\bfy_2}\int_{\bbR^d}\df{\bfy_1}\; P_{\ga_k\gb_1}(\bfy_2) P_{\gb_1\gb_2}(\bfy_1) \pd_{\gc_2,\bfxp_k}\pd_{\gc_1,\bfx_k}\\
\times\oavg\biggl[ \prod_{\substack{l=1\\ l\neq k}}^n w_{\ga_l} (\bfX_l) \biggr] w_{\gb_2}(\bfx_k-\bfy_1-\bfy_2, \bfxp_k-\bfy_1-\bfy_2)\cpha \\ \opha \times w_{\gc_1}(\bfx_k-\bfy_1-\bfy_2, \bfx_k-\bfy_2) w_{\gc_2}(\bfxp_k-\bfy_2, \bfs_2)  \cavg
\label{eq:interaction-double-integral-four}
\end{multline}
The other two contributions $\pd_{\gc_2,\bfx_k}\pd_{\gc_1,\bfxp_k}$ and $\pd_{\gc_2,\bfxp_k}\pd_{\gc_1,\bfxp_k}$ give identical locality arguments upon a symmetric exchange $\bfx_k\leftrightarrow \bfxp_k$, so we shall not consider them explicitly. 

To establish locality, we stress that the integrals are done one at a time. Consequently, once the $\bfy_1$  integral is shown to be local, the major contribution to the $\bfy_1$  integral originates from the velocity differences situated at some inertial range length scale $R_1$, and given that, the locality of the $\bfy_2$ integral is then considered. For UV locality, we consider the separate limits $\bfy_1 \to 0^+$ and $\bfy_2\to 0^+$,   where the projection functions become singular. All other UV limits are less singular,  owing to the regularity of the projection functions, so they do not require special consideration. First,  we note that $\df{\bfy_1}$ contributes $\df{\rho_1}\rho_1^{d-1}$  and  $P_{\gb_1\gb_2}(\bfy_1)$ contributes $\rho_1^{-d}$, so the combination $\df{\bfy_1} P_{\gb_1\gb_2}(\bfy_1)$ makes no $\rho_1$-dependent contribution to the integral, when $\rho_1=\nrm{\bfy_1}\to 0^+$. Likewise, no $\rho_2$-dependent contribution is expected from $\df{\bfy_2} P_{\ga_k\gb_1}(\bfy_2)$, when $\rho_2=\nrm{\bfy_2}\to 0^+$. With no loss of generality, let us consider the limit $\rho_2\to 0^+$. In general, there are three possibilities for the geometric configuration of velocity differences in the velocity product under the UV limit $\rho_1 \to 0^+$ or the UV limit $\rho_2\to 0^+$.

\textbf{Case 1:} It is possible that there are no fusions with no velocity difference endpoints approaching each other, under the $\rho_1\to  0^+$ or $\rho_2\to 0^+$ limits. In this case, the velocity ensemble average combined with the derivatives with respect to $\bfx_k$ or $\bfxp_k$ can be Taylor expanded, similarly to the situation in Case 1 of section~\ref{sec:uv-locality}. The same argument is repeated where the first two terms of the expansion vanish and the third term respects UV locality unconditionally.

\textbf{Case 2:} There may be a fusion where one velocity difference's endpoints are brought together while it is attached to one or two other velocity differences. Let us assume, with no loss of generality, that this velocity difference corresponds to the limit $\rho_1\to 0^+$. Then, the ensemble average of the velocity product gives an $\rho_1^{\xi_{n+2,1}}$ contribution. Similarly to Case 2 of section~\ref{sec:uv-locality}, the corresponding $\rho_1$-dependent factor is locally homogeneous (in the sense of velocity increments) with respect to $\bfx_k$ shifting. As a result, the derivatives make no contributions to the scaling exponent of $\rho_1$, and the integral will scale as $\rho_1^{\xi_{n+2,1}}$, leading to $\xi_{n+2,1}>0$ as a necessary UV locality scenario. With two or more nonlinear interactions operators, it is possible to encounter new velocity difference geometries such as the ones shown in Fig.~\ref{fig:p-one-double-attachment-one} or Fig.~\ref{fig:p-one-multiple-attachment}. As we explained in section II.B, these will still yield the same fusion rule as the one corresponding to Fig.~\ref{fig:p-one-case-two-fusion-rule}, so the argument of section III.B case 1 continues to carry through with no additional considerations. 

\textbf{Case 3:} A new possibility that arises in the UV multilocality integrals, but not in the original integrals for $\cO_n F_{n+1}$, is a fusion in which the velocity difference whose endpoints are brought together is not attached to any of the other velocity differences. Again, with no loss of generality, let us assume that the fused velocity difference corresponds to the limit $\rho_1\to 0^+$, and let $R_{min}$ be the closest distance between an endpoint of the $\rho_1$ velocity difference and an endpoint of another nearest velocity difference. If $\rho_1 \ll R_{min}$, then the ensemble average of the velocity product will scale as $F_{n+2}\sim (\rho_1/R_{min}) S_{n+2}(R)$. In the absence of fractional scaling with respect to $\rho_1$, the derivative of the ensemble average of the velocity product leads to a Taylor expansion with respect to $\rho_1$. UV locality can be then established unconditionally by repeating the previous argument of Case 1. If $\rho_2\gg R_{min}$, then we can use the general property of velocity differences $w_\ga (\bfx,\bfy) = w_\ga (\bfx, \bfz)+ w_\ga (\bfz,\bfy)$ to reattach the $\rho_1$ velocity difference to its nearest neighbor (see Fig.~\ref{fig:p-one-case-two-detached-fusion-rule}, for a similar situation). The attached velocity difference separations also follow $\rho_1$ scaling and the problem of UV locality reduces now to the previous argument of Case 2. 

Now let us consider the fusion events in the integrals $I_1, I_2, I_3, I_4$ given by Eq.~\eqref{eq:interaction-double-integral-one}, Eq.~\eqref{eq:interaction-double-integral-two}, Eq.~\eqref{eq:interaction-double-integral-three}, Eq.~\eqref{eq:interaction-double-integral-four} in view of the above 3 cases. 

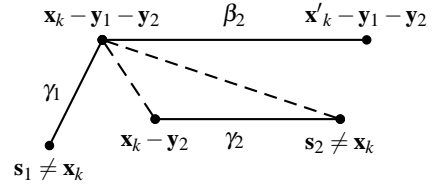
\begin{figure}[tb]\begin{center}
\begin{pspicture}[shift=*](-1,-2)(14,6)
\psline{*-*}(0,0)(2,4)
\uput[d](0,0){$\bfs_1\neq\bfx_k$}
\uput[l](1,2){$\gc_1$}
\uput[u](2,4){$\bfx_k-\bfy_1-\bfy_2$}
\psline{*-*}(2,4)(12,4)
\uput[u](12,4){$\bfxp_k-\bfy_1-\bfy_2$}
\uput[u](7,4){$\gb_2$}
\psline{*-*}(4,1)(11,1)
\uput[d](4,1){$\bfx_k-\bfy_2$}
\uput[d](7,1){$\gc_2$}
\uput[d](11,1){$\bfs_2\neq\bfx_k$}
\psline[linestyle=dashed]{*-*}(4,1)(2,4)
\psline[linestyle=dashed]{*-*}(11,1)(2,4)
\end{pspicture}
\caption{\label{fig:interaction-double-integral-one}\small $I_1$ integral with $\bfs_1\neq\bfx_k$ and $\bfs_2\neq\bfx_k$. There is no fusion event when $\rho_2\to 0^+$, but when $\rho_1\to 0^+$ the endpoints $\bfx_k-\bfy_2$ and $\bfx_k-\bfy_1-\bfy_2$ come together.}
\end{center}\end{figure}

For the integral $I_1$ with $\bfs_1\neq\bfx_k$ and $\bfs_2\neq\bfx_k$, we show the velocity difference geometry for the product $w_{\gb_2} w_{\gc_1} w_{\gc_2}$ in Fig.~\ref{fig:interaction-double-integral-one}. We see that in the limit $\rho_2\to 0^+$ there are no fusion events, so UV locality follows unconditionally under Case 1, above. In the limit $\rho_1\to 0^+$ the endpoint $\bfx_k-\bfy_2$ of $w_{\gc_2}$ and the endpoint $\bfx_k-\bfy_1-\bfy_2$ of $w_{\gb_2}$ will fuse together. The simplest way to handle the limit is to reattach $w_{\gc_2}$ onto $w_{\gb_2}$ by writing 
\begin{align}
w_{\gc_2}(\bfx_k-\bfy_2, \bfs_2) &= w_{\gc_2}(\bfx_k-\bfy_2, \bfx_k-\bfy_1-\bfy_2) \\ &\qquad + w_{\gc_2}(\bfx_k-\bfy_1-\bfy_2, \bfs_2)
\end{align}
 and breaking the integral $I_1$ into two corresponding contributions. The decomposition is shown via the dashed lines of Fig.~\ref{fig:interaction-double-integral-one}. For the first contribution, we expect UV locality according to the argument of Case 2. The second contribution has unconditional UV locality according to the argument of Case 1.

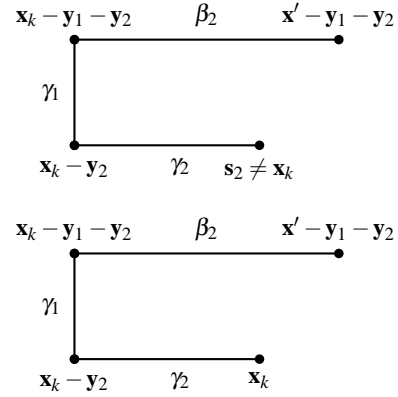
\begin{figure}[tb]\begin{center}
\begin{pspicture}[shift=*](-1,-2)(14,6)
\psline{*-*}(0,0)(7,0)
\uput[d](0,0){$\bfx_k-\bfy_2$}
\uput[d](4,0){$\gc_2$}
\uput[d](7,0){$\bfs_2\neq\bfx_k$}
\psline{*-*}(0,0)(0,4)
\uput[l](0,2){$\gc_1$}
\uput[u](0,4){$\bfx_k-\bfy_1-\bfy_2$}
\psline{*-*}(0,4)(10,4)
\uput[u](5,4){$\gb_2$}
\uput[u](10,4){$\bfxp-\bfy_1-\bfy_2$}
\end{pspicture}

\begin{pspicture}[shift=*](-1,-2)(14,6)
\psline{*-*}(0,0)(7,0)
\uput[d](0,0){$\bfx_k-\bfy_2$}
\uput[d](4,0){$\gc_2$}
\uput[d](7,0){$\bfx_k$}
\psline{*-*}(0,0)(0,4)
\uput[l](0,2){$\gc_1$}
\uput[u](0,4){$\bfx_k-\bfy_1-\bfy_2$}
\psline{*-*}(0,4)(10,4)
\uput[u](5,4){$\gb_2$}
\uput[u](10,4){$\bfxp-\bfy_1-\bfy_2$}
\end{pspicture}
\caption{\label{fig:interaction-double-integral-two}\small $I_2$ integral with $\bfs_2\neq\bfx_k$ or $\bfs_2=\bfx_k$. In both cases, the velocity difference $w_{\gc_1}$ fuses in the limit $\rho_1\to 0^+$. For $\bfs_2=\bfx_k$, $w_{\gc_2}$ fuses in the limit $\rho_2\to 0^+$. For $\bfs_2\neq\bfx_k$, there is no fusion when $\rho_2\to 0^+$.}
\end{center}\end{figure}

For the integral $I_2$ we distinguish between the cases $\bfs_2\neq\bfx_k$ and $\bfs_2 = \bfx_k$. The corresponding velocity difference geometries are shown in Fig.~\ref{fig:interaction-double-integral-two}. In both cases, under the limit $\rho_1\to 0^+$, the velocity difference $w_{\gc_1}$ fuses, while remaining attached on both sides to $w_{\gb_2}$ and $w_{\gc_2}$, and the integral is UV local via the argument of Case 2. For $\bfs_2\neq\bfx_k$, there is no fusion when $\rho_2 \to 0^+$, so the integral is unconditionally local via the argument of Case 1. For $\bfs_2 =\bfx_k$, the velocity difference $w_{\gc_2}$ fuses under the limit $\rho_2\to 0^+$ while being attached to $w_{\gc_1}$, so the integral is UV local via the argument of Case 2.

\begin{figure}[tb]\begin{center}
\begin{pspicture}[shift=*](-1,-2)(14,6)
\psline{*-*}(2,0)(0,4)
\uput[d](2,0){$\bfs_1\neq\bfx_k$}
\uput[l](1,2){$\gc_1$}
\uput[u](0,4){$\bfx_k-\bfy_1-\bfy_2$}
\psline{*-*}(0,4)(10,4)
\uput[u](5,4){$\gb_2$}
\uput[u](10,4){$\bfxp_k-\bfy_1-\bfy_2$}
\psline{*-*}(7,0)(12,0)
\uput[d](7,0){$\bfxp_k-\bfy_2$}
\uput[d](12,0){$\bfs_2$}
\uput[u](10,0){$\gc_2$}
\psline[linestyle=dashed]{*-*}(7,0)(10,4)
\psline[linestyle=dashed]{*-*}(12,0)(10,4)
\end{pspicture}
\caption{\label{fig:interaction-double-integral-three}\small
$I_3$ integral with $\bfs_1\neq\bfx_k$. It is possible that $\bfs_2=\bfxp_k$ or $\bfs_2\neq\bfxp_k$. In the limit $\rho_1\to 0^+$ the only possible fusion is between the points $\bfxp_k-\bfy_2$ and $\bfxp_k-\bfy_1-\bfy_2$. If we assume $\bfs_2\neq\bfxp_k$, then there is no fusion when $\rho_2\to 0^+$. However, if we assume that $\bfs_2=\bfxp_k$, then the velocity difference $w_{\gc_2}$ fuses in the limit $\rho_2\to 0^+$.}
\end{center}\end{figure}
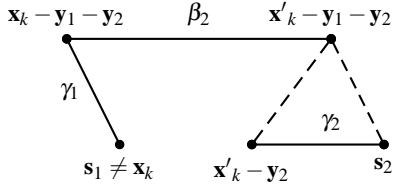

For the $I_3$ integral we assume that $\bfs_1\neq\bfx_k$ and distinguish between the cases $\bfs_2=\bfxp_k$ and $\bfs_2\neq\bfxp_k$. In both cases, under the limit $\rho_1\to 0^+$ the only possible fusion is between the endpoints $\bfxp_k-\bfy_2$ of $w_{\gc_2}$ and $\bfxp_k-\bfy_1-\bfy_2$ of $w_{\gb_2}$. To handle this, we reattach the velocity difference $w_{\gc_2}$ onto $w_{\gb_2}$ by writing
\begin{align}
w_{\gc_2}(\bfxp_k-\bfy_2, \bfs_2) &= w_{\gc_2}(\bfxp_k-\bfy_2, \bfxp_k-\bfy_1-\bfy_2) \\ &\qquad + w_{\gc_2}(\bfxp_k-\bfy_1-\bfy_2, \bfs_2)
\label{eq:interaction-double-integral-three-decomposition}
\end{align}
 and, similarly to the situation in Fig.~\ref{fig:interaction-double-integral-one}, the integral breaks into two contributions that are UV local (see Fig.~\ref{fig:interaction-double-integral-three}); the first contribution via the argument of Case 2 and the second contribution via the argument of Case 1. Under the limit $\rho_2 \to 0^+$ there is no fusion event when $\bfs_2 \neq \bfxp_k$, so UV locality follows unconditionally via the argument of Case 1. However, when $\bfs_2=\bfxp_k$, under the limit $\rho_2\to 0^+$ the velocity difference $w_{\gc_2}$ fuses.  If it happens to be near another velocity difference, we can reattach it to that velocity difference and deduce UV locality using the argument of Case 2. If, on the other hand, it is not near other velocity differences, in order to deduce UV locality, it becomes necessary to employ the argument of Case 3.

\begin{figure}[tb]\begin{center}
\begin{pspicture}[shift=*](-1,-2)(15,6)
\psline{*-*}(2,0)(0,4)
\uput[d](2,0){$\bfx_k-\bfy_2$}
\uput[l](1,2){$\gc_1$}
\uput[u](0,4){$\bfx_k-\bfy_1-\bfy_2$}
\psline{*-*}(0,4)(8,4)
\uput[d](4,4){$\gb_2$}
\uput[u](8,4){$\bfxp_k-\bfy_1-\bfy_2$}
\psline{*-*}(10,0)(13,0)
\uput[d](10,0){$\bfxp_k-\bfy_2$}
\uput[d](13,0){$\bfs_2$}
\uput[u](11,0){$\gc_2$}
\psline[linestyle=dashed]{*-*}(10,0)(8,4)
\psline[linestyle=dashed]{*-*}(13,0)(8,4)
\end{pspicture}
\caption{\label{fig:interaction-double-integral-four}\small $I_4$ integral with $\bfs_2=\bfxp_k$ or $\bfs_2\neq\bfxp_k$. In both cases, under the limit $\rho_1\to 0^+$, the velocity difference $w_{\gc_1}$ fuses and simultaneously the endpoint $\bfxp-\bfy_2$ of $w_{\gc_2}$ fuses with the endpoint $\bfxp_k-\bfy_1-\bfy_2$ of $w_{\gb_2}$. Fusion events of this type require special consideration, as discussed in the text. For $\bfs_2\neq\bfxp_k$, there is no fusion in the limit $\rho_2\to 0^+$. However, for $\bfs_2=\bfxp_k$, in the limit $\rho_2\to 0^+$, the velocity difference $w_{\gc_2}$ fuses.}
\end{center}\end{figure}
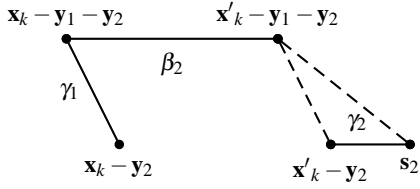

For the $I_4$ integral we distinguish between the cases $\bfs_2=\bfxp_k$ and $\bfs_2\neq\bfxp_k$. The corresponding velocity difference geometry is shown in Fig.~\ref{fig:interaction-double-integral-four}. For $\bfs_2\neq\bfxp_k$, there is no fusion in the limit $\rho_2 \to 0^+$, consequently $I_4$ is UV local via the argument of Case 1. For $\bfs_2=\bfxp_k$, the velocity difference $w_{\gc_2}$ fuses as $\rho_2\to 0^+$. This is exactly the same situation we encountered previously for the integral $I_3$ under the same limit, and locality can be established via the argument of Case 3 or Case 2, depending on the relative position of $w_{\gc_2}$ with respect to other velocity differences. 

The limit $\rho_1\to 0^+$ however presents us with a special challenge and requires the following careful consideration. In both cases $\bfs_2=\bfxp_k$ and $\bfs_2\neq\bfxp_k$ we see that, under the limit $\rho_1\to 0^+$, there are two simultaneous fusion events. The velocity difference $w_{\gc_1}$ fuses and separately, the $\bfxp-\bfy_1-\bfy_2$ endpoint of $w_{\gb_2}$ fuses with the $\bfxp_k-\bfy_2$ endpoint of $w_{\gc_2}$. We reattach the velocity difference $w_{\gc_2}$ onto $w_{\gb_2}$ by rewriting it according to Eq.~\eqref{eq:interaction-double-integral-three-decomposition} and breaking the integral $I_4$ into two corresponding contributions. In the second contribution, there is only one fusion event, namely the fusion of the velocity difference $w_{\gc_1}$, consequently locality is easily established via the argument of Case 2. In the first contribution, however we have two fusing velocity differences each attached on either side of the $w_{\gb_2}$ velocity difference. The scaling is governed by the $p=2$ fusion rule, and it is very tempting to invoke the argument that was previously given under Case 2. The problem is that in that argument we disregard any contribution of the spatial derivative to the $\rho_1$ scaling exponent and can justify doing so because the $p=1$ fusion rule that is used in that argument results in a velocity difference geometry where the spatial derivative does not act on the $\bfy_1$ separation between the velocity difference endpoints. Here, the velocity difference geometry is different. Fortunately, we note from Fig.~\ref{fig:interaction-double-integral-four}, that in spite of the involvement of two fused velocity differences, the 4 velocity difference endpoints involved form a parallelogram in which the separation at $w_{\gc_1}(\bfx_k-\bfy_1-\bfy_2, \bfx_k-\bfy_2)$ and $w_{\gc_2}(\bfxp-\bfy_2, \bfxp-\bfy_1-\bfy_2)$ have the same magnitude and direction. Consequently the tensor structure of the $\rho_1$ dependent factor depends exclusively on the vector $\bfy_1$. We also note that when the spatial derivatives ``shake'' $\bfx_k$ or $\bfxp_k$, the two fused velocity differences may drift closer or further away but remain parallel and maintain their orientation. The separation $\bfy_1$ remains unaffected; the derivatives only impact the large-scale separation in the $w_{\gb_2}$ velocity difference. It follows that the derivatives still do not contribute to the scaling exponent of $\rho_2$, and the argument of Case 2 carries through.

\begin{figure}[tb]\begin{center}
\begin{pspicture}[shift=*](-1,-1)(15,14)
\psline{*-*}(10,0)(10,3)
\uput[r](10,0){$\bfx_k$}
\uput[l](10,1.5){$\gc_p$}
\psline{*-*}(10,3)(10,6)
\uput[r](10,3){$\bfx_k-\bfy_p$}
\uput[l](10,4.5){$\gc_{p-1}$}
\psline[linestyle=dashed]{*-*}(10,6)(10,9)
\uput[r](10,6){$\bfx_k-\bfy_{p-1}-\bfy_p$}
\psline{*-*}(10,9)(12,12)
\uput[r](10,9){$\bfx_k-\bfy_2-\cdots -\bfy_p$}

\uput[r](12,12){$\bfx_k-\bfy_1-\cdots -\bfy_p$}
\psline{*-*}(2,12)(12,12)
\uput[u](2,12){$\bfxp_k-\bfy_1-\cdots-\bfy_p$}
\uput[d](7,12){$\gb_p$}
\psline[linestyle=dashed]{*-*}(2,12)(0,9)
\psline[linestyle=dashed]{*-*}(0,9)(0,6)
\psline[linestyle=dashed]{*-*}(0,6)(0,3)
\psline[linestyle=dashed]{*-*}(0,3)(0,0)
\end{pspicture}
\caption{\label{fig:general-cross-term-visualization}\small Velocity difference geometry for the $\bfy$ dependent velocity differences in the cross-term of $\cO_n \cO_{n+1}\cdots\cO_{n+p-1}F_{n+p}$ given by Eq.~\eqref{eq:SpecialCrossTerm}. The $w_{\gc_l}$ velocity differences are hanging, like a chain, from one end of the $w_{\gb_p}$ velocity difference. A ``phantom'' chain that is piecewise parallel to the real chain hanging from the other endpoint of the velocity difference $w_{\gb_p}$ is shown, using unmarked endpoints.}
\end{center}\end{figure}
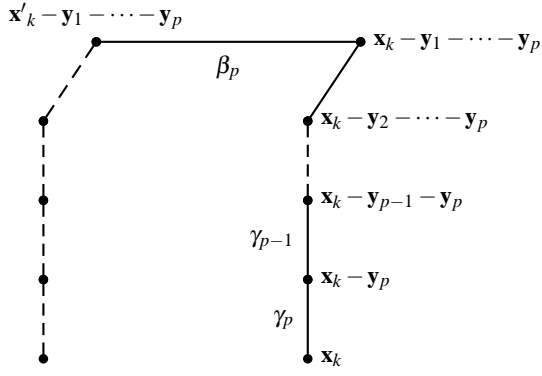

\subsection{UV multilocality proof for multiple operators}
\label{sec:UV-multilocality-proof-multiple-operators}

Cross-terms for $\cO_n \cO_{n+1}\cdots\cO_{n+p-1} F_{n+p}$ are more complicated and come in various combinations. Many of them take the form: 
\begin{multline}
I_1 = \int_{\bbR^d}\df{\bfy_p}\int_{\bbR^d}\df{\bfy_{p-1}}\cdots\int_{\bbR^d}\df{\bfy_1}\; P_{\ga_k\gb_1}(\bfy_p) \\ \times
\left[ \prod_{l=1}^{p-1} P_{\gb_l \gb_{l+1}} (\bfy_{p-l}) \right] 
\times \left[ \prod_{l=1}^{p} \pd_{\gc_l, \bfx_k} \right] \\ \times
\oavg
\biggl[ \prod_{\substack{l=1\\ l\neq k}}^n w_{\ga_l} (\bfX_l) \biggr]
w_{\gb_p} \left(\bfx_k-\sum_{l=1}^p \bfy_l, \bfxp_k-\sum_{l=1}^p \bfy_l  \right) \cpha \\
\opha \times\left[ \prod_{l=1}^p w_{\gc_l}\left(\bfx_k-\sum_{q=l}^p \bfy_q, \bfs_l \right)\right] \cavg \label{eq:GeneralCrossTerm}
\end{multline}
where we assume that $\bfs_l\neq\bfx_k$ for all $l\in\{1,2,\ldots, p\}$. The integrals over $\bfy_1, \bfy_2,\ldots,\bfy_p$ correspond to the operators $\cO_{n+p-1}, \cO_{n+p-2}, \ldots, \cO_{n+1}, \cO_n$. Note that if $\bfs_l=\bfx_k$ for some $l\in\{1,\ldots,p\}$, then the integrals over $\bfy_{l+1},\ldots,\bfy_p$ will shift it into $\bfx_k-\bfy_{l+1}-\bfy_{l+2}-\cdots -\bfy_p$, so there are many combinations with such modified velocity difference geometries.

One way to account for all combinations is to redefine $\bfs_l$ more generally as a function of $\bfy_1,\bfy_2,\ldots,\bfy_p$ via 
\begin{multline}
\bfs_l (\bfy_1,\ldots,\bfy_p|\{\bfX\}_n,\gs, \tau)  = \\ \casethree{\bfx_{\gs (l)}}{\gs (l) \neq k \land \tau (l)=1}{\bfxp_{\gs (l)}}{\tau (l)=2}{\bfx_k - \sum_{q=l+1}^p \bfy_q}{\gs (l) = k \land \tau (l) = 1}
\end{multline}
where $\gs$ is any arbitrary mapping $\gs : \{1,\ldots, p\}\to \{1,\ldots, n\}$ and $\tau$ is any arbitrary mapping $\tau : \{1,\ldots, p\}\to \{1,2\}$. Going through all possible mappings $\gs, \tau$ accounts for all cross-terms that involve spatial derivatives with respect to $\bfx_1,\bfx_2,\ldots,\bfx_n$. For the special case $\gs (l)=k$ and $\tau (l)=1$ for all $l\in\{1,\ldots, p\}$, we obtain a cross-term of the form 
\begin{multline}
I_2 = \int_{\bbR^d}\df{\bfy_p}\int_{\bbR^d}\df{\bfy_{p-1}}\cdots\int_{\bbR^d}\df{\bfy_1}\; P_{\ga_k\gb_1}(\bfy_p) \\ \times
\left[ \prod_{l=1}^{p-1} P_{\gb_l \gb_{l+1}} (\bfy_{p-l}) \right] 
\times \left[ \prod_{l=1}^{p} \pd_{\gc_l, \bfx_k} \right] \\ \times
\oavg
\biggl[ \prod_{\substack{l=1\\ l\neq k}}^n w_{\ga_l} (\bfX_l) \biggr]
w_{\gb_p} \left(\bfx_k-\sum_{l=1}^p \bfy_l, \bfxp_k-\sum_{l=1}^p \bfy_l  \right) \cpha \\
\opha \times\left[ \prod_{l=1}^p w_{\gc_l}\left(\bfx_k-\sum_{q=l}^p \bfy_q, \bfx_k-\sum_{q=l+1}^p \bfy_q \right)\right] \cavg \label{eq:SpecialCrossTerm}
\end{multline}

A much wider set of cross-terms can be constructed if we account for the terms where some of the spatial derivatives $\pd_{\gc_l,\bfx_k}$ are replaced with $\pd_{\gc_l,\bfxp_k}$. The corresponding terms can be obtained from Eq.~\eqref{eq:GeneralCrossTerm} by replacing $\bfx_k$ with $\bfxp_k$ in the first argument of $w_{\gc_l}$ and by replacing $\bfs_l$ with $\bfsp_l$, defined as: 
\begin{multline}
\bfsp_l (\bfy_1,\ldots,\bfy_p|\{\bfX\}_n, \gs, \tau) = \\ \casethree{\bfx_{\gs (l)}}{\tau (l)=1}{\bfxp_{\gs (l)}}{\tau (l)=2 \land \gs (l) \neq k}{\bfxp_k-\sum_{q=l+1}^p \bfy_q}{\tau (l)=2 \land \gs (l) = k}
\end{multline}
Visualizing the velocity difference geometry of those terms in general can be challenging, but a good point of departure is to begin with the integral $I_2$ given by Eq~\eqref{eq:SpecialCrossTerm}. The corresponding velocity difference geometry is shown in Fig.~\ref{fig:general-cross-term-visualization}. It consists of the velocity difference $w_{\gb_p}$ with the velocity differences $w_{\gc_1},\ldots, w_{\gc_p}$ forming a chain that is hanging from the $\bfx_k-\bfy_1-\cdots -\bfy_p$ endpoint. Fig.~\ref{fig:general-cross-term-visualization} also shows a phantom chain with unlabeled vertices that is hanging from the other endpoint of the velocity difference $w_{\gb_p}$ in a piecewise parallel fashion. Replacing all spatial derivatives $\pd_{\gc_l,\bfx_k}$ with $\pd_{\gc_l,\bfxp_k}$ for all $l\in\{ 1,\ldots, p\}$ corresponds to replacing the hanging chain with the phantom chain shown in Fig.~\ref{fig:general-cross-term-visualization}. Likewise, replacing only some of the spatial derivatives corresponds to replacing some of the velocity differences with their parallel counterparts from the phantom chain, resulting in isolated ``island chains'' or isolated velocity differences. Of course, these geometries correspond to the special case $\gs (l)=k$ for all $l\in\{1,\ldots, p\}$. Once we have $\gs (l)\neq k$ for some $l\in\{1,\ldots, p\}$, it corresponds to breaking the hanging chain at the velocity difference $w_{\gb_l}$ into two pieces. Choosing $\gs (l)\neq k$ for multiple values of $l$ results in multiple chain interruptions, giving a velocity difference geometry with many islands of velocity difference chains and possibly one or more isolated velocity differences. Taking all this under consideration, we can now give both the UV and IR locality proofs for all possible cross-terms that emerge from $\cO_n \cO_{n+1}\cdots\cO_{n+p-1}F_{n+p}$.

To establish UV locality, we consider the limit $\rho_l = \nrm{\bfy_l}\to 0^+$ for some $l\in\{1,\ldots, p\}$ and reuse the arguments given in section~\ref{sec:UV-multilocality-proof-two-operators} as Case 1,2, or 3. If there are no fusion events, UV locality follows unconditionally from the argument of Case 1. If there is a fusing velocity difference that is part of a velocity difference chain, then locality can be established by the argument of Case 2. If the fusing velocity difference is isolated from all other velocity differences, then we use the argument of Case 3. Another possibility is that instead of having one fusing geometry, we have two velocity difference endpoints from two distinct velocity differences approach each other. This could happen within the $\bfy$-dependent group of $w_{\gb_p}, w_{\gc_1},\ldots, w_{\gc_p}$ but it can also happen between one velocity difference from within that group and another velocity difference outside the group. In this situation, decomposing the approaching velocity difference can allow us to reattach it to the other velocity difference, and by making it once again part of some velocity difference chain allows us to reuse our previous repertoire of arguments. Last, but not least, in the case where spatial derivatives with respect to one of $\bfx_1,\bfx_2,\ldots, \bfx_n$ are mixed with spatial derivatives with respect to one of $\bfxp_1,\bfxp_2,\ldots, \bfxp_n$ we could have a situation where there are two parallel fusing velocity differences. We have seen previously, in Section~\ref{sec:UV-multilocality-proof-multiple-operators}, that even in this situation, the partial derivatives have no effect on the scaling exponent of $\rho_l$, so the argument of Case 2 can still be used to establish UV locality. In spite of the combinatorial complexity of the general cross-terms, the same arguments that were used to establish the UV locality of $\cO_n \cO_{n+1} F_{n+2}$ remain applicable to the most general case.

\subsection{IR multilocality proof}

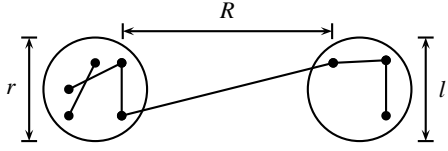
\begin{figure}[tb]\begin{center}
\begin{pspicture}[shift=*](-3,-3)(13,3)
\pscircle(0,0){2}
\psline{*-*}(-1,-1)(0,1)
\psline{*-*}(-1,0)(1,1)
\psline{*-*}(1,1)(1,-1)
\psline{|<->|}(-2.5,-2)(-2.5,2)
\uput[l](-2.5,0){$r$}
\psline{*-*}(1,-1)(9,1)
\psline{|<->|}(1,2.2)(9,2.2)
\uput[u](5,2.2){$R$}
\pscircle(10,0){2}
\psline{*-*}(9,1)(11,1.1)
\psline{*-*}(11,1.1)(11,-1)
\psline{|<->|}(12.5,-2)(12.5,2)
\uput[r](12.5,0){$l$}
\end{pspicture}
\caption{\label{fig:IR-multilocality-proof-geometry}\small
An example of the velocity difference geometry encountered in the IR limit. One of the velocity difference chains has one of its links stretched as $\rho_l$. As a result, part of the chain is pushed into a separate $l$-blob while all other velocity differences remain behind in the main blob. This results in the generalized two-blob fusion rule discussed earlier. }
\end{center}\end{figure}

To establish IR locality, we can now offer a better and more general argument than was given previously \cite{article:Procaccia:1996:2,article:Procaccia:1996:3,article:Gkioulekas:2008:1}. Considering the case of the cross-terms given by Eq.~\eqref{eq:SpecialCrossTerm} for $\cO_n\cO_{n+1}\cdots\cO_{n+p-1} F_{n+p}$, taking the IR limit $R =\nrm{\bfy_m}\to +\infty$ results in the fusion geometry shown in Fig.~\ref{fig:IR-multilocality-proof-geometry}:  one small blob has $n-1$ velocity differences from the $w_{\ga_l}$ factors, and $p-m$ velocity differences from the $w_{\gc_l}$ factors with $l\in\{p,p-1,\ldots, m+1\}$ with a total of $n+p-m-1$ velocity differences. The other small blob has $m$ velocity differences, including the factors $w_{\gc_l}$ with $l\in\{m-1,\ldots, 1\}$. The $w_{\gc_m}$ factor straddles across between the two small blobs over the large scale $R$. Other  types of cross-terms will still yield the same two-blob geometry, except the number of velocity differences on the two blobs may be $n+p-m'-1$ and $m'$, with $m'\neq m$, but still $1\leq m'\leq p$. As a result, there is no loss of generality in limiting our attention to the two-blob geometry of Fig.~\ref{fig:IR-multilocality-proof-geometry}. Similarly to the UV locality proof, the integrals are done one at a time, and we note that the integral differential $\df{\bfy_l}$ together with the corresponding projection function do not contribute to the scaling exponent of $R$. The derivative may or may not contribute an $R^{-1}$ factor. If it does, it is helping the IR locality argument, since we are looking at an $R\to +\infty$ limit. At this point, in order to proceed with the argument, it is necessary to distinguish between downscale and upscale cascades, and treat them separately, due to technical complications with the two-blob geometry in upscale cascades.

For the case of a downscale cascade, the two-blob geometry fusion rule gives the scaling $R^{\gD (m|n,p)}$ with $\gD (m|n,p)=\gz_{n+p}-\xi_{n+p, n+p-m}-\xi_{n+p,m+1}$, so for IR locality, a sufficient condition is $\gD (m|n,p)<0$ for all $1\leq m<p$. If we assume the fusion rules hypothesis, then since the cascade is downscale, we use $\xi_{np}=\gz_p$, and therefore 
\begin{align}
\gD (m|n,p) &= \gz_{n+p}-\gz_{n+p-m}-\gz_{m+1} \\
&\leq (\gz_{n+p-m}+\gz_m)-\gz_{n+p-m}-\gz_{m+1} \\
&= \gz_m-\gz_{m+1} < 0
\end{align}
Here we have used the Holder inequalities for the scaling exponents $\gz_n$ (see appendix D of Ref.~\cite{article:Gkioulekas:2008:1}) to obtain $\gz_{n+p}\leq\gz_{n+p-m}+\gz_m$. We have also used the well known result that the scaling exponents $\gz_n$ form an increasing sequence with respect to $n$ \cite{article:Frisch:1991,book:Frisch:1995}. Finally, we note that we do not encounter the special cases $\xi_{3,1}=\gz_3$ and $\xi_{3,2}=\gz_2$, because at minimum $n\geq 2$ and $p>1$ which implies that $n+p>3$. This establishes IR locality for downscale cascades. 

For upscale cascades, a technical difficulty, that was overlooked by my previous paper \cite{article:Gkioulekas:2008:1} regarding the two-blob velocity difference geometry, is that it represents one fusion event, unlike with downscale cascades where it represents two fusion events. This was previously discussed in section~\ref{sec:fusion-rule-two-blob-geometry}, where we show that the corresponding scaling of a generalized structure function with a total of $n$ velocity differences under a two-blob geometry should be $r^{\gz_n+\ga}R^{-\ga}$. If we disregard any contribution from the derivative, this scaling is  good enough to ensure IR locality under the corresponding limit $R\to +\infty$. However, if the derivative contributes an $R^{-1}$ factor, then the overall scaling with respect to $R$ will make the nonlinear interactions integral IR local, even if we do not account for the cancellation in Fig.~\ref{fig:separation-of-one-vel-d-from-one-blob} and assume that $\ga=0$. We will now argue that the condition $\ga>0$ is a necessary condition for IR locality and multilocality that cannot be removed.

\begin{figure}[tb]\begin{center}
\begin{pspicture}[shift=*](-3,-3.3)(13,3.3)
\pscircle(0,0){3}
\pscircle(10,0){3.3}
\psline{*-*}(-1,0)(-1,2)
\psline{*-*}(-2,0)(0,2)
\psline{*-*}(9,2)(9,-2)
\uput[r](9,2){$\bfx_k-\bfy$}
\uput[r](9,-2){$\bfxp_k-\bfy$}
\psline{*-*}(0,-2)(2,1)
\uput[l](2,1){$\bfs$}
\uput[l](0,-2){$\bfsp$}
\psline{*-*}(2,1)(9,2)
\end{pspicture}
\caption{\label{fig:IR-multilocality-geometry}\small
Velocity difference geometry for the case $m=1$ in the IR limit, resulting in a two-blob geometry.}
\end{center}\end{figure}
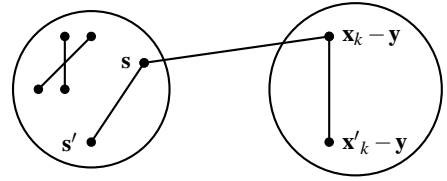

\begin{figure}[tb]\begin{center}
\begin{pspicture}[shift=*](-3,-3.3)(10,4)
\pscircle(0,0){3}
\psline{*-*}(-1,0)(-1,2)
\psline{*-*}(-2,0)(0,2)
\psline{*-*}(1,1)(10,1)
\uput[u](1,1){$\bfs$}
\uput[ul](10,1){$\bfx_k-\bfy$}
\psline{|<->|}(0,3.3)(10,3.3)
\uput[u](5,3.3){$R$}
\end{pspicture}
\caption{\label{fig:IR-multilocality-zero-case-one}\small 
One of the simplest velocity differences for $m=0$ where we assume that no other velocity difference is attached to the point $\bfs$.}
\end{center}\end{figure}
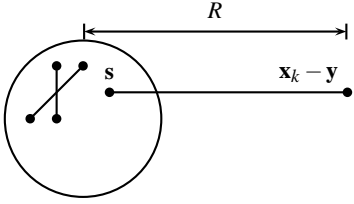

\begin{figure}[tb]\begin{center}
\begin{pspicture}[shift=*](-3,-3.3)(10,4)
\pscircle(0,0){3}
\psline{*-*}(-1,0)(-1,2)
\psline{*-*}(-2,0)(0,2)
\psline{*-*}(1,1)(10,1)
\uput[u](1,1){$\bfs$}
\uput[ul](10,1){$\bfx_k-\bfy$}
\psline{|<->|}(0,3.3)(10,3.3)
\uput[u](5,3.3){$R$}
\psline{*-*}(1,1)(0,-2)
\uput[l](0,-2){$\bfsp$}
\end{pspicture}
\caption{\label{fig:IR-multilocality-zero-case-two}\small 
Another $m=0$ velocity difference geometry, however one that becomes problematic with regards to the IR locality argument.}
\end{center}\end{figure}
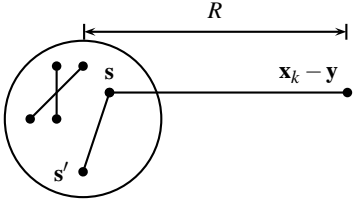

 The challenge we face if we attempt to derive IR locality and multilocality under the assumption  $\ga = 0$ is that in many velocity difference configurations, a derivative with respect to $\bfx_k$ can affect multiple velocity differences at the same time, and by the product rule of differentiation we are expecting a sum of terms where some will have the additional $R^{-1}$ scaling factor, and others will not. For example, for the case $m=1$, let us consider the velocity difference geometry shown in Fig.~\ref{fig:IR-multilocality-geometry}, and assume that $\bfs\neq\bfx_k$. The derivative with respect to $\bfx_k$will shake the point $\bfx_k-\bfy$ which is shared by the straddling velocity difference between $\bfs$ and $\bfx_k-\bfy$ and the velocity difference between $\bfx_k-\bfy$ and $\bfxp_k-\bfy$ situated on the right blob. Shaking the separation in the straddling velocity difference will indeed yield  a term with $R^{-1}$ overall scaling. However, the simultaneous shaking of the other velocity difference on the right blob will yield an additional term that will scale as $R^0$. If we assume that $\bfs=\bfx_k$, the situation gets seemingly worse. Now, the length of the straddling velocity difference does not even shake, so we do not even get the $R^{-1}$ term.

To gain some further insight into the mathematics of the IR limit, it is worth making the following additional observations. First, if IR locality and multilocality can be shown for all velocity difference geometries with $m=1$, it automatically follows that all velocity difference geometries with $m>1$ will also be IR local. This can be shown by an induction proof where we use the argument corresponding to Fig.~\ref{fig:two-blob-upscale-induction-general-step} for the inductive step. So, it is only necessary to establish IR locality and multilocality  for the $m=1$ velocity difference geometries. As a matter of fact, although none of the interaction integrals correspond to an $m=0$ geometry, if the induction argument can be initialized at $m=0$, and we show that for all $m=0$ velocity difference geometries the derivative with respect to $\bfx_k$ introduces an overall $R^{-1}$ factor, we should expect it to do the same for all the relevant velocity difference geometries with $m\geq 1$. 

The simplest possible velocity difference corresponding to $m=0$ is shown in Fig.~\ref{fig:IR-multilocality-zero-case-one}. If we assume that $\bfs\neq\bfx_k$, then a derivative with respect to $\bfx_k$ will only affect the large velocity difference separation between $\bfs$ and $\bfx_k-\bfy$, and will thus yield only one term with $R^{-1}$ scaling. If we assume that $\bfs = \bfx_k$, then when the derivative ``shakes'' $\bfx_k$, this shakes the entire velocity difference back and forth without changing the separation length. Since the cascade is upscale, universal local homogeneity gives invariance when shifting all of the large velocity differences simultaneously by the same displacement vector. It follows that, via universal local homogeneity, the derivative annihilates the velocity difference geometry of Fig.~\ref{fig:IR-multilocality-zero-case-one}  when $\bfs=\bfx_k$. Both are favorable outcomes with regards to establishing IR locality. 

Problems begin when we consider velocity difference geometries such as the one shown in Fig.~\ref{fig:IR-multilocality-zero-case-two}  where there are both a large and a small velocity difference attached at the point $\bfs$. For the case $\bfs=\bfx_k$, with $\bfsp$ independent of $\bfx_k$, we have once again a problem when differentiating with respect to $\bfx_k$. The derivative shakes the large velocity difference back and forth without affecting the separation between its endpoints, so we cannot be expecting an $R^{-1}$ factor. Furthermore, the velocity difference separation between $\bfs$ and $\bfsp$ is shaken instead, resulting in a problematic contribution with $R^0$ scaling. The attachment of the large velocity difference to a small one prevents us from using universal local homogeneity to have the derivative with respect to $\bfx_k$ annihilate the entire term. Worse, these types of contributions will turn up even when attempting to establish IR locality for the simplest case of $\cO_n F_{n+1}$ by attempting an inductive argument initiated from $m=0$. With multilocality integrals the situation becomes even worse as we can have entire chains of velocity differences passing through endpoints of the form $\bfx_k, \bfx_k-\bfy_1, \bfx_k-\bfy_1-\bfy_2,\ldots$. A derivative with respect to $\bfx_k$ will shake these chains as a whole. The corresponding terms cannot be annihilated by invoking universal local homogeneity as these chains will involve either a combination of one large velocity difference with some small velocity differences or islands made entirely of small velocity differences, situated entirely within the small $r$-blob. These geometries will still be there if we attempt an IR multilocality proof with the aforementioned inductive argument initialized at $m=0$, and they will result in contributions that scale as $R^0$. 

Taking into account the above considerations, it is relatively safe to conclude that the condition $\ga >0$ is necessary in order to establish IR locality and multilocality, for the case of an upscale cascade, i.e. the inverse energy cascade of two-dimensional turbulence.

\section{Conclusion}

This paper has focused primarily on the question of multilocality. As was explained in the introduction, the goal is to investigate whether the nonlinear interactions terms, that arise from the balance equations governing higher-order time derivatives of generalized structure functions, involve integrals that are local in the UV and IR limits, where these terms are evaluated inside the inertial range of a turbulence cascade. The locality of these terms is a gateway to employing the balance equations to investigate a number of interesting questions, such as bridge relations between scaling exponents \cite{article:Procaccia:1996:3} and the transition to the dissipation range \cite{article:Procaccia:1996:2,article:Procaccia:1996:3,article:Gkioulekas:p14}. In this concluding discussion, we will begin by summarizing our main results. We will then compare our argument with an interesting investigation of locality by Eyink and Aluie \cite{article:Eyink:2005,article:Aluie:2009,article:Eyink:2009}. Finally, we will discuss the limitations of the notion of non-perturbative locality/multilocality, addressing numerical evidence that question the universality and locality of the cascades of two-dimensional turbulence \cite{article:Chekhlov:1999,article:Gurarie:2001,article:Gurarie:2001:1,article:Danilov:2003,article:Vallgren:2011}, in the context of our previous work \cite{article:Tung:2005,article:Tung:2005:1,article:Gkioulekas:2007,article:Gkioulekas:2008:1,article:Gkioulekas:p14}, in order to clarify in physical terms what non-perturbative locality/multilocality  entails and does not entail. 

For downscale cascades, including both the downscale energy cascade of three-dimensional Navier-Stokes turbulence and the downscale enstrophy cascade of two-dimensional Navier-Stokes turbulence, we have shown that both locality and multilocality follow as a consequence of the fusion rules hypothesis, in both the UV and IR limits. We have also seen that the IR multilocality proof makes use of all fusion rules with $p\geq 2$, unlike the locality proof which is only dependent on the $p=2$ fusion rule. This is noteworthy because, in terms of theoretical studies, the fusion rules have been established only for the case $p=2$ for the downscale energy cascade of three-dimensional turbulence \cite{article:Procaccia:1995:1,article:Procaccia:1995:2,article:Procaccia:1996}. This is sufficient for both locality and multilocality in the UV limit, and for locality in the IR limit. However, multilocality in the IR limit, also requires the use of fusion rules with $p>2$. 

The situation is more nebulous with regard to upscale cascades, and specifically the inverse energy cascade of two-dimensional Navier-Stokes turbulence. We have shown that the fusion rules hypothesis continues to yield locality and multilocality in the UV limit. However, in the IR limit, both locality and multilocality would have been  at the cross-over point between holding and failing, but they are salvaged thanks to a cancellation associated with the spatial decorrelation in pulling two groups of velocity differences apart, as shown in Fig.~\ref{fig:separation-of-one-vel-d-from-one-blob}. The underlying culprit is the two-blob geometry fusion rule for the case of upscale cascades where we predict scaling of the form $r^{\gz_n+\ga}R^{-\ga}$  with $r\ll R$. In order to recover both locality and multilocality in the IR limit, the scaling exponent with respect to $R$ needs to be negative, and this hinges entirely on the assumption $\ga >0$ which is both necessary and sufficient. 

It is also worth commenting on the restrictions that must be satisfied by the scaling exponents $\gz_n$ in order to have multilocality. In the UV limit, for both downscale and upscale cascades, the condition $\xi_{n,1}>0$ is sufficient for both locality and multilocality.  Under monoscaling $\gz_n=nh$, in both cascade directions the locality condition reduces to $h>0$, with respect to the Holder exponent $h$. In the IR limit, restricting ourselves to downscale cascades, the multilocality condition for $p$ applications of the nonlinear interactions operator is $\gz_{n+p}-\xi_{n+p, n+p-m}-\xi_{n+p, m+1}<0$ for all $m$ such that $1\leq m <p$. Under monoscaling, this condition also reduces to $h>0$. Notable is the absence of the restriction $h<1$, corresponding to the requirement that the corresponding energy spectrum should not scale steeper than $k^{-3}$. This is a very important point that was previously discussed in detail in the conclusion of my previous paper \cite{article:Gkioulekas:2008:1} on cascade locality. In brief, it is reasonable to expect that a condition $h<1$ is hidden behind the theoretical argument needed to establish the fusion rules from first principles. 

An alternative approach for establishing the locality of cascades was proposed in an earlier paper by Eyink \cite{article:Eyink:2005}  in which it was shown that under monoscaling $\gz_n = nh$, the condition $0<h<1$ is sufficient for locality in the IR and UV limits. Furthermore, his result is applicable to cascades of both three-dimensional and two-dimensional turbulence, and it is mathematically rigorous. The only assumption that needs to be made is \Holder continuity of the velocity field with \Holder exponent $0<h<1$ in the limit of infinite Reynolds number. His result also holds in a multifractal case in which there is a range of \Holder exponents $[h_{\text{min}}, h_{\text{max}}]\subseteq (0,1)$. 

One could argue that Eyink's result is more rigorous than ours because it does not depend on the fusion rules, therefore it is reasonable to want to compare the two arguments. First, we observe that both arguments are rigorous in the sense of connecting assumptions to conclusions. Eyink's argument does not require either a spatial or ensemble average. Our argument requires an ensemble average to the extent that it is needed for stable self-similar scaling and by the fusion rules. Our assumption that the generalized structure functions are self-similar in the inertial range with scaling exponents $\gz_n$, by itself, is weaker than Eyink's assumption of \Holder continuity in the limit of infinite Reynolds number. However, our assumption of the fusion rules combined with universal local homogeneity and universal local isotropy increase the overall assumption load that we bring into the argument, and it is reasonable to inquire about the relative strength of the conclusions. 

Eyink's approach is based on filtering the Navier-Stokes equations with a smooth low-pass filter and writing corresponding balance equations for the energy, enstrophy, and helicity. From these balance equations, he extracts Gallilean-invariant expressions for the energy, enstrophy, and helicity fluxes, and uses them to establish locality. From a mathematical standpoint, this framework is equivalent to working with standard, as opposed to generalized structure functions, and it is limited to the balance equation of the second-order structure function. Our notion of non-perturbative locality is stronger in two ways: it applies to generalized structure functions with arbitrary geometries, and it applies to all balance equations of generalized structure functions for all orders. On top of that, the newly introduced notion of multilocality is an additional generalization that broadens the concept of non-perturbative locality even further. The price that we pay is the need to assume the fusion rules hypothesis, which arises directly from the interaction integrals and is due to our use of generalized structure functions. 

Both arguments are useful for different reasons. Eyink's argument limits the notion of locality to the aspects that admit obvious physical interpretations, and that makes it possible to carry out his argument with a very light array of assumptions. Our notion of non-perturbative locality is useful if one's point of view is to build a broader theory that is capable of accounting for the presence or absence of intermittency corrections to monoscaling \cite{article:Procaccia:1998,article:Procaccia:1998:1,article:Procaccia:1998:2}. It also allows us to envision the very concept of an inertial range as a multidimensional region, which can give some new insights on understanding the locality of the downscale enstrophy cascade in two-dimensional turbulence \cite{article:Gkioulekas:p14} and the transition to the dissipation range \cite{article:Procaccia:1996:2,article:Procaccia:1996:3,article:Gkioulekas:p14}. 

It is interesting to note that a combination of the fusion rules and Eyink's argument gives stronger results \cite{article:Aluie:2009,article:Eyink:2009} that reconciles them with predictions from closure models. However, one should bear in mind the distinction between perturbative and non-perturbative locality (see discussion at the conclusion of Ref.~\cite{article:Gkioulekas:2008:1}). Any study of locality based on closure models gives us only perturbative locality. The relation between these concepts is that perturbative locality combined with some additional requirements leads to the fusion rules, and the fusion rules in turn yield non-perturbative locality. We believe that the condition $0<h<1$ is needed during the very first step of establishing perturbative locality. We also see that non-perturbative locality only requires $h>0$ combined with the fusion rules hypothesis. This can become relevant to the case of a downscale enstrophy cascade where $h=1$. Even though perturbative locality could fail, in a borderline fashion, if the fusion rules survive, then non-perturbative locality survives, and that in turn can account for the possible observability  of the enstrophy cascade, under certain conditions, in spite of its apparent nonlocality. 

It should be noted that even if non-perturbative locality is satisfied, the downscale enstrophy cascade, due to its steep scaling, can be crashed both from the forcing range and from the dissipation range. We have shown previously \cite{article:Gkioulekas:2008:1} that even if the forcing spectrum is limited to a finite band of large scales, the corresponding forcing term $Q_n$ of the balance equations can still creep itself into the inertial range, due to its dependence on the generalized structure function $F_{n-2}$. This creeping effect depends on the magnitude of the small downscale energy flux that accompanies the downscale enstrophy cascade. In the limit of large Reynolds number, this downscale energy flux tends to zero, and the forcing range of the $Q_n$ term, will recede into the same range of large scales as the original forcing spectrum. Furthermore, when the downscale enstrophy cascade is dissipated by a standard Laplacian viscosity term, the dissipation range tends to creep into the inertial range from the other side, but the actual multidimensional region that corresponds to the enstrophy cascade inertial range becomes inflated, and thus salvaged, via the logarithmic correction to the power-law scaling \cite{article:Gkioulekas:p14}. This problem is eliminated when hyperdiffusion is used instead of a standard Laplacian for the small-scale dissipation. 

Even though our argument has shown that the fusion rules imply both the IR and UV locality and multilocality of the inverse energy cascade, we believe that, unlike with downscale cascades, trouble can arise from the sweeping term $I_n$ of the balance equations for the generalized structure functions, which cannot be safely ignored  \cite{article:Gkioulekas:2007}. This relates to extensive numerical evidence that may seem to indicate the strong involvement of nonlocal interactions driving the inverse energy cascade \cite{article:Gurarie:2001,article:Gurarie:2001:1,article:Danilov:2003}.  This apparent nonlocality was especially highlighted  by Danilov \cite{article:Danilov:2003} where he noted that Kolmogorov $k^{-5/3}$ scaling is achieved when the dissipation of large scales is driven by linear damping that intrudes into the inertial range to the extent that the inverse energy flux is nowhere constant. At the same time, when the large scale dissipation is replaced with hypodiffusion and constant energy flux is achieved, the energy spectrum deviates from Kolmogorov scaling. This departure manifests itself physically in the spontaneous emergence of coherent structures that accumulate vast amounts of energy, overshadowing the $k^{-5/3}$ energy spectrum. These coherent structures take the form of point vortices, and although they are eventually dissipated, new ones arise to take their place \cite{article:Gurarie:2001,article:Gurarie:2001:1,article:Danilov:2003}. This paradoxical behavior of the inverse energy cascade has already been discussed in my previous papers \cite{article:Tung:2005:1,article:Gkioulekas:2007}. The coherent structures were explained as a particular solution of the underlying statistical theory of randomly forced Navier-Stokes equations coexisting linearly with a homogeneous solution corresponding to the inverse energy cascade \cite{article:Tung:2005,article:Tung:2005:1}. Because the vortex spikes associated with the coherent structures intensify the sweeping of the flow around them, we identified the sweeping term $I_n$ as the term primarily forcing the particular solution, and the sensing of the loss of homogeneity by the boundary conditions at large scales as the mechanism jumpstarting the emergence of these coherent structures. 

Some of the more recent numerical results have been consistent with the observations by Danilov \cite{article:Danilov:2003} with regards to the inverse energy cascade. There are particularly two studies that warrant special mention: Boffetta \cite{article:Boffetta:2007}  was able to reproduce a joint inverse energy cascade simultaneously with a downscale enstrophy cascade using very high numerical resolution $16384^2$. The large scales were dissipated using linear dissipation, and although the energy spectrum of the inverse energy cascade gave $k^{-5/3}$ scaling, the corresponding energy flux was not constant. A follow-up simulation at $32768^2$ resolution \cite{article:Musacchio:2010} seems to indicate a small window of constant energy flux. However, the statistics of the energy flux were not collected over a large enough time scale to achieve proper convergence, so it is not clear that constant energy flux was achieved. 

A later study by Vallgren \cite{article:Vallgren:2011} revisited the problem of the non-robustness of the inverse energy cascade.Vallgren showed that nonlocal interactions play an essential role in driving the upscale transfer of energy. He also showed that increasing the strength of linear dissipation at large scales decreases the role of these nonlocal interactions. A recent paper \cite{article:Eyink:2009:1} reports numerical simulations that are able to simultaneously  reproduce both the inverse energy cascade spectrum $k^{-5/3}$ and a wide range of scales with constant upscale energy flux, regardless of whether at large scales the dissipation mechanism is linear damping or hypodiffusion. A careful reading of the reported results shows that the common feature of all of the reported simulations is a very wide dissipation range that begins at length scales that have considerable separation from the periodic boundary condition length scale. 

Combining the above observational evidence with our theoretical understanding, our explanation of the overall phenomenology is that the emergence of coherent structures in the inverse energy cascade of two-dimensional turbulence is driven by the sweeping interactions \cite{article:Gkioulekas:2007}, associated with the $I_n$ term of the balance equations for generalized structure functions, which become important over a range of large scales where the loss of homogeneity by the boundary conditions can be sensed by the nonlinear dynamics that transfer energy upscale. The emergence of coherent vortices then should amplify the sweeping term, resulting in a run-away dominance of nonlocal sweeping over the local interactions driving the inverse energy cascade. In order to effectively suppress these coherent structures, it is necessary to arrange forcing and dissipation so that the range of scales that are forced by the sweeping term (we can call them \emph{the sweeping range}) are entirely submerged under the dissipation range at large scales. This is easier done with linear dissipation rather than hypodiffusion, so it creates the impression that we have to trade off between suppressing the coherent structures versus achieving an inertial range dominated by local interactions and constant upscale energy flux. The numerical results by Ref.~\cite{article:Eyink:2009:1} provide with a counterexample where coherent structures have been effectively suppressed in a numerical simulation that uses hypodiffusion at large scales. 

Of course, suppressing the coherent structures does not imply total suppresion of the nonlocality that results from the remaining weakened effect of the sweeping term $I_n$ in the inertial range, which is still evidenced by the results of Vallgren \cite{article:Vallgren:2011}. However, as we pointed out in previous work \cite{article:Tung:2005,article:Tung:2005:1}, due to the linear structure of the exact statistical theory of the randomly forced Navier-Stokes equations, all that sweeping can do is force a \emph{``particular solution''}, manifesting itself as the coherent structures, that coexists linearly with a \emph{``homogeneous solution''}, manifesting as a local inverse energy cascade. Our claim of locality and multilocality for the inverse energy cascade apply only to the homogeneous solution, which is dominant when the coherent structures are suppressed and subdominant when the coherent structures are strong enough to hoard most of the energy and override the $k^{-5/3}$ scaling in the energy spectrum. In realistic situations, both phenomena coexist, creating the appearance that the inverse energy cascade itself is nonlocal.

We would like to conclude this discussion by mentioning that similar considerations apply to the downscale energy cascade of three-dimensional turbulence. The main difference is that, as a result of the downscale cascade direction, the sweeping range coincides, for the most part, with the forcing range. It could be entirely submerged inside the forcing range or the two ranges could possibly intersect but have some disjoint regions. Either way, simply increasing the Reynolds number separates the dissipation region from both forcing and sweeping ranges, enabling an inertial range where neither forcing nor sweeping is dominant. The nice slope of the energy spectrum in the inertial range of the downscale energy cascade also helps to shield it from both the forcing range and the dissipation range. The key difference  between the downscale and upscale  energy cascade is that in the inverse energy cascade the sweeping range needs to be entirely submerged inside and dominated by the dissipation range, requiring careful tuning of forcing and dissipation, whereas in the downscale energy cascade there is no need to submerge the entire sweeping range inside the forcing range. This contributes to  the substantial robustness of the downscale energy cascade of three-dimensional turbulence relative to the inverse energy cascade of two-dimensional turbulence. Again, the sweeping term will still force a subdominant particular solution that wil coexist linearly with the dominant homogeneous solution driving the downscale energy cascade. The particular solution is expected to be nonlocal. Our locality and multilocality proofs are applicable only to the homogeneous solution that is inherent to the $\cO_n$ system of operators of the generalized balance equations, and corresponds to the observed cascade phenomenology. 

\begin{acknowledgements}
Constructive criticism from an anonymous reviewer helped me to improve the introduction of the paper, and  is sincerely appreciated. 
\end{acknowledgements}

\bibliography{references}

\begin{thebibliography}{70}
\expandafter\ifx\csname natexlab\endcsname\relax\def\natexlab#1{#1}\fi
\expandafter\ifx\csname bibnamefont\endcsname\relax
  \def\bibnamefont#1{#1}\fi
\expandafter\ifx\csname bibfnamefont\endcsname\relax
  \def\bibfnamefont#1{#1}\fi
\expandafter\ifx\csname citenamefont\endcsname\relax
  \def\citenamefont#1{#1}\fi
\expandafter\ifx\csname url\endcsname\relax
  \def\url#1{\texttt{#1}}\fi
\expandafter\ifx\csname urlprefix\endcsname\relax\def\urlprefix{URL }\fi
\providecommand{\bibinfo}[2]{#2}
\providecommand{\eprint}[2][]{\url{#2}}

\bibitem[{\citenamefont{Richardson}(1922)}]{book:Richardson:1922}
\bibinfo{author}{\bibfnamefont{L.}~\bibnamefont{Richardson}},
  \emph{\bibinfo{title}{Weather prediction by numerical process}}
  (\bibinfo{publisher}{Cambridge University Press},
  \bibinfo{address}{Cambridge}, \bibinfo{year}{1922}).

\bibitem[{\citenamefont{Kolmogorov}(1941{\natexlab{a}})}]{article:Kolmogorov:1941}
\bibinfo{author}{\bibfnamefont{A.}~\bibnamefont{Kolmogorov}},
  \bibinfo{journal}{Dokl. Akad. Nauk. SSSR} \textbf{\bibinfo{volume}{30}},
  \bibinfo{pages}{301} (\bibinfo{year}{1941}{\natexlab{a}}),
  \bibinfo{note}{english translation published in volume 434 of \textsl{Proc.
  R. Soc. Lond. A}}.

\bibitem[{\citenamefont{Kolmogorov}(1941{\natexlab{b}})}]{article:Kolmogorov:1941:1}
\bibinfo{author}{\bibfnamefont{A.}~\bibnamefont{Kolmogorov}},
  \bibinfo{journal}{Dokl. Akad. Nauk. SSSR} \textbf{\bibinfo{volume}{32}},
  \bibinfo{pages}{16} (\bibinfo{year}{1941}{\natexlab{b}}),
  \bibinfo{note}{english translation published in volume 434 of \textsl{Proc.
  R. Soc. Lond. A}}.

\bibitem[{\citenamefont{Batchelor}(1947)}]{article:Batchelor:1947}
\bibinfo{author}{\bibfnamefont{G.}~\bibnamefont{Batchelor}},
  \bibinfo{journal}{Proc. Camb. Phil. Soc.} \textbf{\bibinfo{volume}{43}},
  \bibinfo{pages}{533} (\bibinfo{year}{1947}).

\bibitem[{\citenamefont{Kraichnan}(1967)}]{article:Kraichnan:1967:1}
\bibinfo{author}{\bibfnamefont{R.}~\bibnamefont{Kraichnan}},
  \bibinfo{journal}{Phys. Fluids} \textbf{\bibinfo{volume}{10}},
  \bibinfo{pages}{1417} (\bibinfo{year}{1967}).

\bibitem[{\citenamefont{Leith}(1968)}]{article:Leith:1968}
\bibinfo{author}{\bibfnamefont{C.}~\bibnamefont{Leith}},
  \bibinfo{journal}{Phys. Fluids} \textbf{\bibinfo{volume}{11}},
  \bibinfo{pages}{671} (\bibinfo{year}{1968}).

\bibitem[{\citenamefont{Batchelor}(1969)}]{article:Batchelor:1969}
\bibinfo{author}{\bibfnamefont{G.}~\bibnamefont{Batchelor}},
  \bibinfo{journal}{Phys. Fluids Suppl. II} \textbf{\bibinfo{volume}{12}},
  \bibinfo{pages}{233} (\bibinfo{year}{1969}).

\bibitem[{\citenamefont{Frisch}(1995)}]{book:Frisch:1995}
\bibinfo{author}{\bibfnamefont{U.}~\bibnamefont{Frisch}},
  \emph{\bibinfo{title}{Turbulence: The legacy of {A.N. K}olmogorov}}
  (\bibinfo{publisher}{Cambridge University Press},
  \bibinfo{address}{Cambridge}, \bibinfo{year}{1995}).

\bibitem[{\citenamefont{Lindborg and Alvelius}(2000)}]{article:Alvelius:2000}
\bibinfo{author}{\bibfnamefont{E.}~\bibnamefont{Lindborg}} \bibnamefont{and}
  \bibinfo{author}{\bibfnamefont{K.}~\bibnamefont{Alvelius}},
  \bibinfo{journal}{Phys. Fluids} \textbf{\bibinfo{volume}{12}},
  \bibinfo{pages}{945} (\bibinfo{year}{2000}).

\bibitem[{\citenamefont{Sukoriansky et~al.}(1999)\citenamefont{Sukoriansky,
  Galperin, and Chekhlov}}]{article:Chekhlov:1999}
\bibinfo{author}{\bibfnamefont{S.}~\bibnamefont{Sukoriansky}},
  \bibinfo{author}{\bibfnamefont{B.}~\bibnamefont{Galperin}}, \bibnamefont{and}
  \bibinfo{author}{\bibfnamefont{A.}~\bibnamefont{Chekhlov}},
  \bibinfo{journal}{Phys. Fluids} \textbf{\bibinfo{volume}{11}},
  \bibinfo{pages}{3043} (\bibinfo{year}{1999}).

\bibitem[{\citenamefont{Danilov and
  Gurarie}(2001{\natexlab{a}})}]{article:Gurarie:2001}
\bibinfo{author}{\bibfnamefont{S.}~\bibnamefont{Danilov}} \bibnamefont{and}
  \bibinfo{author}{\bibfnamefont{D.}~\bibnamefont{Gurarie}},
  \bibinfo{journal}{Phys. Rev. E} \textbf{\bibinfo{volume}{63}},
  \bibinfo{pages}{020203(R)} (\bibinfo{year}{2001}{\natexlab{a}}).

\bibitem[{\citenamefont{Danilov and
  Gurarie}(2001{\natexlab{b}})}]{article:Gurarie:2001:1}
\bibinfo{author}{\bibfnamefont{S.}~\bibnamefont{Danilov}} \bibnamefont{and}
  \bibinfo{author}{\bibfnamefont{D.}~\bibnamefont{Gurarie}},
  \bibinfo{journal}{Phys. Rev. E} \textbf{\bibinfo{volume}{63}},
  \bibinfo{pages}{061208} (\bibinfo{year}{2001}{\natexlab{b}}).

\bibitem[{\citenamefont{Danilov}(2005)}]{article:Danilov:2003}
\bibinfo{author}{\bibfnamefont{S.}~\bibnamefont{Danilov}},
  \bibinfo{journal}{Discrete Contin. Dyn. Syst. Ser. B}
  \textbf{\bibinfo{volume}{5}}, \bibinfo{pages}{67} (\bibinfo{year}{2005}).

\bibitem[{\citenamefont{Vallgren}(2011)}]{article:Vallgren:2011}
\bibinfo{author}{\bibfnamefont{A.}~\bibnamefont{Vallgren}},
  \bibinfo{journal}{J. Fluid. Mech.} \textbf{\bibinfo{volume}{667}},
  \bibinfo{pages}{463} (\bibinfo{year}{2011}).

\bibitem[{\citenamefont{Tabeling}(2002)}]{article:Tabeling:2002}
\bibinfo{author}{\bibfnamefont{P.}~\bibnamefont{Tabeling}},
  \bibinfo{journal}{Phys. Rep.} \textbf{\bibinfo{volume}{362}},
  \bibinfo{pages}{1} (\bibinfo{year}{2002}).

\bibitem[{\citenamefont{Gkioulekas and Tung}(2006)}]{article:Tung:2006}
\bibinfo{author}{\bibfnamefont{E.}~\bibnamefont{Gkioulekas}} \bibnamefont{and}
  \bibinfo{author}{\bibfnamefont{K.}~\bibnamefont{Tung}}, \bibinfo{journal}{J.
  Low Temp. Phys.} \textbf{\bibinfo{volume}{145}}, \bibinfo{pages}{25}
  (\bibinfo{year}{2006}).

\bibitem[{\citenamefont{Kraichnan}(1958)}]{article:Kraichnan:1958}
\bibinfo{author}{\bibfnamefont{R.}~\bibnamefont{Kraichnan}},
  \bibinfo{journal}{Phys. Rev.} \textbf{\bibinfo{volume}{109}},
  \bibinfo{pages}{1407} (\bibinfo{year}{1958}).

\bibitem[{\citenamefont{Kraichnan}(1959)}]{article:Kraichnan:1959}
\bibinfo{author}{\bibfnamefont{R.}~\bibnamefont{Kraichnan}},
  \bibinfo{journal}{J. Fluid. Mech.} \textbf{\bibinfo{volume}{5}},
  \bibinfo{pages}{497} (\bibinfo{year}{1959}).

\bibitem[{\citenamefont{Grant et~al.}(1962)\citenamefont{Grant, Stewart, and
  A.Moilliet}}]{article:A.Moilliet:1962}
\bibinfo{author}{\bibfnamefont{H.}~\bibnamefont{Grant}},
  \bibinfo{author}{\bibfnamefont{R.}~\bibnamefont{Stewart}}, \bibnamefont{and}
  \bibinfo{author}{\bibnamefont{A.Moilliet}}, \bibinfo{journal}{J. Fluid.
  Mech.} \textbf{\bibinfo{volume}{12}}, \bibinfo{pages}{241}
  (\bibinfo{year}{1962}).

\bibitem[{\citenamefont{Gibson}(1962)}]{article:Gibson:1962}
\bibinfo{author}{\bibfnamefont{M.}~\bibnamefont{Gibson}}, \bibinfo{journal}{J.
  Fluid. Mech.} \textbf{\bibinfo{volume}{15}}, \bibinfo{pages}{161}
  (\bibinfo{year}{1962}).

\bibitem[{\citenamefont{Kraichnan}(1964)}]{article:Kraichnan:1964}
\bibinfo{author}{\bibfnamefont{R.}~\bibnamefont{Kraichnan}},
  \bibinfo{journal}{Phys. Fluids} \textbf{\bibinfo{volume}{7}},
  \bibinfo{pages}{1723} (\bibinfo{year}{1964}).

\bibitem[{\citenamefont{Kraichnan}(1965)}]{article:Kraichnan:1965}
\bibinfo{author}{\bibfnamefont{R.}~\bibnamefont{Kraichnan}},
  \bibinfo{journal}{Phys. Fluids} \textbf{\bibinfo{volume}{8}},
  \bibinfo{pages}{575} (\bibinfo{year}{1965}).

\bibitem[{\citenamefont{Kraichnan}(1966)}]{article:Kraichnan:1966}
\bibinfo{author}{\bibfnamefont{R.}~\bibnamefont{Kraichnan}},
  \bibinfo{journal}{Phys. Fluids} \textbf{\bibinfo{volume}{9}},
  \bibinfo{pages}{1728} (\bibinfo{year}{1966}).

\bibitem[{\citenamefont{Leslie}(1972)}]{book:Leslie:1972}
\bibinfo{author}{\bibfnamefont{D.}~\bibnamefont{Leslie}},
  \emph{\bibinfo{title}{Developments in the theory of turbulence}}
  (\bibinfo{publisher}{Clarendon Press}, \bibinfo{address}{Oxford},
  \bibinfo{year}{1972}).

\bibitem[{\citenamefont{Wyld}(1961)}]{article:Wyld:1961}
\bibinfo{author}{\bibfnamefont{H.}~\bibnamefont{Wyld}}, \bibinfo{journal}{Ann.
  Phys.} \textbf{\bibinfo{volume}{14}}, \bibinfo{pages}{143}
  (\bibinfo{year}{1961}).

\bibitem[{\citenamefont{Martin et~al.}(1973)\citenamefont{Martin, Siggia, and
  Rose}}]{article:Rose:1973}
\bibinfo{author}{\bibfnamefont{P.}~\bibnamefont{Martin}},
  \bibinfo{author}{\bibfnamefont{E.}~\bibnamefont{Siggia}}, \bibnamefont{and}
  \bibinfo{author}{\bibfnamefont{H.}~\bibnamefont{Rose}},
  \bibinfo{journal}{Phys. Rev. A} \textbf{\bibinfo{volume}{8}},
  \bibinfo{pages}{423} (\bibinfo{year}{1973}).

\bibitem[{\citenamefont{Phythian}(1977)}]{article:Phythian:1977}
\bibinfo{author}{\bibfnamefont{R.}~\bibnamefont{Phythian}},
  \bibinfo{journal}{J. Phys. A} \textbf{\bibinfo{volume}{10}},
  \bibinfo{pages}{777} (\bibinfo{year}{1977}).

\bibitem[{\citenamefont{Belinicher and L'vov}(1987)}]{article:Lvov:1987}
\bibinfo{author}{\bibfnamefont{V.}~\bibnamefont{Belinicher}} \bibnamefont{and}
  \bibinfo{author}{\bibfnamefont{V.}~\bibnamefont{L'vov}},
  \bibinfo{journal}{Sov. Phys. JETP} \textbf{\bibinfo{volume}{66}},
  \bibinfo{pages}{303} (\bibinfo{year}{1987}).

\bibitem[{\citenamefont{L'vov}(1991)}]{article:Lvov:1991}
\bibinfo{author}{\bibfnamefont{V.}~\bibnamefont{L'vov}},
  \bibinfo{journal}{Phys. Rep.} \textbf{\bibinfo{volume}{207}},
  \bibinfo{pages}{2} (\bibinfo{year}{1991}).

\bibitem[{\citenamefont{Gkioulekas}(2007)}]{article:Gkioulekas:2007}
\bibinfo{author}{\bibfnamefont{E.}~\bibnamefont{Gkioulekas}},
  \bibinfo{journal}{Physica D} \textbf{\bibinfo{volume}{226}},
  \bibinfo{pages}{151} (\bibinfo{year}{2007}).

\bibitem[{\citenamefont{L'vov and Procaccia}(1994)}]{lect:Procaccia:1994}
\bibinfo{author}{\bibfnamefont{V.}~\bibnamefont{L'vov}} \bibnamefont{and}
  \bibinfo{author}{\bibfnamefont{I.}~\bibnamefont{Procaccia}}, in
  \emph{\bibinfo{booktitle}{Fluctuating Geometries in Statistical Mechanics and
  Field Theory, Proceedings of the {L}es {H}ouches 1994 {S}ummer {S}chool of
  Theoretical Physics}}, edited by
  \bibinfo{editor}{\bibfnamefont{F.}~\bibnamefont{David}} \bibnamefont{and}
  \bibinfo{editor}{\bibfnamefont{P.}~\bibnamefont{Ginsparg}}
  (\bibinfo{publisher}{North-Holland}, \bibinfo{address}{Amsterdam},
  \bibinfo{year}{1994}).

\bibitem[{\citenamefont{L'vov and
  Procaccia}(1995{\natexlab{a}})}]{article:Procaccia:1995:1}
\bibinfo{author}{\bibfnamefont{V.}~\bibnamefont{L'vov}} \bibnamefont{and}
  \bibinfo{author}{\bibfnamefont{I.}~\bibnamefont{Procaccia}},
  \bibinfo{journal}{Phys. Rev. E} \textbf{\bibinfo{volume}{52}},
  \bibinfo{pages}{3840} (\bibinfo{year}{1995}{\natexlab{a}}).

\bibitem[{\citenamefont{L'vov and
  Procaccia}(1995{\natexlab{b}})}]{article:Procaccia:1995:2}
\bibinfo{author}{\bibfnamefont{V.}~\bibnamefont{L'vov}} \bibnamefont{and}
  \bibinfo{author}{\bibfnamefont{I.}~\bibnamefont{Procaccia}},
  \bibinfo{journal}{Phys. Rev. E} \textbf{\bibinfo{volume}{52}},
  \bibinfo{pages}{3858} (\bibinfo{year}{1995}{\natexlab{b}}).

\bibitem[{\citenamefont{L'vov and
  Procaccia}(1996{\natexlab{a}})}]{article:Procaccia:1996}
\bibinfo{author}{\bibfnamefont{V.}~\bibnamefont{L'vov}} \bibnamefont{and}
  \bibinfo{author}{\bibfnamefont{I.}~\bibnamefont{Procaccia}},
  \bibinfo{journal}{Phys. Rev. E} \textbf{\bibinfo{volume}{53}},
  \bibinfo{pages}{3468} (\bibinfo{year}{1996}{\natexlab{a}}).

\bibitem[{\citenamefont{L'vov and
  Procaccia}(1996{\natexlab{b}})}]{article:Procaccia:1996:1}
\bibinfo{author}{\bibfnamefont{V.}~\bibnamefont{L'vov}} \bibnamefont{and}
  \bibinfo{author}{\bibfnamefont{I.}~\bibnamefont{Procaccia}},
  \bibinfo{journal}{Phys. Rev. Lett.} \textbf{\bibinfo{volume}{76}},
  \bibinfo{pages}{2898} (\bibinfo{year}{1996}{\natexlab{b}}).

\bibitem[{\citenamefont{L'vov and
  Procaccia}(1996{\natexlab{c}})}]{article:Procaccia:1996:2}
\bibinfo{author}{\bibfnamefont{V.}~\bibnamefont{L'vov}} \bibnamefont{and}
  \bibinfo{author}{\bibfnamefont{I.}~\bibnamefont{Procaccia}},
  \bibinfo{journal}{Phys. Rev. Lett.} \textbf{\bibinfo{volume}{77}},
  \bibinfo{pages}{3541} (\bibinfo{year}{1996}{\natexlab{c}}).

\bibitem[{\citenamefont{L'vov and
  Procaccia}(1996{\natexlab{d}})}]{article:Procaccia:1996:3}
\bibinfo{author}{\bibfnamefont{V.}~\bibnamefont{L'vov}} \bibnamefont{and}
  \bibinfo{author}{\bibfnamefont{I.}~\bibnamefont{Procaccia}},
  \bibinfo{journal}{Phys. Rev. E} \textbf{\bibinfo{volume}{54}},
  \bibinfo{pages}{6268} (\bibinfo{year}{1996}{\natexlab{d}}).

\bibitem[{\citenamefont{L'vov et~al.}(1997)\citenamefont{L'vov, Podivilov, and
  Procaccia}}]{article:Procaccia:1997}
\bibinfo{author}{\bibfnamefont{V.}~\bibnamefont{L'vov}},
  \bibinfo{author}{\bibfnamefont{E.}~\bibnamefont{Podivilov}},
  \bibnamefont{and}
  \bibinfo{author}{\bibfnamefont{I.}~\bibnamefont{Procaccia}},
  \bibinfo{journal}{Phys. Rev. E} \textbf{\bibinfo{volume}{55}},
  \bibinfo{pages}{7030} (\bibinfo{year}{1997}).

\bibitem[{\citenamefont{L'vov and Procaccia}(1998)}]{article:Procaccia:1998}
\bibinfo{author}{\bibfnamefont{V.}~\bibnamefont{L'vov}} \bibnamefont{and}
  \bibinfo{author}{\bibfnamefont{I.}~\bibnamefont{Procaccia}},
  \bibinfo{journal}{Physica A} \textbf{\bibinfo{volume}{257}},
  \bibinfo{pages}{165} (\bibinfo{year}{1998}).

\bibitem[{\citenamefont{Belinicher
  et~al.}(1998{\natexlab{a}})\citenamefont{Belinicher, L'vov, and
  Procaccia}}]{article:Procaccia:1998:1}
\bibinfo{author}{\bibfnamefont{V.}~\bibnamefont{Belinicher}},
  \bibinfo{author}{\bibfnamefont{V.}~\bibnamefont{L'vov}}, \bibnamefont{and}
  \bibinfo{author}{\bibfnamefont{I.}~\bibnamefont{Procaccia}},
  \bibinfo{journal}{Physica A} \textbf{\bibinfo{volume}{254}},
  \bibinfo{pages}{215} (\bibinfo{year}{1998}{\natexlab{a}}).

\bibitem[{\citenamefont{Belinicher
  et~al.}(1998{\natexlab{b}})\citenamefont{Belinicher, L'vov, Pomyalov, and
  Procaccia}}]{article:Procaccia:1998:2}
\bibinfo{author}{\bibfnamefont{V.}~\bibnamefont{Belinicher}},
  \bibinfo{author}{\bibfnamefont{V.}~\bibnamefont{L'vov}},
  \bibinfo{author}{\bibfnamefont{A.}~\bibnamefont{Pomyalov}}, \bibnamefont{and}
  \bibinfo{author}{\bibfnamefont{I.}~\bibnamefont{Procaccia}},
  \bibinfo{journal}{J. Stat. Phys.} \textbf{\bibinfo{volume}{93}},
  \bibinfo{pages}{797} (\bibinfo{year}{1998}{\natexlab{b}}).

\bibitem[{\citenamefont{L'vov and Procaccia}(2000)}]{article:Procaccia:2000}
\bibinfo{author}{\bibfnamefont{V.}~\bibnamefont{L'vov}} \bibnamefont{and}
  \bibinfo{author}{\bibfnamefont{I.}~\bibnamefont{Procaccia}},
  \bibinfo{journal}{Phys. Rev. E} \textbf{\bibinfo{volume}{62}},
  \bibinfo{pages}{8037} (\bibinfo{year}{2000}).

\bibitem[{\citenamefont{Kolmogorov}(1962)}]{article:Kolmogorov:1962}
\bibinfo{author}{\bibfnamefont{A.}~\bibnamefont{Kolmogorov}},
  \bibinfo{journal}{J. Fluid. Mech.} \textbf{\bibinfo{volume}{13}},
  \bibinfo{pages}{237} (\bibinfo{year}{1962}).

\bibitem[{\citenamefont{Oboukhov}(1962)}]{article:Oboukhov:1962}
\bibinfo{author}{\bibfnamefont{A.}~\bibnamefont{Oboukhov}},
  \bibinfo{journal}{J. Fluid. Mech.} \textbf{\bibinfo{volume}{13}},
  \bibinfo{pages}{77} (\bibinfo{year}{1962}).

\bibitem[{\citenamefont{Zhou}(2010)}]{article:Zhou:2010}
\bibinfo{author}{\bibfnamefont{Y.}~\bibnamefont{Zhou}}, \bibinfo{journal}{Phys.
  Rep.} \textbf{\bibinfo{volume}{488}}, \bibinfo{pages}{1}
  (\bibinfo{year}{2010}).

\bibitem[{\citenamefont{Giles}(2001)}]{article:Giles:2001}
\bibinfo{author}{\bibfnamefont{M.}~\bibnamefont{Giles}}, \bibinfo{journal}{J.
  Phys. A} \textbf{\bibinfo{volume}{34}}, \bibinfo{pages}{4389}
  (\bibinfo{year}{2001}).

\bibitem[{\citenamefont{Yakhot}(1981)}]{article:Yakhot:1981}
\bibinfo{author}{\bibfnamefont{V.}~\bibnamefont{Yakhot}},
  \bibinfo{journal}{Phys. Rev. A} \textbf{\bibinfo{volume}{23}},
  \bibinfo{pages}{1486} (\bibinfo{year}{1981}).

\bibitem[{\citenamefont{Adzhemyan et~al.}(2002)\citenamefont{Adzhemyan,
  Antonov, Kompaniets, and Vasil'ev}}]{article:Vasilev:2002}
\bibinfo{author}{\bibfnamefont{L.}~\bibnamefont{Adzhemyan}},
  \bibinfo{author}{\bibfnamefont{N.}~\bibnamefont{Antonov}},
  \bibinfo{author}{\bibfnamefont{M.}~\bibnamefont{Kompaniets}},
  \bibnamefont{and} \bibinfo{author}{\bibfnamefont{A.}~\bibnamefont{Vasil'ev}},
  \bibinfo{journal}{Acta Phys. Slovaka} \textbf{\bibinfo{volume}{52}},
  \bibinfo{pages}{565} (\bibinfo{year}{2002}).

\bibitem[{\citenamefont{Adzhemyan et~al.}(2003)\citenamefont{Adzhemyan,
  Antonov, Kompaniets, and Vasil'ev}}]{article:Vasilev:2003}
\bibinfo{author}{\bibfnamefont{L.}~\bibnamefont{Adzhemyan}},
  \bibinfo{author}{\bibfnamefont{N.}~\bibnamefont{Antonov}},
  \bibinfo{author}{\bibfnamefont{M.}~\bibnamefont{Kompaniets}},
  \bibnamefont{and} \bibinfo{author}{\bibfnamefont{A.}~\bibnamefont{Vasil'ev}},
  \bibinfo{journal}{Int. J. Mod. Phys. B} \textbf{\bibinfo{volume}{17(10)}},
  \bibinfo{pages}{2137} (\bibinfo{year}{2003}).

\bibitem[{\citenamefont{Canet and Chate}(2007)}]{article:Chate:2007}
\bibinfo{author}{\bibfnamefont{L.}~\bibnamefont{Canet}} \bibnamefont{and}
  \bibinfo{author}{\bibfnamefont{H.}~\bibnamefont{Chate}}, \bibinfo{journal}{J.
  Phys. A} \textbf{\bibinfo{volume}{40}}, \bibinfo{pages}{1937}
  (\bibinfo{year}{2007}).

\bibitem[{\citenamefont{Canet et~al.}(2011)\citenamefont{Canet, Chate, and
  Delamotte}}]{article:Delamotte:2011}
\bibinfo{author}{\bibfnamefont{L.}~\bibnamefont{Canet}},
  \bibinfo{author}{\bibfnamefont{H.}~\bibnamefont{Chate}}, \bibnamefont{and}
  \bibinfo{author}{\bibfnamefont{B.}~\bibnamefont{Delamotte}},
  \bibinfo{journal}{J. Phys. A} \textbf{\bibinfo{volume}{44}},
  \bibinfo{pages}{495001} (\bibinfo{year}{2011}).

\bibitem[{\citenamefont{Canet et~al.}(2015)\citenamefont{Canet, Delamotte, and
  Wschebor}}]{article:Wschebor:2015}
\bibinfo{author}{\bibfnamefont{L.}~\bibnamefont{Canet}},
  \bibinfo{author}{\bibfnamefont{B.}~\bibnamefont{Delamotte}},
  \bibnamefont{and} \bibinfo{author}{\bibfnamefont{N.}~\bibnamefont{Wschebor}},
  \bibinfo{journal}{Phys. Rev. E} \textbf{\bibinfo{volume}{91}},
  \bibinfo{pages}{053004} (\bibinfo{year}{2015}).

\bibitem[{\citenamefont{Canet et~al.}(2016)\citenamefont{Canet, Delamotte, and
  Wschebor}}]{article:Wschebor:2016}
\bibinfo{author}{\bibfnamefont{L.}~\bibnamefont{Canet}},
  \bibinfo{author}{\bibfnamefont{B.}~\bibnamefont{Delamotte}},
  \bibnamefont{and} \bibinfo{author}{\bibfnamefont{N.}~\bibnamefont{Wschebor}},
  \bibinfo{journal}{Phys. Rev. E} \textbf{\bibinfo{volume}{93}},
  \bibinfo{pages}{063101} (\bibinfo{year}{2016}).

\bibitem[{\citenamefont{Eyink}(1993{\natexlab{a}})}]{article:Eyink:1993:1}
\bibinfo{author}{\bibfnamefont{G.}~\bibnamefont{Eyink}},
  \bibinfo{journal}{Phys. Rev. E} \textbf{\bibinfo{volume}{48}},
  \bibinfo{pages}{1823} (\bibinfo{year}{1993}{\natexlab{a}}).

\bibitem[{\citenamefont{Antonov}(2006)}]{article:Antonov:2006}
\bibinfo{author}{\bibfnamefont{N.}~\bibnamefont{Antonov}}, \bibinfo{journal}{J.
  Phys. A} \textbf{\bibinfo{volume}{39}}, \bibinfo{pages}{7825}
  (\bibinfo{year}{2006}).

\bibitem[{\citenamefont{Antonov and Kostenko}(2015)}]{article:Kostenko:2015}
\bibinfo{author}{\bibfnamefont{N.}~\bibnamefont{Antonov}} \bibnamefont{and}
  \bibinfo{author}{\bibfnamefont{M.}~\bibnamefont{Kostenko}},
  \bibinfo{journal}{Phys. Rev. E} \textbf{\bibinfo{volume}{92}},
  \bibinfo{pages}{053013} (\bibinfo{year}{2015}).

\bibitem[{\citenamefont{Gkioulekas and
  Tung}(2005{\natexlab{a}})}]{article:Tung:2005}
\bibinfo{author}{\bibfnamefont{E.}~\bibnamefont{Gkioulekas}} \bibnamefont{and}
  \bibinfo{author}{\bibfnamefont{K.}~\bibnamefont{Tung}},
  \bibinfo{journal}{Discrete Contin. Dyn. Syst. Ser. B}
  \textbf{\bibinfo{volume}{5}}, \bibinfo{pages}{79}
  (\bibinfo{year}{2005}{\natexlab{a}}).

\bibitem[{\citenamefont{Gkioulekas and
  Tung}(2005{\natexlab{b}})}]{article:Tung:2005:1}
\bibinfo{author}{\bibfnamefont{E.}~\bibnamefont{Gkioulekas}} \bibnamefont{and}
  \bibinfo{author}{\bibfnamefont{K.}~\bibnamefont{Tung}},
  \bibinfo{journal}{Discrete Contin. Dyn. Syst. Ser. B}
  \textbf{\bibinfo{volume}{5}}, \bibinfo{pages}{103}
  (\bibinfo{year}{2005}{\natexlab{b}}).

\bibitem[{\citenamefont{Gkioulekas}(2008)}]{article:Gkioulekas:2008:1}
\bibinfo{author}{\bibfnamefont{E.}~\bibnamefont{Gkioulekas}},
  \bibinfo{journal}{Phys. Rev. E} \textbf{\bibinfo{volume}{78}},
  \bibinfo{pages}{066302} (\bibinfo{year}{2008}).

\bibitem[{\citenamefont{Gkioulekas}(2010)}]{article:Gkioulekas:p14}
\bibinfo{author}{\bibfnamefont{E.}~\bibnamefont{Gkioulekas}},
  \bibinfo{journal}{Phys. Rev. E} \textbf{\bibinfo{volume}{82}},
  \bibinfo{pages}{046304} (\bibinfo{year}{2010}).

\bibitem[{\citenamefont{Frisch}(1991)}]{article:Frisch:1991}
\bibinfo{author}{\bibfnamefont{U.}~\bibnamefont{Frisch}},
  \bibinfo{journal}{Proc. R. Soc. Lond. A} \textbf{\bibinfo{volume}{434}},
  \bibinfo{pages}{89} (\bibinfo{year}{1991}).

\bibitem[{\citenamefont{Eyink}(1993{\natexlab{b}})}]{article:Eyink:1993}
\bibinfo{author}{\bibfnamefont{G.}~\bibnamefont{Eyink}},
  \bibinfo{journal}{Phys. Lett. A} \textbf{\bibinfo{volume}{172}},
  \bibinfo{pages}{335} (\bibinfo{year}{1993}{\natexlab{b}}).

\bibitem[{\citenamefont{Eyink}(1995)}]{article:Eyink:1995:3}
\bibinfo{author}{\bibfnamefont{G.}~\bibnamefont{Eyink}},
  \bibinfo{journal}{Chaos, Solitons and Fractals} \textbf{\bibinfo{volume}{5}},
  \bibinfo{pages}{1465} (\bibinfo{year}{1995}).

\bibitem[{\citenamefont{Falkovich and
  Zamolodchikov}(2015)}]{article:Zamolodchikov:2015}
\bibinfo{author}{\bibfnamefont{G.}~\bibnamefont{Falkovich}} \bibnamefont{and}
  \bibinfo{author}{\bibfnamefont{A.}~\bibnamefont{Zamolodchikov}},
  \bibinfo{journal}{J. Phys. A} \textbf{\bibinfo{volume}{48}},
  \bibinfo{pages}{18FT02} (\bibinfo{year}{2015}).

\bibitem[{\citenamefont{Eyink}(2005)}]{article:Eyink:2005}
\bibinfo{author}{\bibfnamefont{G.}~\bibnamefont{Eyink}},
  \bibinfo{journal}{Physica D} \textbf{\bibinfo{volume}{207}},
  \bibinfo{pages}{91} (\bibinfo{year}{2005}).

\bibitem[{\citenamefont{Eyink and Aluie}(2009)}]{article:Aluie:2009}
\bibinfo{author}{\bibfnamefont{G.}~\bibnamefont{Eyink}} \bibnamefont{and}
  \bibinfo{author}{\bibfnamefont{H.}~\bibnamefont{Aluie}},
  \bibinfo{journal}{Phys. Fluids} \textbf{\bibinfo{volume}{21}},
  \bibinfo{pages}{115107} (\bibinfo{year}{2009}).

\bibitem[{\citenamefont{Aluie and Eyink}(2009)}]{article:Eyink:2009}
\bibinfo{author}{\bibfnamefont{H.}~\bibnamefont{Aluie}} \bibnamefont{and}
  \bibinfo{author}{\bibfnamefont{G.}~\bibnamefont{Eyink}},
  \bibinfo{journal}{Phys. Fluids} \textbf{\bibinfo{volume}{21}},
  \bibinfo{pages}{115108} (\bibinfo{year}{2009}).

\bibitem[{\citenamefont{Boffetta}(2007)}]{article:Boffetta:2007}
\bibinfo{author}{\bibfnamefont{G.}~\bibnamefont{Boffetta}},
  \bibinfo{journal}{J. Fluid. Mech.} \textbf{\bibinfo{volume}{589}},
  \bibinfo{pages}{253} (\bibinfo{year}{2007}).

\bibitem[{\citenamefont{Boffetta and Musacchio}(2010)}]{article:Musacchio:2010}
\bibinfo{author}{\bibfnamefont{G.}~\bibnamefont{Boffetta}} \bibnamefont{and}
  \bibinfo{author}{\bibfnamefont{S.}~\bibnamefont{Musacchio}},
  \bibinfo{journal}{Phys. Rev. E} \textbf{\bibinfo{volume}{82}},
  \bibinfo{pages}{016307} (\bibinfo{year}{2010}).

\bibitem[{\citenamefont{Xiao et~al.}(2009)\citenamefont{Xiao, Wan, Chen, and
  Eyink}}]{article:Eyink:2009:1}
\bibinfo{author}{\bibfnamefont{Z.}~\bibnamefont{Xiao}},
  \bibinfo{author}{\bibfnamefont{M.}~\bibnamefont{Wan}},
  \bibinfo{author}{\bibfnamefont{S.}~\bibnamefont{Chen}}, \bibnamefont{and}
  \bibinfo{author}{\bibfnamefont{G.}~\bibnamefont{Eyink}}, \bibinfo{journal}{J.
  Fluid. Mech.} \textbf{\bibinfo{volume}{619}}, \bibinfo{pages}{1}
  (\bibinfo{year}{2009}).

\end{thebibliography}
\bibliographystyle{apsrev}
\end{document}